\documentclass[a4paper,10pt]{elsarticle}
\usepackage{amsmath,amssymb,amsfonts}
\usepackage{bm}
\usepackage{algorithm}
\usepackage{algpseudocode}
\usepackage{jcomp}
\usepackage{graphicx}
\usepackage{float}
\usepackage{xcolor}
\usepackage{subcaption}
\usepackage{mathrsfs}
\newtheorem{remark}{\bf Remark}[section]
\usepackage{lineno}
\usepackage{hyperref}

\hypersetup{
  colorlinks=true,
  linkcolor=blue,      
  citecolor=green,     
  urlcolor=cyan,       
  filecolor=magenta}   

\begin{document}
 \begin{frontmatter}
  \title{Wave-appropriate reconstruction of compressible flows: physics-constrained acoustic dissipation and rank-1 entropy wave correction}
  \author{Amareshwara Sainadh Chamarthi\corref{cor1}}
  \ead{sainath@caltech.edu}
  \cortext[cor1]{Corresponding author.}
  \address{Division of Engineering and Applied Science, California Institute of Technology, Pasadena, CA, 91125, USA}
\begin{abstract}
Within the finite volume framework, the wave-appropriate reconstruction approach \cite{chamarthi2023wave,hoffmann2024centralized,chamarthi2025wave} decomposes the reconstruction procedure into characteristic wave families,  centralizing non-acoustic waves to minimize dissipation while retaining an  upwind bias for acoustic waves. In all previous implementations, the acoustic upwind parameter $\eta_a$ was fixed at its maximum value of $1.0$; however, this choice is conservative and can be further improved, motivating a systematic search for \textcolor{black}{a lower feasible value of $\eta_a$ that preserves robust stability across flow regimes while minimizing the stated accuracy objective}. To this end, the CFD solver is treated as a black box within Brent's bounded scalar minimization, which minimizes an accuracy objective for the subsonic inviscid Taylor--Green vortex subject to a stability constraint enforced by the supersonic viscous Taylor--Green vortex. Because the wave-appropriate framework leaves $\eta_a$ as the sole degree of freedom, the optimization converges in approximately 25 evaluations, with the objective function plateauing after roughly 12 iterations. The resulting optimal values,  $\eta_a^* = 0.54$ for the third-order scheme and $\eta_a^* = 0.6010$ for  the fifth-order scheme, generalize without retuning across the full range from subsonic turbulence to hypersonic flows with shocks and contact  discontinuities. \textcolor{black}{Critically, the optimized \emph{nonlinear} $N$th-order scheme consistently matches or outperforms the standard \emph{linear} $(N\!+\!2)$th-order scheme at full upwinding, where order refers to the one-dimensional face-normal reconstruction order.}

The second contribution focuses on eliminating the need for an explicit contact-discontinuity detector, which is commonly required in flows involving both shock waves and contact discontinuities. In such cases, the reconstruction deficiency appears solely within the entropy characteristic wave and can be corrected by a rank-1 update along the entropy right eigenvector. The proposed algorithm, WA-CR, relies only on the Ducros sensor and is limiter-agnostic, facilitating direct use in other schemes, such as WENO, while maintaining the same $\eta_a^*$. This approach reduces wall time by $29$--$41\%$ compared to full characteristic decomposition. To further demonstrate the method's generality, introducing a controlled acoustic bias exclusively to the normal momentum in a kinetic-energy-preserving scheme eliminates spurious vortices in periodic shear layers, confirming that the acoustic stability mechanism operates independently of the discretization framework.
\end{abstract}
 \end{frontmatter}

\section{Introduction}
\label{sec:intro}

The simulation of compressible turbulent flows demands a careful balance between two competing requirements: resolving turbulent structures with minimal artificial dissipation while maintaining numerical stability near discontinuities, such as shocks and contact discontinuities. High-order shock-capturing schemes typically address stability by applying upwind reconstruction uniformly across all flow variables, but this uniform treatment imposes unnecessary dissipation where none is required. The fundamental question is therefore not \emph{only} how much dissipation to add, but \emph{where} and \emph{to which flow structures} it should be directed.
\subsection{Background and motivation}
\label{sec:background}

\begin{figure}[H]
\centering
\includegraphics[width=1.0\textwidth]{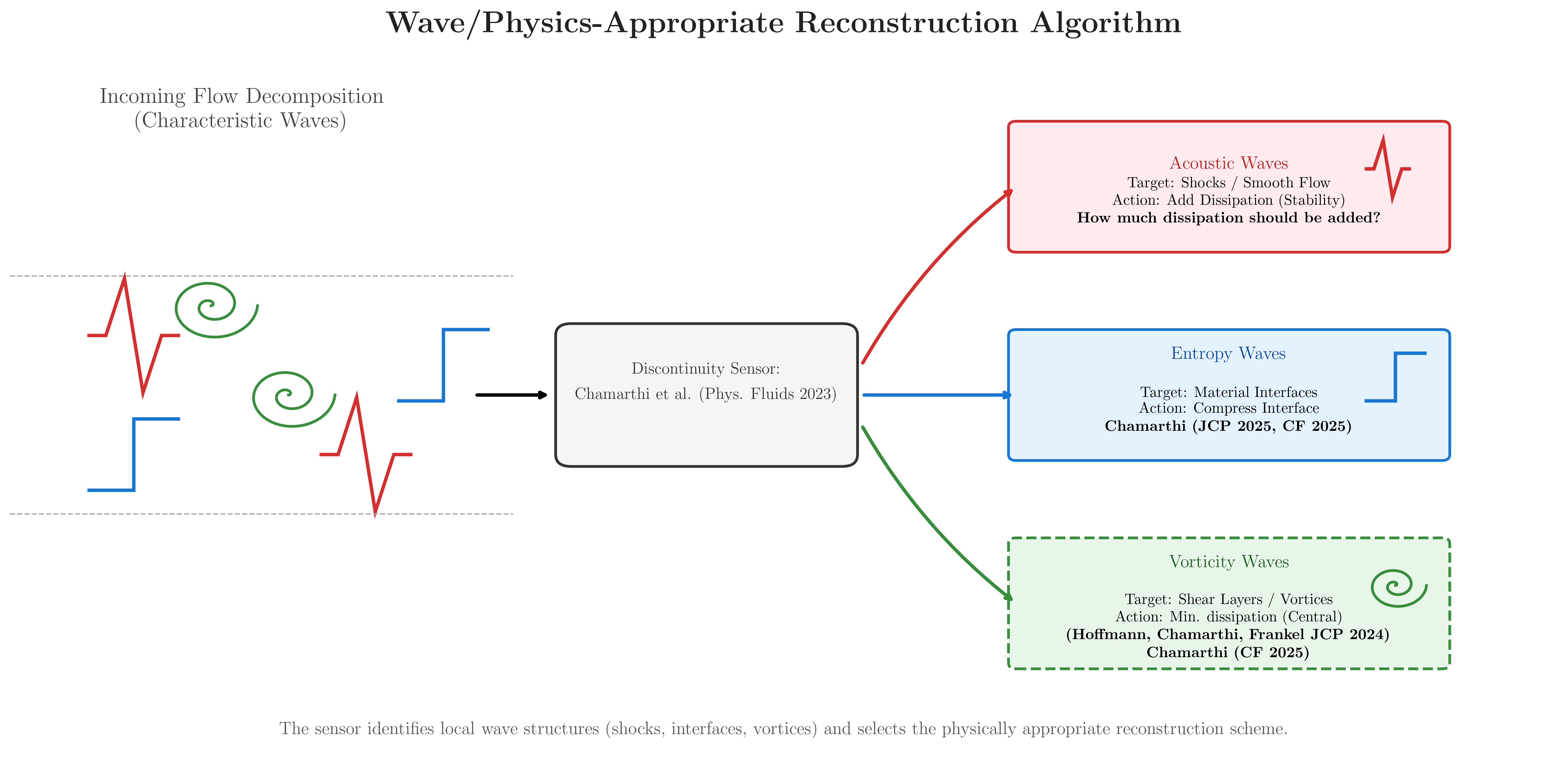}
\caption{Wave-appropriate reconstruction algorithm.}
\label{fig:physics}
\end{figure}

The wave-appropriate reconstruction framework answers this question by exploiting the characteristic structure of the governing equations. The compressible flow field is decomposed into its characteristic wave families, and each family is treated by the reconstruction scheme most appropriate to its physical character, as illustrated in Figure~\ref{fig:physics}.  For the three-dimensional compressible Euler equations, there are five characteristic waves: two acoustic waves that carry pressure fluctuations and are responsible for shocks, one entropy/contact wave that carries density jumps across material interfaces, and two shear/vortical waves that carry transverse momentum and are responsible for turbulent structures.  This decomposition is not merely a mathematical convenience; it  reflects the distinct physical roles that each wave family plays  in the flow. Each family also has distinct numerical  characteristics that require a specific reconstruction approach. The framework was developed over a series of papers~\cite{chamarthi2023wave,hoffmann2024centralized,chamarthi2024generalized,chamarthi2025physics,chamarthi2025wave}, with each contribution identifying one more wave family whose treatment could be made more physically consistent, as summarized below and in Figure~\ref{fig:physics}.

\subsubsection*{Part 1 (2023): Wave-appropriate discontinuity
sensor~\cite{chamarthi2023wave}}

The first contribution established the basic wave-appropriate framework. The key observation was that shock capturing should be performed selectively by wave type rather than uniformly across all characteristic fields. The Ducros sensor \cite{ducros1999large} detects shocks and activates nonlinear MP limiting for the acoustic and shear/vortical waves; a separate density-based criterion detects contact discontinuities, carried by the entropy wave, and activates limiting for that wave independently, regardless of the Ducros sensor state. Despite this selective limiting, all five characteristic waves were still reconstructed with the same fully upwind interpolation ($\eta_a = 1$), leaving the dissipation level of non-acoustic waves unnecessarily high. This contribution is represented by the grey box in Figure~\ref{fig:physics}.

\subsubsection*{Part 2 (2024): Centralization of vortical
waves~\cite{hoffmann2024centralized}}

The second contribution~\cite{hoffmann2024centralized} showed that shear and vortical waves, which travel at the local advection velocity and carry only transverse momentum perturbations, can be reconstructed using a central scheme ($\eta_a = 0.5$) in smooth regions. The Ducros sensor activates MP limiting for these waves near shocks, as in Part~1, but outside shock regions, the central scheme is used for the shear/vorticity waves. This reduced unnecessary dissipation on vortical structures and had a significant practical consequence: in the oblique shock impingement case of Sandham et al.~\cite{sandham2014transitional}, only the centralized scheme achieved full laminar-to-turbulent transition on wall-modeled LES grids. This work is represented by the green box in Figure~\ref{fig:physics}.

\subsubsection*{Part 3 (2025): Extension to multicomponent and multiphase
flows~\cite{chamarthi2025physics,chamarthi2025wave}}

The third contribution extended the wave-appropriate framework to multicomponent and multiphase flows, where material interfaces introduce an additional wave family that requires specialized treatment. The entropy/contact wave carries jumps in species volume fraction across material interfaces. Rather than applying a standard polynomial reconstruction to this wave, the Tangent of Hyperbola for INterface Capturing (THINC) method \textcolor{black}{\cite{xiao2005simple,xiao2011revisit}} was applied selectively to the entropy characteristic, sharpening the interface representation without introducing oscillations in the acoustic or vortical fields. This work also established a key physical constraint for 2-D and 3-D flows: the tangential velocity components are continuous across a contact discontinuity (in viscous flows \cite{batchelor1967introduction} or if there is artificial viscosity \cite{meng2018numerical}), which means the shear characteristic waves carry no jump at the interface, and a central scheme is therefore appropriate. This contribution is represented by the blue and green boxes in Figure~\ref{fig:physics}.

\subsubsection*{The remaining open question}

The preceding work in this series~\cite{chamarthi2023wave,hoffmann2024centralized,chamarthi2025wave, chamarthi2024generalized,chamarthi2025physics} established the wave-appropriate reconstruction framework: acoustic waves are reconstructed with an upwind scheme, shear and vortical waves with a central scheme, and the entropy wave with selective limiting near contact discontinuities. Notably, in all prior implementations, the acoustic upwind bias, $\eta_a \in [0.5,\,1]$, was typically set to $1.0$. \textcolor{black}{Since the non-acoustic dissipation has already been reduced by the wave-specific treatment, reducing $\eta_a < 1$ could be stable; yet, no theoretical analysis precisely predicts the practically useful lower acoustic bias.} This is due to the complex nonlinear interactions among the limiter, the Ducros sensor, the Riemann solver, and the characteristic projection, which collectively determine stability in a way that linear theory cannot capture. Consequently, $\eta_a$ must be found empirically \textcolor{black}{for the stated objective and stability tests}.

A key advantage of the wave-appropriate framework is that only a \emph{single} parameter, $\eta_a$, requires empirical determination; all other dissipation sources are fixed during design. As a result, tuning focuses solely on $\eta_a$. The present optimization process reduces to a bounded scalar minimization, which converges in about 25 CFD evaluations: the objective function typically plateaus after approximately 12 iterations, and subsequent evaluations serve only to refine $\eta_a^*$ to four decimal places.

Building on this, the present paper addresses the open question for both the third- and fifth-order schemes. It identifies the physical mechanism by which optimized nonlinear $N$th-order schemes can outperform standard $(N\!+\!2)$th-order linear schemes. In addition, it introduces a second contribution: the rank-1 entropy correction, which further reduces the computational cost of the wave-appropriate framework by removing the need for explicit contact detection. Both contributions are highlighted by the red box in Figure~\ref{fig:physics}.

A related but distinct issue concerns the stability of kinetic-energy-preserving (KEP) and entropy-stable schemes, which employ a non-dissipative central discretization of the convective flux in smooth flow regions and switch to a characteristic-based upwind or WENO scheme near shocks via the Ducros sensor. This approach has been employed in several high-fidelity compressible flow solvers~\cite{de2020sharp}, including the formulation of Subbareddy and Candler~\cite{subbareddy2009fully} and the CHAMPS immersed boundary solver of van Noordt et al.~\cite{van2022immersed}. The total numerical flux in such formulations is written as
\begin{equation}
\hat{f}^{conv} = (1 - \alpha)\hat{f}^{cent} + \alpha\hat{f}^{diss},
\end{equation}
where $\hat{f}^{cent}$ is a non-dissipative kinetic-energy-preserving flux computed directly in physical space, $\hat{f}^{diss}$ is the dissipative flux computed using characteristic-based upwind or WENO reconstruction, and $\alpha$ is the Ducros sensor output. When the Ducros sensor is inactive ($\alpha = 0$), as is the case in smooth shear-layer flows, the flux reduces to the purely central kinetic-energy-preserving scheme with zero dissipation across all wave types. Because the central component is computed in physical space without characteristic decomposition, the characteristic decomposition is invoked only through the dissipative path $\hat{f}^{diss}$ and is therefore available only when the Ducros sensor is active. The Ducros threshold has been reported to require case-by-case adjustment in practice; van Noordt et al.~\cite{van2022immersed} found that values of $\alpha < 0.0125$ generate noise in the solution while higher values cause the flow to deviate from the expected behavior, ultimately selecting $\alpha = 0.0125$ empirically. Furthermore, De Vanna et al.~\cite{de2023uranos} combined a kinetic-energy-preserving scheme with a WENO scheme using a modified Ducros sensor and similarly reported that the threshold value may be case-dependent.

Ghate and Lele~\cite{ghate2023finite} identified two important limitations of such KEP schemes. First, the resulting solutions may suffer from energy pile-up near Nyquist scales in the absence of a subgrid-scale closure, since the kinetic energy transfer associated with the inviscid fluxes is non-dissipative. Second, kinetic energy conservation holds only in a semi-discrete sense at low Mach numbers, with many past applications requiring additional de-aliasing filtering for robustness and most numerical investigations conducted at CFL numbers below 0.1~\cite{ghate2023finite}. A further limitation, not explicitly discussed in the literature to the authors' knowledge, is that the fully central acoustic treatment can be unstable in shear-layer flows, even in the absence of shocks. As demonstrated in  Section~\ref{sec:shear_layer}, the unmodified KEP scheme produces  significant oscillations in the periodic shear layer despite the flow  being entirely shock-free and the Ducros sensor remaining inactive  throughout. These limitations motivate applying the wave-appropriate principle to KEP schemes. Section~\ref{sec:WA-KEP} demonstrates that  introducing a controlled acoustic bias exclusively through the normal-momentum  component of the Riemann solver's dissipative flux, while leaving the  non-dissipative KEP treatment intact for all non-acoustic variables,  eliminates the shear-layer instability without compromising energy  conservation. This extends the wave-appropriate framework beyond reconstruction-based schemes to encompass energy-preserving formulations.

\begin{remark}[On the Ducros sensor]
 \normalfont A related point concerns the formulation of the Ducros sensor itself. The original sensor of Ducros et al.~\cite{ducros1999large} is the product of a pressure-based shock indicator and a dilatation-to-vorticity ratio. In much of the subsequent literature, only the dilatation-to-vorticity component has been retained, while the pressure-based indicator has been dropped. This simplification is consequential: the dilatation-to-vorticity ratio alone can misidentify intense vortical regions as shocked~\cite{sciacovelli2021assessment}, which may partly explain the need for threshold tuning in practice. Sciacovelli et al.~\cite{sciacovelli2021assessment} demonstrated this deficiency directly, showing that the dilatation-to-vorticity ratio alone is insufficient for reliable shock detection and that incorporating Jameson's pressure-based indicator substantially improves sensor specificity, as illustrated in Figure~\ref{passio}. Notably, this combined formulation is precisely what Ducros et al.~\cite{ducros1999large} originally proposed; the pressure-based component appears to have been inadvertently dropped in much of the subsequent literature. The present work, consistent with the authors' earlier contributions~\cite{chamarthi2023wave,hoffmann2024centralized,chamarthi2025physics}, employs the complete two-component sensor with a fixed threshold $\Omega_d > 0.01$, which Sciacovelli et al.~\cite{sciacovelli2021assessment} independently identify as appropriate. The robustness of this fixed threshold, applied uniformly across all test cases from subsonic turbulence to Mach~10 flows without case-by-case adjustment, may be attributed in part to the use of the complete two-component formulation, as detailed in Section~\ref{sec:wac}.

\begin{figure}[H]
\centering
\includegraphics[width=0.7\textwidth]{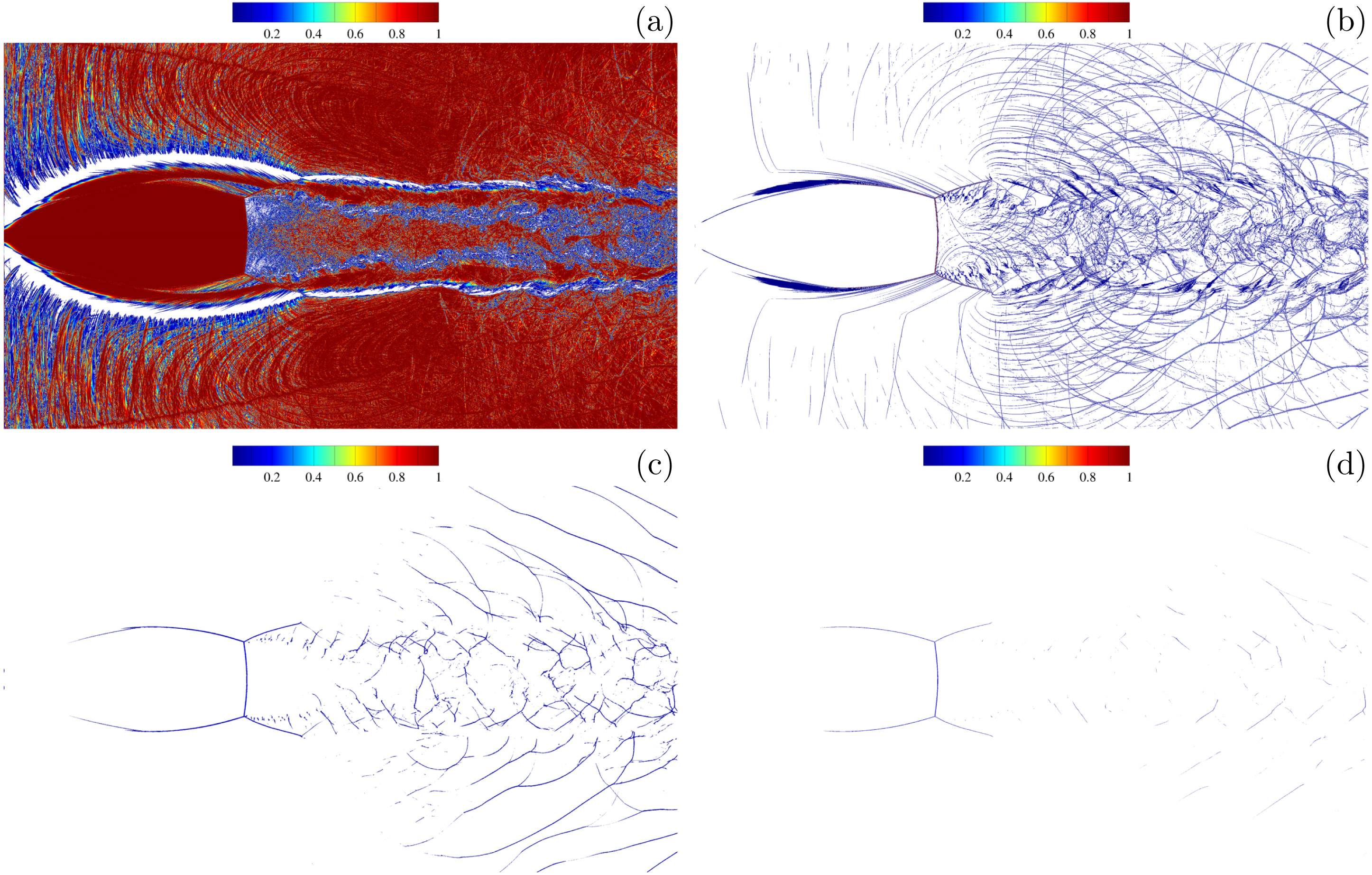}
\caption{Shock sensor components for underexpanded jet, reproduced
from Sciacovelli et al.~\cite{sciacovelli2021assessment}: (a)~dilatation-to-vorticity
ratio alone, (b)~Bhagatwala and Lele's indicator,
(c)~Jameson's pressure-based indicator, and (d)~combination of all three
components. Panel~(a) clearly misidentifies intense vortical regions as
shocked, while the combined sensor in panel~(d) correctly localizes
activation in the shock regions.}
\label{passio}
\end{figure}
\end{remark}

\subsection{Prior work on scheme optimization}
Prior work has explored data-driven optimization of numerical scheme parameters for compressible flows using surrogate-based Bayesian optimization~\cite{schranner2013physically,schranner2016optimization, winter2016iterative} and multi-objective frameworks~\cite{feng2022multi,feng2024general,feng2023deep}. The most relevant is Feng et al.~\cite{feng2024general}, which optimized four free parameters—a Ducros cutoff threshold $C_D$ and three upwind bias coefficients $(\eta_{\mathrm{eno}}, \eta_{\mathrm{lnr}}, \eta_v)$—using 100 CFD evaluations with Gaussian process surrogates. Examining the requirement for four parameters is instructive from the point of view of the current work. First, the Ducros sensor implementation in~\cite{feng2024general} employs only the dilatation-to-vorticity ratio, omitting the pressure-based shock indicator in the original formulation of Ducros et al.~\cite{ducros1999large} (see Remark~\ref{rem:ducros}). As discussed in Section~\ref{sec:wac}, the complete two-component sensor admits a fixed threshold ($\Omega_d > 0.01$) requiring no tuning, indicating $C_D$ may lack practical significance if the sensor is implemented in full. Second, the three upwind bias coefficients $(\eta_{\mathrm{eno}}, \eta_{\mathrm{lnr}}, \eta_v)$ regulate dissipation uniformly across all characteristic wave families instead of distinguishing by physical character. The wave-appropriate framework addresses two of these issues by design: shear/vortical waves are centralized ($\eta = 0.5$), and the entropy wave is treated with selective limiting, both of which are informed by eigenvector structure. Feng et al. thus empirically recovered what the wave-appropriate framework derives from first principles: their optimized $\eta_v$ converges to a small value aligning with the central treatment of vortical structures—the assignment determined by physical reasoning in the present framework. Only the acoustic upwind bias $\eta_a$ remains to be determined. Consequently, the degrees of freedom are reduced from four empirically optimized parameters requiring 100 surrogate-assisted evaluations to a single bounded scalar addressed with approximately 25 direct evaluations via Brent's method. This reduction is due to resolving three degrees of freedom through physical reasoning rather than to the optimizer's efficiency. Additionally, Feng et al. calibrate with three configurations—the inviscid Taylor-Green vortex and both 2D and 3D implosion problems—because their dispersion objective cannot be isolated with the Taylor-Green vortex alone, requiring data-driven dissipation learning for wave specificity. The present optimization uses only Taylor-Green vortex variants: the subsonic inviscid case supplies the accuracy objective, while the supersonic viscous case sets the stability constraint. This sufficiency results from addressing all non-acoustic waves physically, narrowing the remaining task to determining the acoustic stability boundary.

\textcolor{black}{A related class of optimization methods improves the spectral behavior of the discrete operator~\cite{tam1993dispersion,lele1992compact}. Dispersion-relation-preserving, modified-wavenumber, and adaptive-dissipation approaches tune the stencil to reduce numerical dispersion and dissipation errors while maintaining robustness near shocks~\cite{pirozzoli2006spectral, li2022class,huang2023fivepoint}. Such analyses are useful, but they are usually based on the scalar linear-advection equation. The resulting optimized stencil is then used for the Euler or Navier--Stokes equations, where additional characteristic-wave constraints are present. As discussed in the author's earlier wave-appropriate studies~\cite{chamarthi2025wave,chamarthi2025physics}, the Euler system contains acoustic, entropy/contact, and shear/vortical waves, and the effective accuracy of a scheme depends not only on formal order or spectral behavior, but also on which wave families receive numerical dissipation. The present work follows this viewpoint: rather than optimizing a linearized stencil in Fourier space, the full nonlinear CFD solver is treated as a black box and only the remaining acoustic bias $\eta_a$ is calibrated using a kinetic-energy objective and nonlinear stability tests. Thus, $\eta_a^*$ should be interpreted as a physics-constrained solver-level calibration, not as a replacement for spectral analysis. Including spectral properties as an additional constraint or objective function may be possible, but is beyond the scope of this paper.}

\textcolor{black}{A recent claim in the machine-learning-for-CFD literature is that classical shock-capturing schemes lack the integration of physical knowledge in cell-face reconstruction and require learned operators~\cite{feng2022multi,bezgin2025ml}. A complementary view from the classical numerical methods community is provided by Van Leer~\cite{van2006upwind}, who stated that a numerical method should use upwinding for advection and a central scheme for subsonic acoustic propagation, thus identifying the need for wave-specific treatment well before the application of machine learning to CFD. Roe~\cite{roe1986discrete} observed that the two-dimensional Euler equations involve eight distinct reconstruction parameters corresponding to various wave families and sweep directions, parameters inherently absent from schemes designed for the scalar linear advection equation~\cite{chamarthi2025wave}. The wave-appropriate reconstruction framework~\cite{chamarthi2023wave,chamarthi2025wave} was developed from physical reasoning regarding the Euler equations’ characteristic structure, and it was subsequently noted that it closely aligns with Roe's multidimensional upwinding concepts~\cite{roe1986discrete}. The advantage here is clear: since all wave-specific dissipation requirements are addressed at the design stage, only $\eta_a$ remains to be optimized, reducing CFD evaluations compared to end-to-end DNS training.} All comparisons and discussions in this paper refer to the published descriptions of the cited methods. The author has aimed to describe prior work accurately and objectively; any differences highlighted here reflect technical distinctions only, not judgments of the quality of the respective work.

\subsection{Contributions of this work}

The first contribution identifies an optimized acoustic upwind bias $\eta_a^*$ for the wave-appropriate framework. The optimization yields $\eta_a^*=0.54$ for the third-order scheme and $\eta_a^*=0.6010$ for the fifth-order scheme. \textcolor{black}{These values minimize the stated kinetic-energy objective subject to the stability tests used in this work.} Both values transfer without retuning from subsonic turbulence to supersonic flows at no additional computational cost.

\begin{remark}[On $\eta_a^*$]
\normalfont The value $\eta_a = 0.5$ (fully central acoustics) was tested during the development of the original wave-appropriate framework~\cite{chamarthi2023wave} and found to be unstable, though this result was not reported therein. This prompted the present optimization, aimed at identifying \textcolor{black}{a low-dissipation feasible value in the interval  $(0.5,\, 1]$ for the stated accuracy objective}. A further motivation arises in hypersonic flows, where boundary-layer transition is dominated by acoustic instability modes (Mack modes), and the acoustic upwind bias directly controls the numerical dissipation acting on these modes. Lowering $\eta_a^*$ \textcolor{black}{relative to the fully upwind value} could therefore enhance the fidelity of hypersonic transition simulations such as those in~\cite{hoffmann2024centralized}. This potential improvement remains an avenue for future work.
\end{remark}
The second contribution shows that the contact discontinuity detector used in the prior approach~\cite{chamarthi2024generalized} is unnecessary. Near a contact discontinuity, the reconstruction of a conservative variable suffers from an error in the entropy characteristic amplitude. This error can be corrected by performing a rank-1 update along the entropy right eigenvector at a cost of one dot product and five multiply-accumulate operations per interface. The resulting algorithm uses only the Ducros sensor and resolves contacts algebraically rather than through explicit detection. This approach reduces wall time by 29--41\% across various benchmark configurations. Notably, the correction is limiter-agnostic and can be seamlessly integrated into any scheme that utilizes conservative-characteristic reconstruction. Unlike previous optimization efforts that modify existing dissipation operators, this rank-1 entropy wave correction introduces a novel algebraic mechanism for contact resolution. Consequently, any conservative-characteristic scheme can now inherit wave-appropriate contact treatment without the need for explicit detection or additional empirical parameters.

The third contribution builds on this by demonstrating that the wave-appropriate principle extends beyond reconstruction-based schemes. For instance, the kinetic-energy-preserving (KEP) scheme~\cite{chandrashekar2013kinetic} employs first-order piecewise-constant states and constructs a non-dissipative flux using logarithmic-mean density and arithmetic velocity and pressure averages. By construction, this approach ensures zero numerical dissipation across all wave families (fluxes). However, acoustic dissipation is reintroduced at the Riemann solver level through the normal-momentum component of the dissipative flux. This targeted correction eliminates the spurious vortices produced by the unmodified KEP scheme in shear-layer flows. It confirms that the acoustic stability mechanism identified by the optimization is independent of the discretization framework.

\subsection{Organization}

The paper proceeds as follows. Section~\ref{sec:governingEquations} states the governing equations. Section~\ref{sec:num} presents the numerical methods, including the wave-appropriate reconstruction framework and the parameterized acoustic upwind bias $\eta_a$. Section~\ref{sec:optimization} presents the physics-constrained optimization formulation and results. Section~\ref{sec:megcc} presents the conservative variable reconstruction with rank-1 entropy wave correction. Section~\ref{sec:WA-KEP} extends the wave-appropriate acoustic dissipation principle to kinetic-energy-preserving schemes. Section~\ref{sec:results} validates the proposed schemes across a suite of benchmark test cases. Concluding remarks are given in Section~\ref{sec:conclusions}.

\section{Governing Equations} \label{sec:governingEquations}

In this study, the three-dimensional compressible Navier-Stokes equations are solved in Cartesian coordinates:
\begin{equation}\label{eqn:cns}
\frac{\partial \mathbf{U}}{\partial t}+\frac{\partial \mathbf{F}^c}{\partial x}+\frac{\partial \mathbf{G}^c}{\partial y}+\frac{\partial \mathbf{H}^c}{\partial z}=\frac{\partial \mathbf{F}^v}{\partial x}+\frac{\partial \mathbf{G}^v}{\partial y}+\frac{\partial \mathbf{H}^v}{\partial z},
\end{equation}
where  $\mathbf{U}$ is the conservative variable vector, $\mathbf{F}^c$, $\mathbf{G}^c$,  $\mathbf{H}^c$ and $\mathbf{F}^v$, $\mathbf{G}^v$, $\mathbf{H}^v$, are the convective (superscript $c$) and viscous (superscript $v$) flux vectors in each coordinate direction, respectively.  The conservative variable, convective, and viscous flux vectors are given as:

\begin{subequations}
    \begin{gather}
        \mathbf{U} = \begin{pmatrix}
        \rho \\
        \rho u \\
        \rho v \\
        \rho w \\
        \rho E
        \end{pmatrix},
        \quad
        \mathbf{F}^c = \begin{pmatrix}
\rho u \\
\rho u^2 + p \\
\rho u v \\
\rho u w \\
\rho u H
\end{pmatrix}, \quad
\mathbf{G}^c = \begin{pmatrix}
\rho v \\
\rho v u \\
\rho v^2 + p \\
\rho v w \\
\rho v H
\end{pmatrix}, \quad
\mathbf{H}^c = \begin{pmatrix}
\rho w \\
\rho w u \\
\rho w v \\
\rho w^2 + p \\
\rho w H
\end{pmatrix},
        \tag{\theequation a--\theequation d}
    \end{gather}
\end{subequations}

\begin{equation}
\begin{array}{l}\label{eqn-visc}
\mathbf{F}^v=\left[0, \tau_{x x}, \tau_{x y}, \tau_{x z}, u \tau_{x x}+v \tau_{x y}+w \tau_{x z}-q_{x}\right]^{T}, \\
\mathbf{G}^v=\left[0, \tau_{x y}, \tau_{y y}, \tau_{y z}, u \tau_{y x}+v \tau_{y y}+w \tau_{y z}-q_{y}\right]^{T}, \\
\mathbf{H}^v=\left[0, \tau_{x z}, \tau_{y z}, \tau_{z z}, u \tau_{z x}+v \tau_{z y}+w \tau_{z z}-q_{z}\right]^{T},
\end{array}
\end{equation}
\noindent where $\rho$ is density, $u$, $v$, and $w$ are the velocities in the $x$, $y$, and $z$ directions, respectively, $p$ is the pressure, $E = e + \left(u^2 + v^2 + w^2 \right)/2$ is the specific total energy, and $H = E + p/\rho$ is the specific total enthalpy. The equation of state is for a calorically perfect gas, so that $e = \frac{p}{\rho(\gamma - 1)}$ is the internal energy, where $\gamma$ is the ratio of specific heats. The components of the viscous stress tensor $\tau$ and the heat flux $q$ are defined in tensor notations as:
\begin{equation}\label{eqn:5-stress}
\tau_{i j}=\frac{\mu}{\operatorname{Re}}\left(\frac{\partial u_{i}}{\partial x_{j}}+\frac{\partial u_{j}}{\partial x_{i}}-\frac{2}{3} \frac{\partial u_{k}}{\partial x_{k}} \delta_{i j}\right),
\end{equation}
\begin{equation}\label{eqn:6-heat}
\begin{aligned}
\mathrm{q}_{i}=-\frac{\mu}{\operatorname{Re Pr Ma}(\gamma-1)} \frac{\partial T}{\partial x_{i}}, \quad T= \text{Ma}^{2} \gamma \frac{p}{\rho},
\end{aligned}
\end{equation}
where $\mu$ is the dynamic viscosity, $T$ is the temperature, $\mathrm{Ma}$ and $Re$ are the Mach number and Reynolds number, and Pr is the Prandtl number.

\section{Numerical methods}\label{sec:num}

Using a conservative numerical method, the governing equations cast in semi-discrete form for a Cartesian cell $I_{i,j,k} = \left[ x_{i-\frac{1}{2}}, x_{i+\frac{1}{2}} \right] \times \left[ y_{j-\frac{1}{2}}, y_{j+\frac{1}{2}} \right] \times \left[ z_{k-\frac{1}{2}}, z_{k+\frac{1}{2}} \right]$ can be expressed via the following ordinary differential equation: 

\begin{align}
    \begin{aligned}
        \frac{\text{d}}{\text{d} t} \check{\mathbf{U}}_{i,j,k} = \mathbf{Res}_{i,j,k} = &- \left. \frac{\text{d} \check{\mathbf{F}}^c}{\text{d} x} \right|_{i,j,k} - \left. \frac{\text{d} \check{\mathbf{G}}^c}{\text{d} y} \right|_{i,j,k} - \left. \frac{\text{d} \check{\mathbf{H}}^c}{\text{d} z} \right|_{i,j,k} \\ 
        &+ \left. \frac{\text{d} \check{\mathbf{F}}^v}{\text{d} x} \right|_{i,j,k} + \left. \frac{\text{d} \check{\mathbf{G}}^v}{\text{d} y} \right|_{i,j,k} + \left. \frac{\text{d} \check{\mathbf{H}}^v}{\text{d} z} \right|_{i,j,k},
    \end{aligned}
\end{align}

\noindent where the check accent, $\check{(\cdot)}$, indicates a numerical approximation of physical quantities, $\mathbf{Res}_{i,j,k}$ is the residual function, and the remaining terms are cell centre numerical flux derivatives of the physical fluxes in Eq.~(\ref{eqn:cns}). For brevity, we continue with only the $x$-direction as it is straightforward to extend to all three dimensions in a dimension-by-dimension manner. \textcolor{black}{The labels WA-3 and WA-5 used later in the paper refer to the formal one-dimensional face-normal reconstruction order used in each coordinate direction, as is often done in the literature~\cite{deng2019fifth,fu2019low}. The resulting multidimensional finite-volume discretization should not be interpreted as strictly third- or fifth-order accurate, since high-order Gaussian quadrature is not used for the flux integration.} The indices  $j$ and $k$ are also dropped for simplicity. In the following sections, viscous and convective flux discretization will be presented.

\subsection{Viscous flux spatial discretization}\label{sec:visc}

Viscous fluxes, $\check{\mathbf{F}}^v$, are computed using the fourth-order $\alpha$-damping scheme of Nishikawa \cite{Nishikawa2010}. In the one-dimensional scenario, the cell center numerical viscous flux derivative is:

\begin{equation}
    \left. \frac{\mathrm{d} \check{\mathbf{F}}^v}{\mathrm{d} x} \right|_{i} = \frac{1}{\Delta x} \left( \check{\mathbf{F}}^v_{i+\frac{1}{2}} - \check{\mathbf{F}}^v_{i-\frac{1}{2}} \right).
\end{equation}

\noindent The cell interface numerical viscous flux is computed as,

\begin{equation}
    \check{\mathbf{F}}^v_{i+\frac{1}{2}} = 
    \begin{pmatrix}
        0 \\
        \tau_{i+\frac{1}{2}} \\
        \tau_{i+\frac{1}{2}} u_{i+\frac{1}{2}} + q_{i+\frac{1}{2}} \\
    \end{pmatrix},         \tau_{i+\frac{1}{2}} = \frac{4}{3} {\mu}_{i+\frac{1}{2}} \left. \frac{\partial u}{\partial x} \right|_{i+\frac{1}{2}},
        \quad
        q_{i+\frac{1}{2}} = -{\kappa}_{i+\frac{1}{2}} \left. \frac{\partial T}{\partial x} \right|_{i+\frac{1}{2}}.
\end{equation}

\noindent For an arbitrary variable, $\phi$, the $\alpha$-damping approach computes cell interface gradients as:

\begin{equation}
    \left. \frac{\partial \phi}{\partial x} \right|_{i+ \frac{1}{2}} = \frac{1}{2} \left( \left. \frac{\partial \phi}{\partial x} \right|_{i} + \left. \frac{\partial \phi}{\partial x} \right|_{i+1} \right) + \frac{\alpha}{2 \Delta x} \left( \phi_R - \phi_L \right),         \phi_L = \phi_i + \left. \frac{\partial \phi}{\partial x} \right|_{i} \frac{\Delta x}{2}, \phi_R = \phi_{i+1} - \left. \frac{\partial \phi}{\partial x} \right|_{i+1} \frac{\Delta x}{2},
\end{equation}

\noindent where, in this work, $\alpha = 8/3$. The gradients at cell centers are computed using the second-order central-difference approximation, as in \cite{Nishikawa2010}, which is as follows:
\begin{equation}
    \left. \frac{\partial \phi}{\partial x} \right|_i = \frac{\phi_{i+1} - \phi_{i-1}}{2 \Delta x}.
    \label{eqn:firstDerivative}
\end{equation}

\subsection{Convective flux spatial discretization}
Similar to the viscous flux discretization, the cell centre numerical convective flux derivative is expressed as:
\begin{equation}
    \left. \frac{\text{d} \check{\mathbf{F}}^c}{\text{d} x} \right|_{i} = \frac{1}{\Delta x} \left( \check{\mathbf{F}}^{c}_{i+\frac{1}{2}} - \check{\mathbf{F}}^{c}_{i-\frac{1}{2}} \right),
\end{equation}

\noindent where $i \pm \frac{1}{2}$ indicates right and left cell interface values, respectively. $\check{\mathbf{F}}^c_{i \pm \frac{1}{2}}$ are computed using an approximate Riemann solver. This work uses the HLLC \cite{toro1994restoration} or HLL \cite{harten1983upstream} approximate Riemann solvers unless otherwise explicitly stated. The numerical fluxes at cell boundaries computed using a Riemann solver can be expressed in the following standard form:

\begin{equation}
    \check{\mathbf{F}}^c_{i \pm \frac{1}{2}} = \frac{1}{2} \left[ \check{\mathbf{F}}^c \left( {\mathbf{U}}^{L}_{i \pm \frac{1}{2}} \right) + \check{\mathbf{F}}^c \left( {\mathbf{U}}^{R}_{i \pm \frac{1}{2}} \right) \right] - \frac{1}{2} \left| \mathbf{A}_{i \pm \frac{1}{2}} \right| \left( {\mathbf{U}}^{R}_{i \pm \frac{1}{2}} - {\mathbf{U}}^{L}_{i \pm \frac{1}{2}} \right),
\end{equation}

\noindent where the $L$ and $R$ superscripts denote the left- and right-biased states, respectively, and $\left| \mathbf{A}_{i \pm \frac{1}{2}} \right|$ denotes the convective flux Jacobian. The objective is to obtain the left- and right-biased states, $\mathbf{U}^{L}_{i \pm \frac{1}{2}}$ and $\mathbf{U}^{R}_{i \pm \frac{1}{2}}$. The procedure to obtain these interface values is described in the following sections. 

\subsubsection{Linear and nonlinear schemes}

In this subsection, we provide the details of the calculations of candidate polynomials that can be used to approximate the values of $\mathbf{U}^{L}_{i \pm \frac{1}{2}}$ and $\mathbf{U}^{R}_{i \pm \frac{1}{2}}$. While the original wave-appropriate reconstruction approach of Chamarthi et al. \cite{chamarthi2023wave, hoffmann2024centralized} used a gradient-based reconstruction approach \cite{chamarthi2023gradient,chamarthi2023efficient,chamarthi2023implicit}, here we consider the widely used standard third and fifth-order schemes \cite{shu2006essentially}, which are as follows:

\textbf{Third-order linear and nonlinear schemes:} The third-order upwind schemes for obtaining the values of the left and right interfaces are as follows:
\begin{equation}
\begin{aligned}
& \phi_{i+1/2}^{L3,Linear} = -\frac{1}{6} \phi_{i-1} + \frac{5}{6} \phi_{i+0} + \frac{1}{3} \phi_{i+1}, \\
& \phi_{i+1/2}^{R3,Linear} = -\frac{1}{6} \phi_{i+2} + \frac{5}{6} \phi_{i+1} + \frac{1}{3} \phi_{i+0},
\end{aligned}
\label{eqn:third-linear}
\end{equation}
where $\phi$ is an arbitrary variable, either conservative ($\mathbf{U}$) or characteristic ($\mathbf{C}$) variables are used in the present paper. The superscripts $L3$ and $R3$ denote left-biased and right-biased third-order formulas, respectively. The $Linear$ superscript indicates that the scheme is linear; for brevity, the third-, fifth-, and seventh-order linear upwind schemes will be referred to as U-3, U-5, and U-7, respectively, throughout the remainder of the paper. As widely recognized from Godunov’s barrier theorem \cite{godunov1959}, any linear scheme leads to oscillations near discontinuities and necessitates a nonlinear approach \cite{van1977towards}. The third-order nonlinear scheme for the linear scheme mentioned above is the MUSCL scheme \cite{van1979towards}, which is briefly outlined below:
\begin{align}\label{eq:muscl}
\phi_{i+1/2}^{L,MUSCL} &= \phi_{i+0} + \frac{1}{4} \left[ (1 - \bar{\kappa}) \ddot{\Delta}_{i-1/2} \phi + (1 + \bar{\kappa}) \tilde{\Delta}_{i+1/2} \phi \right]\\
\phi_{i+1/2}^{R,MUSCL} &= \phi_{i+1} - \frac{1}{4} \left[ (1 - \bar{\kappa}) \tilde{\Delta}_{i+3/2} \phi + (1 + \bar{\kappa}) \ddot{\Delta}_{i+1/2} \phi \right]
\end{align}

\begin{equation}
\begin{aligned}
\tilde{\Delta}_{i+1/2} \phi &= \operatorname{minmod}\left( \Delta_{i+1/2} \phi, 2 \Delta_{i-1/2} \phi \right), \\
\ddot{\Delta}_{i+1/2} \phi &= \operatorname{minmod}\left( \Delta_{i+1/2} \phi, 2 \Delta_{i+3/2} \phi \right),
\end{aligned}
\end{equation}
where $\text{minmod} \left( a,b \right) = \frac{1}{2} \left[ \text{sgn}(a) + \text{sgn}(b) \right] \min \left( \left| a \right|, \left| b \right| \right)$. The coefficient $\bar{\kappa}$ determines the accuracy of the approximation, and when $\bar{\kappa}=\frac{1}{3}$, the approximation has third-order accuracy.

\textbf{Fifth-order linear and nonlinear schemes:} Similarly, the fifth-order upwind schemes for obtaining the values of the left and right interfaces are as follows:
\begin{equation}
\begin{aligned}
& \phi_{i+1 / 2}^{L5, Linear}=\frac{1}{30} \phi_{i-2}-\frac{13}{60} \phi_{i-1}+\frac{47}{60} \phi_{i+0}+\frac{9}{20} \phi_{i+1}-\frac{1}{20} \phi_{i+2}, \\
& \phi_{i+1 / 2}^{R5, Linear}=\frac{1}{30} \phi_{i+3}-\frac{13}{60} \phi_{i+2}+\frac{47}{60} \phi_{i+1}+\frac{9}{20} \phi_{i+0}-\frac{1}{20} \phi_{i-1},
\end{aligned}
\label{eqn:fifth-linear}
\end{equation}
The superscripts $L5$ and $R5$ denote left- and right-biased fifth-order formulas, respectively. Similar to the third-order scheme, the fifth-order linear scheme, Eq.~\ref{eqn:fifth-linear}, leads to oscillations, and we utilized the MP limiting approach of Suresh and Huynh \cite{suresh1997accurate}. The following details the MP limiting procedure specifically for the left-biased state, although the procedure is identical for the right-biased state.
\begin{equation}
    \phi^{L,MP5}_{i+\frac{1}{2}} = 
    \begin{cases}
        \phi^{L,\mathrm{Linear}}_{i+\frac{1}{2}} & \text{if } \left( \phi^{L,\mathrm{Linear}}_{i+\frac{1}{2}} - \phi_i \right) \left( \phi^{L,\mathrm{Linear}}_{i+\frac{1}{2}} - \phi^{L,\mathrm{MP}}_{i+\frac{1}{2}} \right) \leq 10^{-40}, \\[5pt]
        \phi^{L,\mathrm{NL}}_{i+\frac{1}{2}} & \text{otherwise},
    \end{cases}
    \label{eqn:mpLimitingCriterion}
\end{equation}

\noindent where ${\phi}^{L,Linear}_{i+\frac{1}{2}}$ corresponds to Eq.~\ref{eqn:fifth-linear} and the remaining terms are:

\begin{equation}
\begin{aligned}
\phi^{L,\mathrm{NL}}_{i+\frac{1}{2}} &= \phi^{L,\mathrm{lin}}_{i+\frac{1}{2}} + \operatorname{minmod} \left( \phi^{L,\mathrm{MIN}}_{i+\frac{1}{2}} - \phi^{L,\mathrm{lin}}_{i+\frac{1}{2}}, \phi^{L,\mathrm{MAX}}_{i+\frac{1}{2}} - \phi^{L,\mathrm{lin}}_{i+\frac{1}{2}} \right), \\
\phi^{L,\mathrm{MP}}_{i+\frac{1}{2}} &= \phi_i + \operatorname{minmod} \left[ \phi_{i+1}-\phi_{i}, 4 \left( \phi_{i}-\phi_{i-1} \right) \right], \\
\phi^{L,\mathrm{MIN}}_{i+\frac{1}{2}} &= \max \left[ \min \left( \phi_{i}, \phi_{i+1}, \phi^{L,\mathrm{MD}}_{i+\frac{1}{2}} \right), \min \left( \phi_{i}, \phi^{L,\mathrm{UL}}_{i+\frac{1}{2}}, \phi^{L,\mathrm{LC}}_{i+\frac{1}{2}} \right) \right], \\
\phi^{L,\mathrm{MAX}}_{i+\frac{1}{2}} &= \min \left[ \max \left( \phi_{i}, \phi_{i+1}, \phi^{L,\mathrm{MD}}_{i+\frac{1}{2}} \right), \max \left( \phi_{i}, \phi^{L,\mathrm{UL}}_{i+\frac{1}{2}}, \phi^{L,\mathrm{LC}}_{i+\frac{1}{2}} \right) \right], \\
\phi^{L,\mathrm{MD}}_{i+\frac{1}{2}} &= \frac{1}{2} \left( \phi_{i} + \phi_{i+1} \right) - \frac{1}{2} d^{L,M}_{i+\frac{1}{2}}, \quad \phi^{L,\mathrm{UL}}_{i+\frac{1}{2}} = \phi_{i} + 4 \left( \phi_{i} - \phi_{i-1} \right), \\
\phi^{L,\mathrm{LC}}_{i+\frac{1}{2}} &= \frac{1}{2} \left( 3 \phi_{i} - \phi_{i-1} \right) + \frac{4}{3} d^{L,M}_{i-\frac{1}{2}}, \\
d^{L,M}_{i+\frac{1}{2}} &= \operatorname{minmod}\left(d_{i},d_{i+1}\right), \\
d_i  &= \phi_{i-1}-2 \phi_i+\phi_{i+1}.
\end{aligned}
\end{equation}
The fifth-order upwind method with MP limiting is referred to as MP5 in this paper, as in \cite{suresh1997accurate}.

\textbf{Central schemes:} The above-mentioned schemes are left- and right-biased upwind schemes. They provide necessary dissipation in some regions, but are unsuitable for turbulent simulations because they require a low-dissipation central scheme. A central scheme can be obtained by averaging the left- and right-biased upwind reconstructions (they can also be derived in other possible ways).
\begin{subequations}\label{eqn:centralScheme}
{\color{black}
    \begin{align}
        \phi^{L}_{i+\frac{1}{2}}
        &= \eta\,\phi^{L,\mathrm{Linear}}_{i+\frac{1}{2}}
        + (1-\eta)\,\phi^{R,\mathrm{Linear}}_{i+\frac{1}{2}},
        \\[5pt]
        \phi^{R}_{i+\frac{1}{2}}
        &= (1-\eta)\,\phi^{L,\mathrm{Linear}}_{i+\frac{1}{2}}
        + \eta\,\phi^{R,\mathrm{Linear}}_{i+\frac{1}{2}}.
    \end{align}
}
\end{subequations}
{\color{black}
For $\eta=0.5$, the two interface states coincide:
\begin{equation*}
\phi^{L}_{i+\frac{1}{2}}
=\phi^{R}_{i+\frac{1}{2}}
=\phi^{\mathrm{cen}}_{i+\frac{1}{2}}
=\frac{1}{2}\left(
\phi^{L,\mathrm{Linear}}_{i+\frac{1}{2}}
+\phi^{R,\mathrm{Linear}}_{i+\frac{1}{2}}
\right).
\end{equation*}
}
An Upwind scheme implies that the value of $\eta$ will be one for all the waves, and for $\eta = 0.5$ it will be a central scheme. The well-known fourth-order central scheme can be obtained by averaging the left- and right-biased reconstruction formulas given by Equation (\ref{eqn:third-linear}), and is as follows:
\begin{equation}\label{eqn:fourth}
\phi^{\textcolor{black}{\mathrm{cen}4},Linear}_{i+\frac{1}{2}} = \frac{1}{2}\left(\phi_{i+1/2}^{L3,Linear} + \phi_{i+1/2}^{R3,Linear}\right) = \frac{1}{12}\left(-\phi_{i-1} + 7\phi_i + 7\phi_{i+1} - \phi_{i+2}\right).
\end{equation}
Similarly, the sixth-order central scheme can be obtained by averaging the left- and right-biased reconstruction formulas given by Equation (\ref{eqn:fifth-linear}), and is as follows:
\begin{equation}\label{eqn:sixth}
\phi^{\textcolor{black}{\mathrm{cen}6},Linear}_{i+\frac{1}{2}}=\frac{1}{2}\left(\phi_{i+1 / 2}^{L5, Linear} + \phi_{i+1 / 2}^{R5, Linear}\right)=\frac{1}{60}\left(\phi_{i-2}-8 \phi_{i-1}+37 \phi_i+37 \phi_{i+1}-8 \phi_{i+2}+\phi_{i+3}\right).
\end{equation}
How these linear and nonlinear schemes are utilized in the wave-appropriate reconstruction approach of Hoffmann, Chamarthi, and Frankel \cite{hoffmann2024centralized} will be presented in the following section.

\begin{remark}
\normalfont The numerical schemes presented in this paper are described in a self-contained manner. Standard formulas for the third- and fifth-order upwind reconstructions are reproduced for completeness and to establish the notation used throughout.
\end{remark}

\subsubsection{Wave-appropriate reconstruction approach}\label{sec:wac}

In Ref.~\cite{chamarthi2023wave}, the authors took advantage of the wave structure of the Euler equations. Once the variables are transformed from physical ($\mathbf{U}$) to characteristic space ($\mathbf{C}$), the characteristic variables have specific properties. The first and last variables are acoustic waves; the second variable is the entropy wave; and the rest are shear waves. Together, the entropy and shear waves are known as linearly degenerate waves. The density varies across the entropy wave in characteristic space, and the rest of the variables remain unchanged. Ref. \cite{chamarthi2023wave} took advantage of this and reduced the frequent activation of the MP criterion, Equation (\ref{eqn:mpLimitingCriterion}). Shock waves are detected by the Ducros sensor, which is used for waves (characteristic variables) other than entropy waves. \textcolor{black}{However, all characteristic waves were reconstructed using the fully upwind left- and right-biased formulas, corresponding to $\eta=1$ in Eq.~\eqref{eqn:centralScheme}. Consequently, every wave family received the dissipation associated with the upwind reconstruction.}

\textcolor{black}{In~\cite{hoffmann2024centralized}, two different mechanisms were used for the linearly degenerate waves. The shear/vortical waves were switched to central reconstruction in smooth regions according to the Ducros shock sensor. The entropy/contact wave was centralized using the MP-based criterion for detecting contact discontinuities; this was possible because the MP criterion could still be applied to the centralized entropy/contact stencil and could fall back to limiting when needed. Thus, a central entropy/contact treatment is possible for MP-type high-order schemes. A related observation was made in~\cite{chamarthi2025wave}, where central entropy-wave reconstruction improved the Shu--Osher shock--entropy wave case. In the present work, however, WA-3 is MUSCL based whereas WA-5 is MP based. For the lower-order MUSCL scheme, no equivalent MP-type criterion is available to check a centralized entropy/contact stencil and safely fall back to limiting. To keep both schemes under the same algorithmic structure and to isolate the optimization of the acoustic bias $\eta_a$, no separate central entropy-wave algorithm is considered here. The entropy/contact wave is therefore kept on the same upwind-biased reconstruction for both WA-3 and WA-5, while the shear/vortical waves are centralized in smooth regions when the Ducros criterion permits it.} The algorithm is outlined below:

\begin{description}
\item[Step 1.]Compute Roe-averaged variables at the interface to construct the left, $\mathbf{L}_n$, and right, $\mathbf{R}_n$, eigenvectors of the normal convective flux Jacobian. \\
    
\item[Step 2.] 
    
 For the fifth-order upwind scheme transform $\mathbf{U}_{i}$ to characteristic space by multiplying them by $\mathbf{L}_n$
    
        \begin{subequations}
        \begin{gather}
            \mathbf{C}_{i+m,b} = \mathbf{L}_{n,i+\frac{1}{2}} \mathbf{U}_{i+m},
        \end{gather}
    \end{subequations}

    for $m = -2,-1,0,1,2,3$ and $b = 1,2,3,4,5$, representing the vector of characteristic variables which are defined as follows in the current implementation: 

    \begin{table}[H]
        \centering
        \caption{Characteristic wave types.}
        \begin{tabular}{c c c}
            \hline
            \hline
            $b = 1,5$ & $b = 2$ & $b = 3,4$ \\
            \hline
            Acoustic & Entropy/Contact & Shear/Vortical  \\
            \hline
            \hline
        \end{tabular}
        \label{tab:characteristicWaveStructure}
    \end{table}
    
    For the third-order scheme, $m = -1, 0, 1, 2$ is sufficient, as the stencil is smaller.

 \item[Step 3.]

\noindent The left-biased reconstruction for the fifth-order scheme is then treated by the following algorithm:

\begin{equation}\label{eq:wave_criterion}
    C^{L}_{i+1/2,\,b} =
    \left\{
    \begin{array}{ll}
        \text{if } b = 1,5\text{ (acoustic):} &
        \begin{cases}
            C^{L,\,\text{MP5}}_{i+1/2,\,b}  & \text{if } \Omega_d > 0.01, (\ \text{Refer Eq.~\ref{eqn:mpLimitingCriterion}}) \\[10pt]
            C^{L5,\,\text{Linear}}_{i+1/2,\,b}  &  \text{otherwise}, (\ \text{Refer Eq.~\ref{eqn:fifth-linear}})
        \end{cases}
        \\[30pt]
        \text{if } b = 2\text{ (entropy):} & 
        \begin{cases}
            {C}^{L,\,\text{MP5}}_{i+1/2,\,b} , (\ \text{Refer Eq.~\ref{eqn:mpLimitingCriterion}})
        \end{cases}
        \\[20pt]
        \text{if } b = 3,4\text{ (shear/vortical):} &
        \begin{cases}
            C^{L,\,\text{MP5}}_{i+1/2,\,b}  & \text{if } \Omega_d > 0.01, (\ \text{Refer Eq.~\ref{eqn:mpLimitingCriterion}}) \\[10pt]
            C^{\textcolor{black}{\mathrm{cen}6},\,\text{Linear}}_{i+1/2,\,b} & \text{otherwise}, (\ \text{Refer Eq.~\ref{eqn:sixth}})
        \end{cases}
    \end{array}
    \right.
\end{equation}

where the Ducros sensor is computed as follows: 
\begin{equation}
    \Omega_d = \frac{\left| -p_{i-2} + 16 p_{i-1} - 30 p_{i} + 16 p_{i+1} - p_{i+2} \right|}{\left| p_{i-2} + 16 p_{i-1} + 30 p_{i} + 16 p_{i+1} + p_{i+2} \right|} \frac{ \left( \nabla \cdot \mathbf{u} \right)^2}{ \left( \nabla \cdot \mathbf{u} \right)^2 + \left| \nabla \times \mathbf{u} \right|^2},
    \label{eqn:ducros}
\end{equation}
\noindent where $\mathbf{u}$ represents the velocity vector, and the derivatives of velocities are computed using Equation (\ref{eqn:firstDerivative}), which involves second-order finite-difference approximation of the first derivative. The sensor value is taken as the maximum of $\Omega_{d}$ within a three-cell neighborhood.

Likewise, the algorithm for the third-order MUSCL scheme is as follows:

\begin{equation}\label{eq:wave_criterion_MUSCL}
    C^{L}_{i+1/2,\,b} =
    \left\{
    \begin{array}{ll}
        \text{if } b = 1,5\text{ (acoustic):} &
        \begin{cases}
            C^{L,\,\text{MUSCL}}_{i+1/2,\,b}  & \text{if } \Omega_d > 0.01, (\ \text{Refer Eq.~\ref{eq:muscl}}) \\[10pt]
            C^{L3,\,\text{Linear}}_{i+1/2,\,b}  &  \text{otherwise}, (\ \text{Refer Eq.~\ref{eqn:third-linear}})
        \end{cases}
        \\[30pt]
        \text{if } b = 2\text{ (entropy):} & 
        \begin{cases}
            {C}^{L,\,\text{MUSCL}}_{i+1/2,\,b} , (\ \text{Refer Eq.~\ref{eq:muscl}})
        \end{cases}
        \\[20pt]
        \text{if } b = 3,4\text{ (shear/vortical):} &
        \begin{cases}
            C^{L,\,\text{MUSCL}}_{i+1/2,\,b}  & \text{if } \Omega_d > 0.01, (\ \text{Refer Eq.~\ref{eq:muscl}}) \\[10pt]
            C^{\textcolor{black}{\mathrm{cen}4},\,\text{Linear}}_{i+1/2,\,b} & \text{otherwise}, (\ \text{Refer Eq.~\ref{eqn:fourth}})
        \end{cases}
    \end{array}
    \right.
\end{equation}

\begin{equation}
    \Omega_d = \max \left( \Omega_{i+m} \right), \quad \text{for } m = -1,0,1.    
\end{equation}

A similar procedure is carried out for the right-biased reconstruction.

\item[Step 4.] After obtaining $\mathbf{C}^{L,R}_{i+\frac{1}{2},b}$, the variables are transformed back to physical fields:

\begin{equation}\label{eqn:characteristicToPhysical}
    \mathbf{U}^{L,R}_{i+\frac{1}{2}} = \mathbf{R}_{n,i+\frac{1}{2}} \mathbf{C}^{L,R}_{i+\frac{1}{2}}.
\end{equation}

\end{description}

\begin{remark}[Naming convention of the schemes]
\label{naming}
\normalfont The wave-appropriate algorithm employing the fifth-order (and sixth) scheme, as outlined in Eq.~\ref{eq:wave_criterion}, will be referred to as \textbf{WA-5} in this paper. Similarly, the algorithm using the third-order (and fourth-order) scheme, as presented in Eqs.~\ref{eq:wave_criterion_MUSCL}, will be denoted \textbf{WA-3} in this paper.
\end{remark}

As shown in the above algorithms, for both fifth- and third-order algorithms, the acoustic waves ($b=1,5$) are always computed using an upwind scheme, $C^{L3, Linear}$ or $C^{L5, Linear}$. It implies that in the following equation. 

\begin{equation}\label{eq:eta_blend}
    \phi^{L}_{i+1/2} = \eta_a\,\phi^{L,\,\text{upwind}}_{i+1/2}
                      + (1-\eta_a)\,\phi^{R,\,\text{upwind}}_{i+1/2}, \qquad
    \phi^{R}_{i+1/2} = (1-\eta_a)\,\phi^{L,\,\text{upwind}}_{i+1/2}
                      + \eta_a\,\phi^{R,\,\text{upwind}}_{i+1/2},
\end{equation}
\textcolor{black}{$\eta_a = 1$, acoustic dissipation is at its maximum. In \cite{chamarthi2023wave,hoffmann2024centralized}, the amount of acoustic dissipation is always taken as one due to the stability issues. For $\eta_a = 0.5$, the acoustic waves are also centralized, and the scheme carries no acoustic upwind dissipation.  For compressible flows, this can lead to energy accumulation at the grid scale and eventual divergence.  The optimization procedure that identifies $\eta_a^*$, the optimal feasible value for each scheme under the present objective and stability tests, is described in Section~\ref{sec:optimization}.}

\begin{remark}[On the Ducros sensor formulation]
\label{rem:ducros}
\normalfont The sensor $\Omega_d$ defined in Eq.~\eqref{eqn:ducros} is the product of two components: a pressure-based shock indicator and the dilatation-to-vorticity ratio. This is the formulation originally proposed by Ducros et al.~\cite{ducros1999large}, where the pressure-based component corresponds to the Jameson sensor $\Psi$ (Eqs.~12--13 of~\cite{ducros1999large}) and the dilatation-to-vorticity ratio corresponds to the correction $\Phi$ (Eq.~22 of~\cite{ducros1999large}). The sensor used throughout that reference is the product $\Phi\Psi$, as recorded in Tables~I--III of~\cite{ducros1999large}. In much of the subsequent literature, only the dilatation-to-vorticity component $\Phi$ has been retained and referred to as the Ducros sensor, while the pressure-based component $\Psi$ has been dropped. This simplification appears, for example, in the formulation of Feng et al.~\cite{feng2024general}. The pressure-based component is not redundant: the dilatation-to-vorticity ratio alone can misidentify strong vortical regions as shocked, because intense vortices in compressible flow produce non-negligible dilatation even in the absence of discontinuities~\cite{sciacovelli2021assessment}. The pressure indicator filters such regions because vortices do not produce the sharp pressure jumps characteristic of shocks. The robustness of the fixed threshold $\Omega_d > 0.01$ used throughout this work, applied uniformly across all test cases from subsonic turbulence to Mach~10 flows without case-by-case adjustment, may be attributed in part to the use of the complete two-component formulation. Readers seeking to reproduce the algorithms presented here should note that $\Omega_d$ as defined in Eq.~\eqref{eqn:ducros} includes both components. Using only the dilatation-to-vorticity ratio would yield a different and potentially less robust sensor.
\end{remark}

\section{Physics-Constrained Optimization of {$\eta_a$}}
\label{sec:optimization}

\subsection{Problem formulation}

The stability boundary with respect to $\eta_a$ cannot be determined analytically for the full nonlinear scheme, because the nonlinear interactions between the spatial reconstruction, limiting procedures, shock sensor, and approximate Riemann solver collectively determine stability in a way that is not captured by linear modified wavenumber analysis or even approximate dispersion analysis \cite{pirozzoli2006spectral}. We therefore treat the discrete operator as a black box and seek.
\begin{equation}\label{eq:optimization}
    \eta_a^* = \underset{\eta_a \in \mathcal{S}}{\arg\min}\;
    \mathcal{J}_{\mathrm{acc}}(\eta_a),
\end{equation}
where $\mathcal{S} = \{\eta_a \in [0.5,1] : \text{all stability tests
pass}\}$ is the feasible set and $\mathcal{J}_{\mathrm{acc}}$ is the
accuracy objective defined below.
The lower bound $\eta_a = 0.5$ corresponds to a
perfectly symmetric reconstruction carrying zero acoustic upwind bias.
\textbf{Throughout the optimization (and for all the cases shown later in the paper), shocked cells (identified
by $\sigma > 0.01$) always retain $\eta_a = 1$ via the runtime
selector; the optimization therefore acts exclusively on the
smooth-region acoustic bias.}

\subsubsection{Accuracy objective}

The three-dimensional inviscid Taylor-Green vortex (TGV) at
$\mathrm{Ma}=0.1$ on a $64^3$ grid serves as the accuracy benchmark.
The accuracy objective is the time-integrated absolute error
in volume-averaged turbulent kinetic energy:
\begin{equation}\label{eq:accuracy}
    \mathcal{J}_{\mathrm{acc}}(\eta_a) =
    \int_0^{t_f} \left|
    \overline{E}_k^{\,\mathrm{num}}(t;\,\eta_a)
    - \overline{E}_k^{\,\mathrm{ref}}(t)
    \right| \mathrm{d}t,
\end{equation}
where $t_f = 10$ is the final simulation time. Excess acoustic dissipation manifests as kinetic energy that decays faster than the reference, producing a large $\mathcal{J}_{\mathrm{acc}}$. The reference kinetic energy profile $\overline{E}_k^{\,\mathrm{ref}}(t)$ is obtained from a linear fully upwinded scheme of order $N+2$ on the same $64^3$ grid: a fifth-order linear upwind scheme for optimizing WA-3, and a seventh-order linear upwind scheme for optimizing WA-5. The optimization, therefore, targets the spectral resolution of the next standard scheme in the order hierarchy (the least dissipative linear baseline one would choose).

\begin{remark}
\normalfont Since no DNS exists for the inviscid Taylor-Green vortex, the $(N+2)$th-order  \emph{linear} scheme on the same $64^3$ grid serves as the accuracy reference. The  energy spectra in Figure~\ref{fig:tgv_spectrum} confirm that the optimized scheme  resolves finer scales than this reference, validating the choice of $\mathcal{J}_{\mathrm{acc}}$.  It should further be stressed that WA-3 and WA-5 are \emph{nonlinear} schemes being  judged against a linear baseline; that the optimized nonlinear $N$th-order scheme  matches or outperforms the linear $(N+2)$th-order scheme is therefore a non-trivial result. \textcolor{black}{The resulting $\eta_a^*$ is therefore the optimal feasible value for this reference-matching objective, not necessarily the smallest stable value of $\eta_a$. A different objective function may admit a lower value of $\eta_a$, and in a more general formulation the optimal acoustic bias is expected to be flow- and space-dependent rather than a single constant.}
\end{remark}

\subsubsection{Stability constraint}

A simulation is declared unstable if it encounters non-finite values or terminates prematurely. The stability test suite consists of the three-dimensional supersonic viscous TGV ($Re = 1600$) on both a coarse ($64^3$) and a refined ($128^3$) grid. This problem subjects the scheme to strong compressibility effects, vortex stretching, and under-resolved features at grid scales where the physical viscous dissipation is insufficient to ensure numerical stability. Testing at two resolutions  ensures that apparent stability at the coarser resolution is not an  artifact of excessive grid-scale dissipation.

As documented by Lusher and Sandham~\cite{lusher2021assessment}, the supersonic Taylor-Green vortex at $\mathrm{Ma} = 1.25$ develops eight distinct shock waves due to the symmetries of the initial condition, which propagate across the periodic boundaries and interact with one another to form complex, highly unsteady shock systems, with local Mach numbers peaking at $\mathrm{Ma} \approx 2$ at $t = 6$. This constitutes a genuinely three-dimensional scenario in which multiple strong interacting shock waves coexist with turbulent structures throughout the domain, testing the full wave-appropriate algorithm, including the Ducros sensor, the characteristic reconstruction path in shocked regions, and the conservative reconstruction with rank-1 entropy wave correction in smooth regions (to be presented later).

The subsonic TGV also introduces a second, physically distinct instability criterion: at $\eta_a = 0.5$ the scheme carries no acoustic dissipation and kinetic energy grows monotonically rather than decaying, which at sufficiently small $\eta_a$ eventually produces non-finite values.  A subsonic simulation is therefore declared unstable if $\overline{E}_k^{\,\mathrm{num}}$ increases monotonically after the initial transient or if non-finite values are encountered.

\subsection{Optimization algorithm}

The scalar bounded minimization in Eq.~\eqref{eq:optimization} is solved with Brent's method~\cite{virtanen2020scipy}.\footnote{Implemented as \texttt{scipy.optimize.minimize\_scalar} with \texttt{method=`bounded'} in SciPy.} Each function evaluation proceeds sequentially: the supersonic TGV is run first on the coarse grid, then on the refined grid. If either of the two crashes, the objective function is set to $\mathcal{J}_{\mathrm{acc}}=+1$, and the subsonic TGV is completely skipped. Since all feasible evaluations satisfy $\mathcal{J}_{\mathrm{acc}} \ll 1$, $+1$ serves as an unambiguous indicator of an infeasible region. The subsonic TGV case is run  only if both stability tests pass. This ordering minimizes wall time per function evaluation because a crashed supersonic simulation runs much faster than subsonic runs.

\begin{algorithm}[H]
\caption{Physics-constrained optimization of acoustic upwind bias}
\label{alg:optimization}
\begin{algorithmic}[1]
\Require Bounds $[\eta_{a,\min},\eta_{a,\max}]=[0.5,1.0]$;
         reference $\overline{E}_k^{\,\mathrm{ref}}(t)$
\Ensure Optimal smooth-region acoustic bias $\eta_a^*$

\Function{Objective}{$\eta_a$}
    \State \parbox[t]{0.82\linewidth}{%
        Run supersonic viscous TGV (coarse, $64^3$) with
        smooth-region bias $\eta_a$\\
        \hspace*{1em}(shocked cells $\sigma > 0.01$ use $\eta_a = 1$
        via runtime selector)}
    \If{crashed} \Return $+1$ \EndIf
    \State \parbox[t]{0.82\linewidth}{%
        Run supersonic viscous TGV (refined, $128^3$) with
        smooth-region bias $\eta_a$\\
        \hspace*{1em}(shocked cells $\sigma > 0.01$ use $\eta_a = 1$
        via runtime selector)}
    \If{crashed} \Return $+1$ \EndIf
    \State \parbox[t]{0.82\linewidth}{%
        Run subsonic inviscid TGV ($64^3$) with bias $\eta_a$\\
        \hspace*{1em}($\sigma \leq 0.01$ everywhere, so $\eta_a$
        applies globally)}
    \If{crashed \textbf{or} $\overline{E}_k^{\,\mathrm{num}}$
        increases monotonically} \Return $+1$ \EndIf
    \State \Return $\mathcal{J}_{\mathrm{acc}}(\eta_a)$
\EndFunction

\State $\eta_a^* \gets \textsc{BrentMinimize}
       (\textsc{Objective},\,0.5,\,1.0)$
\end{algorithmic}
\end{algorithm}

\subsection{Results}

\begin{figure}[H]
\centering
\begin{subfigure}{0.49\textwidth}
  \includegraphics[width=\textwidth]{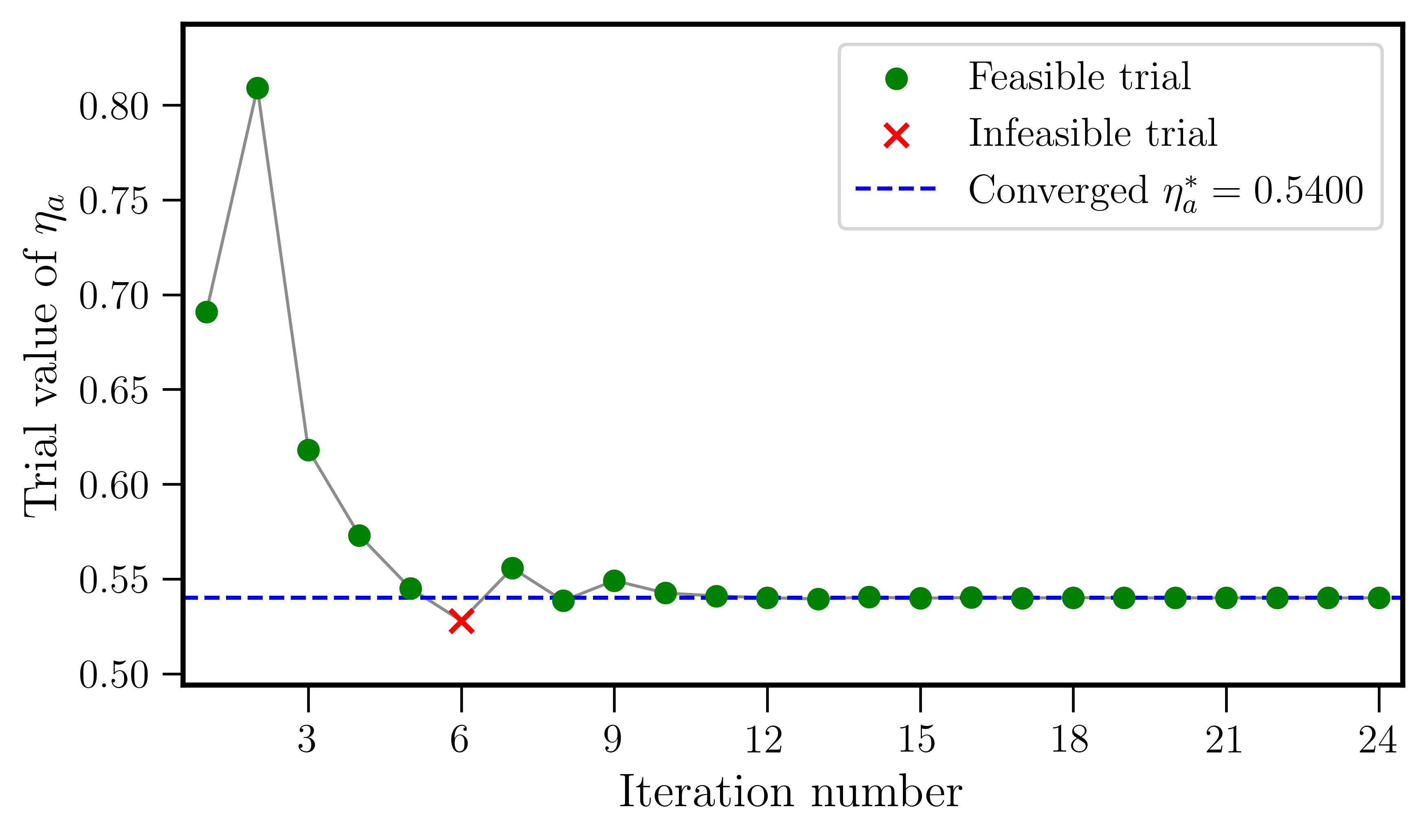}
  \caption{WA-3 (third-order): $\eta_a^*=0.54$.}
  \label{fig:optimizer_3rd}
\end{subfigure}\hfill
\begin{subfigure}{0.49\textwidth}
  \includegraphics[width=\textwidth]{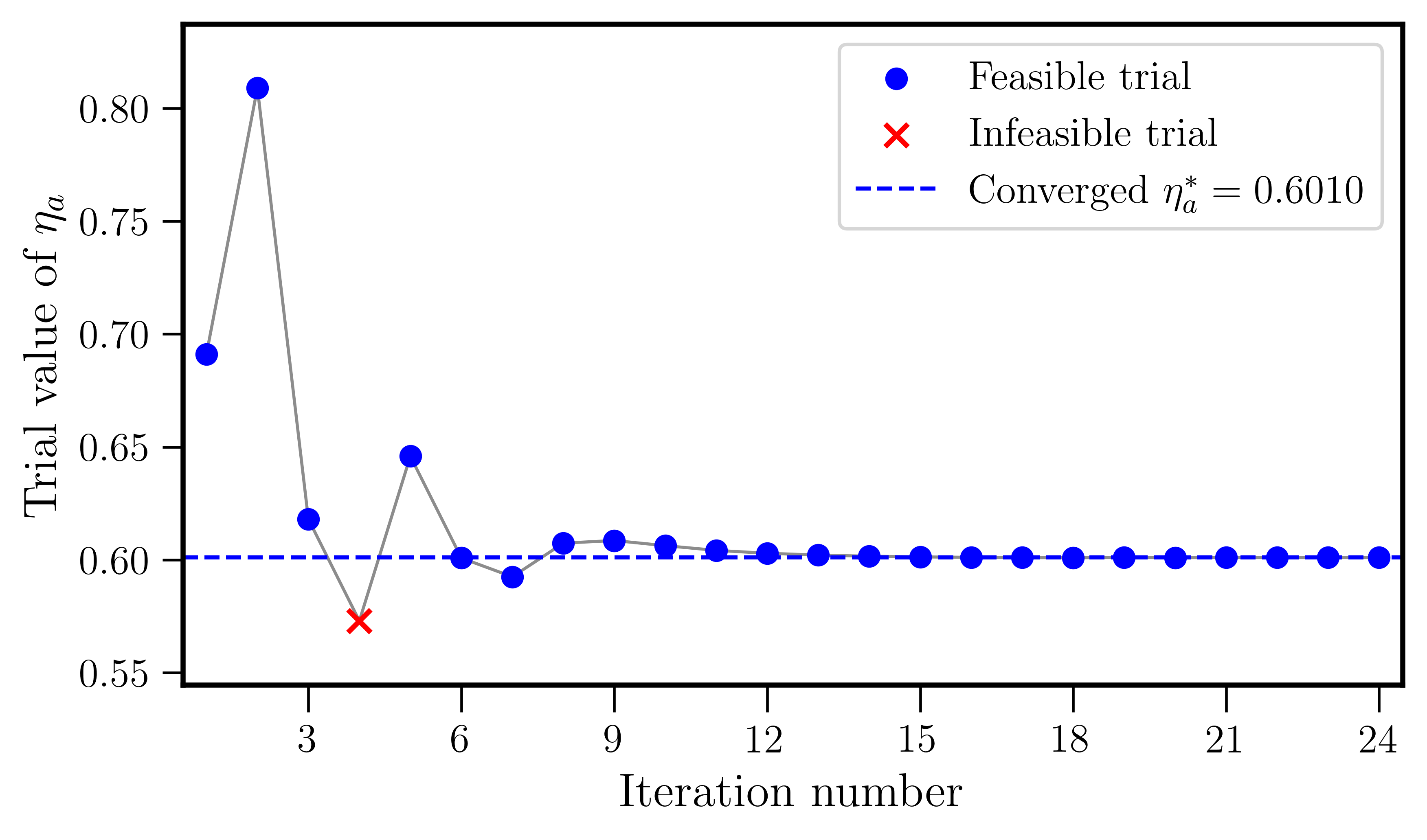}
  \caption{WA-5 (fifth-order): $\eta_a^*=0.6010$.}
  \label{fig:optimizer_5th}
\end{subfigure}
\caption{\textcolor{black}{Optimizer traces for the physics-constrained
         optimization of $\eta_a$. Each point represents one trial value
         evaluated by Brent's bounded method. Feasible and infeasible
         evaluations are marked separately, and the dashed horizontal line
         indicates the converged value $\eta_a^*$.}}
\label{fig:optimizer}
\end{figure}

Figure~\ref{fig:optimizer} shows the trial values of $\eta_a$ evaluated by Brent's method for both reconstruction orders. The horizontal axis is the optimizer iteration number, and the vertical axis is the trial value of $\eta_a$. Feasible and infeasible trials are marked separately, and the converged value is shown by the dashed horizontal line. \textcolor{black}{Both optimizations begin with the same trial value because they use the same bounded interval $[a,b]=[0.5,1.0]$. The first bounded-Brent trial is the golden-section point}
{\color{black}
\begin{align}
\eta_a^{(1)}
&= a+\frac{3-\sqrt{5}}{2}(b-a) \notag\\
&= 0.5+0.381966(1.0-0.5) \notag\\
&= 0.690983 \approx 0.6910. \notag
\end{align}
}
\textcolor{black}{After this common starting value, the two traces differ because the CFD objective response differs for WA-3 and WA-5.}

\subsubsection{Third-order scheme}

For WA-3, the optimizer converges to $\eta_a^*=0.54$ with $\mathcal{J}_{\mathrm{acc}}^*=0.0274$. \textcolor{black}{The optimum lies close to the lower feasible region explored by the optimizer and gives the smallest time-integrated kinetic-energy error relative to the U-5 reference.} Additional acoustic dissipation beyond the optimum increases $\mathcal{J}_{\mathrm{acc}}$ by overdissipating the kinetic energy.

\subsubsection{Fifth-order scheme}

For WA-5, the optimizer converges to $\eta_a^*=0.6010$ with $\mathcal{J}_{\mathrm{acc}}^*=0.0115$. \textcolor{black}{The selected value is again the minimum of the stated reference-matching objective, not a proof that no lower stable value exists.} The fifth-order reconstruction has lower background dissipation than WA-3, so the optimized acoustic bias is shifted upward relative to the third-order case. Nevertheless, the fifth-order scheme achieves a significantly smaller objective value ($\mathcal{J}_{\mathrm{acc}}^*=0.0115$ versus $0.0274$ for WA-3), reflecting its lower dissipation within the useful feasible region.

The values of $\eta_a^*$ determined by the optimizer replace the default upwind value of one for the acoustic waves, as specified in Equation \ref{eq:eta_blend}. The values for the third and fifth-order schemes are summarized in Table~\ref{tab:thresh}.

\begin{table}[H]
\centering
\caption{Optimal acoustic upwind bias $\eta_a^*$ for each scheme,
         determined by physics-constrained optimization.
         \textcolor{black}{$\mathcal{J}_{\mathrm{acc}}^*$ is the minimum
         accuracy objective achieved relative to the selected same-grid
         linear reference.}}
\label{tab:thresh}
\begin{tabular}{lcc}
\hline
Scheme & $\eta_a^*$ 
       & $\mathcal{J}_{\mathrm{acc}}^*$ \\
\hline
WA-3 (third-order)  & $0.54$   & $0.0274$ \\
WA-5 (fifth-order)  & $0.6010$ & $0.0115$ \\
\hline
\end{tabular}
\end{table}

\begin{remark}
\normalfont\textcolor{black}{The resulting $\eta_a^*$ is the
optimal feasible value for the stated reference-matching objective,
not a universal lower stability bound. A different objective or flow class may
lead to a different value. In a more general formulation, the optimal acoustic
bias may be flow- and space-dependent rather than a single constant; wall-bounded
flows, grid stretching, and strongly anisotropic turbulence therefore require
separate analysis.}
\end{remark}

\begin{figure}[H]
\centering
\begin{subfigure}{0.48\textwidth}
  \includegraphics[width=\textwidth]{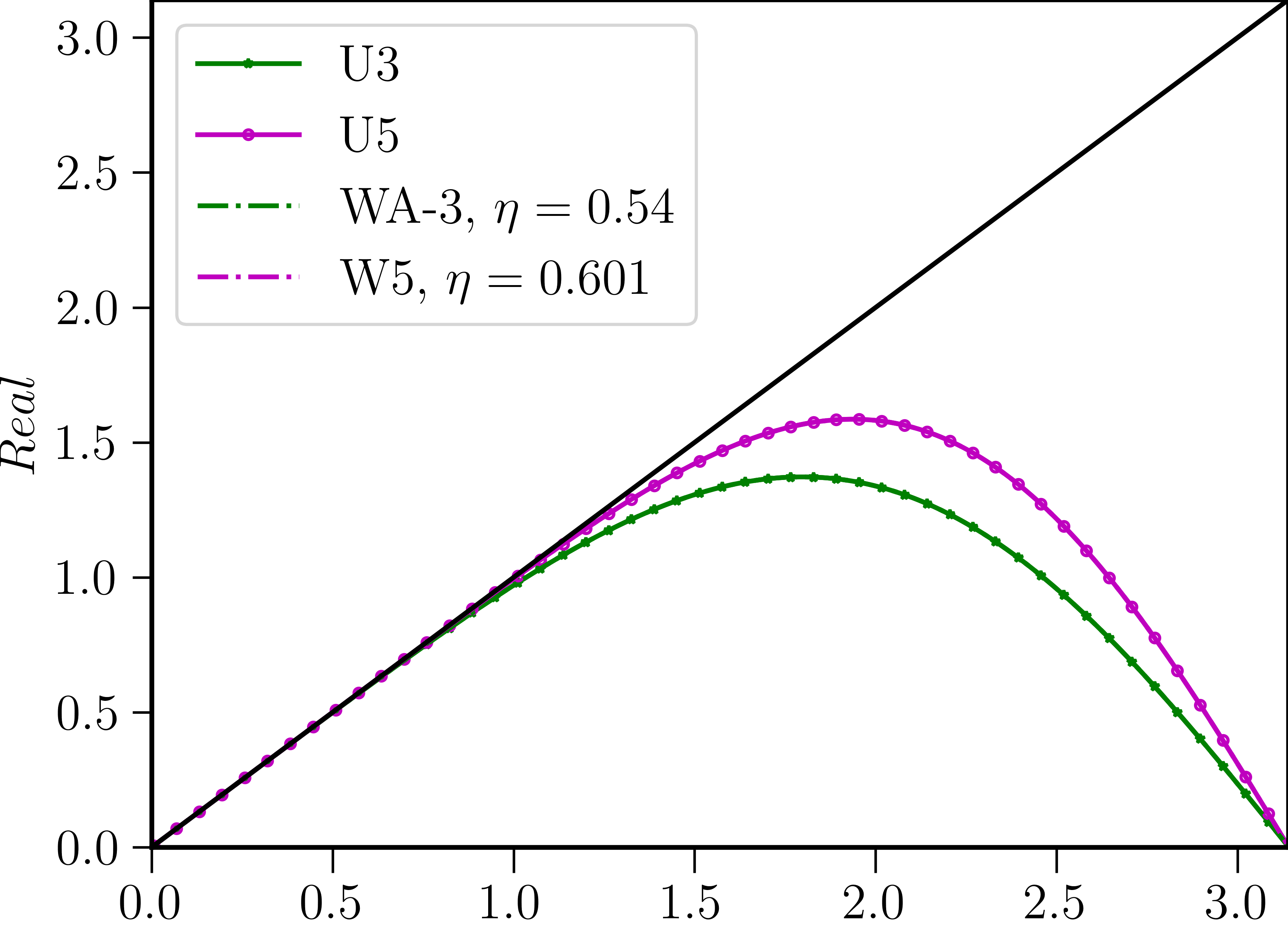}
  \caption{\textcolor{black}{Dispersion}}
  \label{fig:rank_dispersion}
\end{subfigure}\hfill
\begin{subfigure}{0.48\textwidth}
  \includegraphics[width=\textwidth]{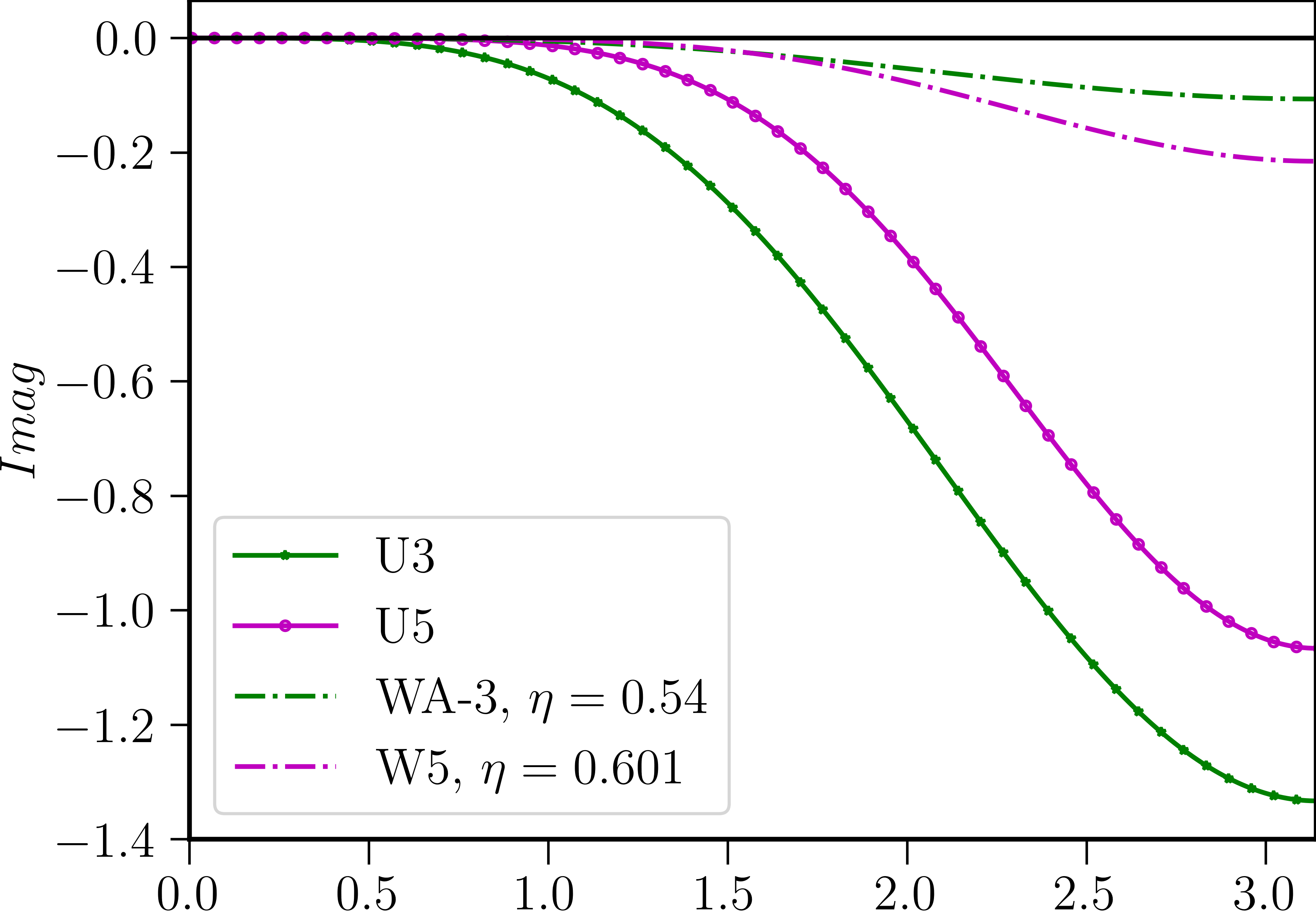}
  \caption{\textcolor{black}{Dissipation}}
  \label{fig:rank_dissipation}
\end{subfigure}
\caption{\textcolor{black}{Linear Fourier analysis of the modified wavenumber for the baseline upwind reconstructions and the optimized wave-appropriate schemes. The real part shows the dispersive branch, while the imaginary part shows the dissipative branch. The optimized values $\eta_a^*=0.54$ and $\eta_a^*=0.6010$ reduce acoustic dissipation relative to full upwinding while preserving the underlying dispersive behavior.}}
\label{fig:rank_spectral}
\end{figure}

\textcolor{black}{The effect of the optimized acoustic bias can also be seen from a linear Fourier analysis. For a Fourier mode, the discrete operator may be written in terms of a complex modified wavenumber, whose real part controls dispersion and whose imaginary part controls numerical dissipation. Figure~\ref{fig:rank_spectral} compares the third- and fifth-order upwind reconstructions with their optimized wave-appropriate counterparts. The optimized blending leaves the dispersive behavior of the underlying reconstruction essentially unchanged while reducing the dissipative part. Thus, the role of $\eta_a^*$ is not to redesign the stencil, but to reduce the unnecessary acoustic dissipation of the upwind reconstruction while retaining enough damping for nonlinear stability. However, this Fourier analysis does not explain why the acoustic wave should use $\eta_a^*=0.54$ while the shear/vortical waves can still use the central value $\eta=0.5$ in smooth regions. The nonlinear effect of reducing $\eta_a$ on kinetic-energy dissipation is therefore assessed separately in the inviscid Taylor--Green vortex results in Figure~\ref{fig:tgv_3}.}

\section{Conservative variable reconstruction with Rank-1 Entropy-Wave Correction}
\label{sec:megcc}

The wave-appropriate scheme of Section~\ref{sec:wac} performs the full characteristic transformation at every interface: the stencil is projected to characteristic space via $\mathbf{L}_n$, each of the five wave families is reconstructed independently, and the result is projected back via $\mathbf{R}_n$. While physically consistent and accurate, it carries a non-trivial cost: two full matrix--vector products per stencil cell are required at every interface regardless of whether a discontinuity is present. The prior conservative--characteristic scheme of~\cite{chamarthi2024generalized} reduced this cost by switching to a cheaper conservative variable reconstruction in smooth regions, but required \emph{two} independent sensors to do so safely: the Ducros shock sensor detected shocks, and a separate density-based criterion detected contact discontinuities.

The primary motivation behind the current algorithm is the observation that explicit contact detection is unnecessary. Near a contact discontinuity, the density varies across the discontinuity while the velocity and pressure remain continuous. In characteristic space, this variation is solely carried by the entropy wave $C_2$, and all other characteristic variables remain smooth. Consequently, the error introduced by a conservative reconstruction near a contact is a rank-1 perturbation along the entropy right eigenvector, which can be corrected algebraically at negligible cost without the need for a sensor. The algorithm presented in this section, Wave-appropriate Conservative Reconstruction (WA-CR), replaces the two-sensor approach with a single Ducros shock threshold and rank-1 entropy wave correction.

Before presenting the algorithm, it is instructive to examine the right eigenvector matrix. The right eigenvector matrix of the 3-D Euler equations in the $x$-direction, ordered as $(\lambda_1,\lambda_2,\lambda_3,\lambda_4,\lambda_5)  = (u-c,\;u,\;u,\;u,\;u+c)$, is
\begin{equation}\label{eq:righteigen}
  \mathbf{R}_n =
  \begin{bmatrix}
    1       & 1  & 0  & 0  & 1      \\
    u-c     & u  & 0  & 0  & u+c    \\
    v       & v  & 1  & 0  & v      \\
    w       & w  & 0  & 1  & w      \\
    H-uc    & \tfrac{1}{2}q^2 & v & w & H+uc
  \end{bmatrix},
\end{equation}
where $q^2=u^2+v^2+w^2$, $c$ is the speed of sound, and $H$ is the specific total enthalpy.  The columns correspond to: left-running acoustic ($\mathbf{r}_1$), entropy ($\mathbf{r}_2$), and two shear waves ($\mathbf{r}_3$, $\mathbf{r}_4$), and right-running acoustic ($\mathbf{r}_5$). Expanding $\mathbf{U}=\mathbf{R}_n\mathbf{C}$ in density gives
\begin{equation}\label{eq:density_decomp}
  \boxed{\rho = C_1 + C_2 + C_5.}
\end{equation}
The shear characteristics $C_3$ and $C_4$ are entirely absent. The zero entries $R_{13}=R_{14}=0$ are not numerical coincidences but the algebraic statement that shear waves are \emph{density-neutral}. This property permits central reconstruction of the shear waves without any adverse effect on the density field. Furthermore, at a contact discontinuity there is no shock, so the Ducros sensor is inactive and the acoustic waves $C_1$ and $C_5$ are reconstructed with the conservative path at the optimized bias $\eta_a^*$. The density error is therefore carried entirely by the entropy wave $C_2$ alone. It is not necessary to detect the contact or deploy a specialized sensor: correcting $C_2$ is both necessary and sufficient, which is the key insight behind WA-CR.

\subsection{Conservative variable reconstruction with rank-1 correction (WA-CR)}
\label{sec:r1smooth}

In smooth regions ($\Omega_d \leq 0.01$), WA-CR avoids the full eigenvector projections and proceeds in two sub-steps.

\paragraph{Step~1: Direction-dependent conservative variable reconstruction.} All five conservative variables are reconstructed directly in physical space, without eigenvector projection.  The upwind bias $\eta_a^*$ is applied to the density, energy, and \emph{normal} momentum; \textcolor{black}{the \emph{tangential} momentum components use the central blend $\eta=0.5$}:
\begin{subequations}\label{eq:cons_smooth_dir}
\begin{alignat}{2}
  x\text{-sweep:} &\quad
    \rho,\;\rho u,\;\rho E \leftarrow \eta_a^*, &\qquad
    \rho v,\;\rho w \leftarrow 0.5, \\
  y\text{-sweep:} &\quad
    \rho,\;\rho v,\;\rho E \leftarrow \eta_a^*, &\qquad
    \rho u,\;\rho w \leftarrow 0.5, \\
  z\text{-sweep:} &\quad
    \rho,\;\rho w,\;\rho E \leftarrow \eta_a^*, &\qquad
    \rho u,\;\rho v \leftarrow 0.5.
\end{alignat}
\end{subequations}
\begin{remark}
\normalfont
\textcolor{black}{In Eqs.~(35a)--(35c), the optimized bias $\eta_a^*$ is applied to
$\rho$, $\rho u_n$, and $\rho E$. This is the general conservative-variable form
used for cases with shocks. For shock-free cases, such as the Taylor--Green
vortex calculations, this additional conservative-variable dissipation is not
necessary; the optimized bias is applied only to the direction-normal momentum
component, namely $\rho u$, $\rho v$, or $\rho w$ depending on the sweep
direction. This distinction avoids assuming an exact equivalence between a
normal-momentum-only correction and the full characteristic acoustic bias in
discontinuous flows. For strong shock problems, the additional
conservative-variable dissipation in $\rho$ and $\rho E$ is retained because
insufficient dissipation across the shock can lead to shock instabilities such
as the carbuncle phenomenon.}
\end{remark}
This direction-dependent biasing mirrors the wave-appropriate dissipation pattern of the full characteristic path: normal momentum behaves like an acoustic variable (requiring upwinding), while tangential momentum behaves like a shear variable (suffices to central differencing).  No eigenvector projection is required.  Denote the resulting interface states $\mathbf{U}^{L,\mathrm{c}}_{i+1/2}$ and $\mathbf{U}^{R,\mathrm{c}}_{i+1/2}$ (superscript ``c'' for \emph{conservative}).
\paragraph{Step~2: Rank-1 entropy wave correction.}
The conservative variable reconstruction of Step~1 carries no guarantee that the entropy wave $C_2$ is oscillation-free near a contact discontinuity.  The following procedure corrects this at minimal cost.

\smallskip\noindent
\emph{Entropy stencil.}  Using only the entropy left eigenvector
$\mathbf{l}_2$, project the six stencil cells onto the entropy wave:
\begin{equation}\label{eq:entropy_chars}
  C_{2,j} = \mathbf{l}_2 \cdot \mathbf{U}_{i+j},
  \qquad j \in \{-2,-1,0,1,2,3\}.
\end{equation}

\noindent\emph{Limited interface value.}
Apply the MP5 limiter to the scalar stencil $\{C_{2,j}\}$:
\begin{equation}\label{eq:entropy_mp5}
  \hat{C}^{L}_{2},\quad \hat{C}^{R}_{2}
  \;=\;
  \mathrm{MP5\text{-}limit}\!\left(C_{2,-2},\ldots,C_{2,3}\right).
\end{equation}

\noindent\emph{Rank-1 update.}
Compute the deficit between the limited entropy value and the entropy
content already present in the conservative state,
\begin{equation}
    \delta^{L} = \hat{C}^{L}_{2}
               - \mathbf{l}_2\cdot\mathbf{U}^{L,\mathrm{c}}_{i+1/2},
    \qquad
    \delta^{R} = \hat{C}^{R}_{2}
               - \mathbf{l}_2\cdot\mathbf{U}^{R,\mathrm{c}}_{i+1/2},
\end{equation}
and apply the rank-1 correction along the entropy right eigenvector
$\mathbf{r}_2 = (1,\;\bar{u},\;\bar{v},\;\bar{w},\;\tfrac{1}{2}\bar{q}^2)^{\top}$:
\begin{equation}\label{eq:rank1_update}
  \mathbf{U}^{L}_{i+1/2}
  = \mathbf{U}^{L,\mathrm{c}}_{i+1/2}
  + \delta^L\,\mathbf{r}_2, \qquad
  \mathbf{U}^{R}_{i+1/2}
  = \mathbf{U}^{R,\mathrm{c}}_{i+1/2}
  + \delta^R\,\mathbf{r}_2.
\end{equation}
In component form:
\begin{equation}\label{eq:rank1_components}
\begin{aligned}
  \rho^{L}          &= \rho^{L,\mathrm{c}}          + \delta^L, \\
  (\rho u)^{L}      &= (\rho u)^{L,\mathrm{c}}      + \delta^L\,\bar{u}, \\
  (\rho v)^{L}      &= (\rho v)^{L,\mathrm{c}}      + \delta^L\,\bar{v}, \\
  (\rho w)^{L}      &= (\rho w)^{L,\mathrm{c}}      + \delta^L\,\bar{w}, \\
  (\rho E)^{L}      &= (\rho E)^{L,\mathrm{c}}      + \delta^L\cdot\tfrac{1}{2}\bar{q}^{2}.
\end{aligned}
\end{equation}
Because $\mathbf{r}_2$ spans the entropy eigenspace, the update leaves all acoustic and shear waves exactly unchanged. Near a contact discontinuity, the MP5 limiter is applied to $C_2$, acting to suppress nonphysical overshoots in this variable. This limiter ensures monotonic behavior and prevents spurious oscillations near discontinuities. The correction from this clipping is then propagated back into all five conservative variables, with adjustments exactly proportional to their entropy eigenvector components, thereby preserving the proper jump relations across the contact. The computational overhead relative to a pure conservative reconstruction is one dot product per stencil cell (six operations) to form the entropy stencil, one scalar MP5 pass, and five multiply-adds per interface side. This is a small fraction of the cost of the full $\mathbf{L}_n$ and $\mathbf{R}_n$ projections it replaces, which require five dot products per stencil cell for the forward projection and a full matrix-vector product for the back projection.


The complete procedure is summarized in Algorithm~\ref{alg:ccr1}. The optimized bias $\eta_a^*$, determined in Section~\ref{sec:optimization}, appears in both paths. On the smooth path, it governs the conservative-variable reconstruction of density, energy, and normal momentum. On the shocked path, it governs the reconstruction of acoustic characteristics. The single value $\eta_a^*$ remains consistent throughout the entire domain. Two sources account for the cost reduction relative to WA-5. The smooth path avoids the full $\mathbf{L}_n$ and $\mathbf{R}_n$ projections at most interfaces. The contact sensor is eliminated entirely.

\begin{algorithm}[H]
\caption{WA-CR reconstruction at interface $i+\tfrac{1}{2}$}
\label{alg:ccr1}
\begin{algorithmic}[1]
\Require Conservative states $\mathbf{U}(j)$, $j=i-2,\ldots,i+3$;
         sensor $\sigma=\max_{m=-1,0,1}\Omega_d(i+m)$
\Ensure  $\mathbf{U}^L_{i+1/2}$, $\mathbf{U}^R_{i+1/2}$
\State Compute Roe-averaged state; form $\mathbf{l}_2$, $\mathbf{r}_2$
\State $C_{2,j}\leftarrow\mathbf{l}_2\cdot\mathbf{U}(j)$
       for $j=i-2,\ldots,i+3$
       \Comment{entropy stencil, shared by both paths}
\State $\hat{C}^{L}_{2},\,\hat{C}^{R}_{2}
       \leftarrow\mathrm{MP5}(C_{2,-2},\ldots,C_{2,3})$
       \Comment{limited entropy values}
\If{$\sigma \leq 0.01$} \Comment{\textit{smooth region - conservative path}}
    \State Reconstruct $\rho,\;\rho u_n,\;\rho E$ with bias $\eta_a^*$;
           \textcolor{black}{$\;\rho u_t,\;\rho u_{t'}$ with $\eta=0.5$}
           $\;\rightarrow\mathbf{U}^{L,\mathrm{c}},\,\mathbf{U}^{R,\mathrm{c}}$
    \State $\delta^{L}\leftarrow
           \hat{C}^{L}_{2}-\mathbf{l}_2\cdot\mathbf{U}^{L,\mathrm{c}}$;\quad
           $\delta^{R}\leftarrow
           \hat{C}^{R}_{2}-\mathbf{l}_2\cdot\mathbf{U}^{R,\mathrm{c}}$
    \State $\mathbf{U}^{L}\leftarrow
           \mathbf{U}^{L,\mathrm{c}}+\delta^{L}\,\mathbf{r}_2$;\quad
           $\mathbf{U}^{R}\leftarrow
           \mathbf{U}^{R,\mathrm{c}}+\delta^{R}\,\mathbf{r}_2$
\Else \Comment{\textit{regions with shocks - full characteristic path}}
    \State $C_{b,j}\leftarrow\mathbf{l}_b\cdot\mathbf{U}(j)$
           for $b=1,3,4,5$;\quad $C_{2,j}$ reused from line~2
    \State $C^{L,R}_1,\,C^{L,R}_5
           \leftarrow$ Algorithm from Eq.~\ref{eq:wave_criterion}
           \Comment{acoustic: upwind bias}
    \State $C^{L,R}_2\leftarrow\hat{C}^{L,R}_{2}$
           \Comment{entropy: reuse from line~3}
    \State $C^{L,R}_3,\,C^{L,R}_4
           \leftarrow$ Algorithm from Eq.~\ref{eq:wave_criterion}
           \Comment{shear: central}
    \State $\mathbf{U}^{L,R}\leftarrow\mathbf{R}_n\,\mathbf{C}^{L,R}$
\EndIf
\end{algorithmic}
\end{algorithm}

\textbf{Generality of the rank-1 correction:} Although the present work uses the MP5 limiter as the baseline reconstruction, the rank-1 entropy wave correction is not tied to any particular nonlinear scheme. The correction requires only that a scalar stencil of entropy characteristic values $\{C_{2,j}\}$ be available and that a limited interface value $\hat{C}_2^{L,R}$ be computed from it. Any monotonicity-preserving or essentially non-oscillatory reconstruction can serve this role. Replacing MP5 with a WENO scheme \cite{Borges2008} throughout yields the variant denoted WA-WENO-CR, which inherits the same rank-1 correction structure and the same cost savings relative to its full characteristic counterpart. Notably, WA-WENO-CR uses the same optimized bias $\eta_a^*=0.6010$ as WA-CR. The correction is equally applicable to the third-order MUSCL variant, though the third-order scheme's higher background dissipation makes the accuracy benefit less pronounced.

The cost of the full characteristic decomposition has been recognized since the earliest high-order schemes. Jiang and Shu~\cite{Jiang1995} noted in their original WENO paper: \emph{``For Euler systems of gas dynamics, we suggest computing the weights from pressure and entropy instead of the characteristic values''} to simplify the costly characteristic procedure. Jiang and Shu denoted their approach as WENO-PS in the corresponding paper. The rank-1 entropy wave correction proposed here directly addresses this cost. Instead of abandoning characteristic reconstruction or approximating it with scalar surrogates, it replaces the complete decomposition with a single entropy-wave correction whenever the Ducros sensor is inactive. This approach significantly reduces the characteristic work to a single dot product and a rank-1 update per interface.

\section{Wave-Appropriate acoustic dissipation for kinetic-energy-preserving Schemes}
\label{sec:WA-KEP}

The kinetic-energy-preserving (KEP) scheme of Chandrashekar~\cite{chandrashekar2013kinetic} constructs the convective flux from piecewise-constant states using logarithmic-mean density and arithmetic velocity and pressure averages, ensuring zero numerical dissipation across all wave families by construction. This property is desirable in smooth turbulent regions, but it implies that no dissipation acts on acoustic waves even when the Ducros sensor is inactive, as is the case throughout shock-free shear-layer flows. \textcolor{black}{The wave-appropriate framework identifies acoustic upwind bias as a targeted stabilizing mechanism: a small bias $\eta_a$ above the central value of $0.5$ adds dissipation to the acoustic content while leaving non-acoustic waves unaffected.} In the KEP scheme, acoustic bias cannot be introduced during the reconstruction step, as in WA-3 or WA-5, because the states are piecewise-constant. Instead, it is introduced through the dissipative component of the numerical flux, applied exclusively to the normal momentum equation. The density, tangential momentum, and energy equations retain the pure KEP flux without modification, preserving the scheme's non-dissipative character across all non-acoustic wave families. The validation of this approach is given in Section~\ref{sec:shear_layer}.

Specifically, the total interface flux is decomposed as
\begin{equation}\label{eq:kep_total}
\hat{\mathbf{F}}_{i+1/2} = \hat{\mathbf{F}}^{\mathrm{KEP}}_{i+1/2} + \hat{\mathbf{D}}_{i+1/2},
\end{equation}
where $\hat{\mathbf{F}}^{\mathrm{KEP}}$ is the non-dissipative kinetic-energy-preserving flux of Chandrashekar~\cite{chandrashekar2013kinetic}, and $\hat{\mathbf{D}}$ is a selective dissipation term that acts only on the normal momentum component.

The dissipation is constructed as follows. For each sweep direction, the normal momentum is reconstructed using the third-order scheme of Eq.~\ref{eqn:third-linear} with an upwind bias $\eta_a$:
\begin{align}
(\rho u_n)^L_{i+1/2} &= \eta_a \, (\rho u_n)^{L3,\mathrm{Linear}}_{i+1/2} + (1-\eta_a) \, (\rho u_n)^{R3,\mathrm{Linear}}_{i+1/2}, \label{eq:kep_recon_L}\\
(\rho u_n)^R_{i+1/2} &= (1-\eta_a) \, (\rho u_n)^{L3,\mathrm{Linear}}_{i+1/2} + \eta_a \, (\rho u_n)^{R3,\mathrm{Linear}}_{i+1/2}, \label{eq:kep_recon_R}
\end{align}
where $u_n$ is the velocity component normal to the interface (i.e., $u$ for the $x$-sweep, $v$ for the $y$-sweep, $w$ for the $z$-sweep). The Rusanov-type dissipation for the normal momentum is then
\begin{equation}\label{eq:kep_diss}
\hat{D}_{n\text{-mom}} = -\frac{1}{2}\lambda_{\max}\left[(\rho u_n)^R_{i+1/2} - (\rho u_n)^L_{i+1/2}\right],
\end{equation}
where $\lambda_{\max} = |\tilde{u}_n| + \tilde{c}$ is the maximum wave speed at the interface, computed from Roe-averaged quantities. All other components of $\hat{\mathbf{D}}$ are identically zero:
\begin{equation}\label{eq:kep_selective}
\hat{D}_k = \begin{cases} \hat{D}_{n\text{-mom}} & \text{if } k = \text{normal momentum equation,} \\ 0 & \text{otherwise (density, tangential momentum, energy).} \end{cases}
\end{equation}
The direction-dependent application mirrors the wave-appropriate principle: normal momentum carries the acoustic wave content, while tangential momentum carries shear wave content, and density carries entropy wave content. Adding dissipation exclusively to the normal momentum, therefore targets acoustic waves without contaminating non-acoustic fields. The procedure is summarized in Algorithm~\ref{alg:wa_kep}.

\begin{remark}[Optimum $\eta_a$ for the KEP scheme]
\label{rem:KEP}
\normalfont \textcolor{black}{An optimum $\eta_a$ for the KEP scheme cannot be determined via the optimization procedure used in this paper for two reasons. First, the KEP scheme is linear and therefore cannot simulate the supersonic Taylor-Green vortex, which serves as the stability constraint in the optimization. Second, $\eta_a = 0.5$ is always stable for the subsonic Taylor-Green vortex since the scheme never crashes for that case. Determining an optimal $\eta_a$ for the KEP scheme would require a suitable nonlinear stability test, which is beyond the scope of the present work. The values $\eta_a = 0.56$ and $\eta_a = 1.0$ are chosen here solely to demonstrate that introducing dissipation exclusively through the normal momentum flux suppresses the dominant spurious vortices produced by the unmodified KEP scheme in the periodic shear layer, and should not be interpreted as optimized values in the sense of Section~\ref{sec:optimization}.}
\end{remark}

\begin{algorithm}[H]
\caption{Wave-appropriate KEP (WA-KEP) flux at interface $i+\frac{1}{2}$, $n$-direction sweep}
\label{alg:wa_kep}
\begin{algorithmic}[1]
\Require Left/right states $\mathbf{U}_L = \mathbf{U}_i$, $\mathbf{U}_R = \mathbf{U}_{i+1}$; bias $\eta_a$;
\Ensure Interface flux $\hat{\mathbf{F}}_{i+1/2}$

\State Compute KEP flux $\hat{\mathbf{F}}^{\mathrm{KEP}}$ using logarithmic-mean (or geometric-mean) density and arithmetic averages

\State Reconstruct normal momentum $(\rho u_n)$ using third-order scheme (Eq.~\ref{eqn:third-linear}) with bias $\eta_a$:
\[
(\rho u_n)^{L,R}_{i+1/2} \quad \text{via Eqs.~(\ref{eq:kep_recon_L})--(\ref{eq:kep_recon_R})}
\]

\State Compute Roe-averaged wave speed: $\lambda_{\max} = |\tilde{u}_n| + \tilde{c}$
 
\State Selective dissipation:
\[
\hat{D}_{n\text{-mom}} = -\tfrac{1}{2}\lambda_{\max}\bigl[(\rho u_n)^R - (\rho u_n)^L\bigr], \qquad \hat{D}_k = 0 \;\;\text{for all other } k
\]

\State $\hat{\mathbf{F}}_{i+1/2} = \hat{\mathbf{F}}^{\mathrm{KEP}}_{i+1/2}$; \quad $\hat{F}_{n\text{-mom}} \mathrel{+}= \hat{D}_{n\text{-mom}}$

\end{algorithmic}
\end{algorithm}


\section{Results}
\label{sec:results}

This section validates the proposed schemes across various test cases spanning smooth turbulence, shear instabilities, and shocked flows. The following schemes are compared: \textbf{WA-3} (third-order MUSCL, full characteristic, $\eta_a^*=0.54$), \textbf{WA-5} (fifth-order MP5, full characteristic, $\eta_a^*=0.6010$), and \textbf{WA-CR} (fifth-order MP5, conservative reconstruction with rank-1 entropy wave correction, $\eta_a^*=0.6010$).  For selected cases, \textbf{WA-WENO-CR} (fifth-order WENO, conservative reconstruction with rank-1 entropy wave correction, $\eta_a^*=0.6010$) is also included. The \textbf{WA-KEP} scheme (Section~\ref{sec:WA-KEP}) is included in the double periodic shear layer test to demonstrate the generality of the wave-appropriate acoustic dissipation principle beyond reconstruction-based schemes. Where relevant, the unoptimized baseline WA-5 at $\eta_a=1$ and the fifth-order TENO5 scheme~\cite{feng2024general} are included for reference. All simulations use the HLLC approximate Riemann solver~\cite{toro1994restoration} and the third-order TVD Runge--Kutta scheme~\cite{Jiang1995} at $\mathrm{CFL}=0.4$, unless otherwise stated.

\subsection{Inviscid Taylor-Green vortex}
\label{sec:tgv}

\begin{figure}[H]
\centering
\begin{subfigure}{0.4\textwidth}
  \includegraphics[width=\textwidth]{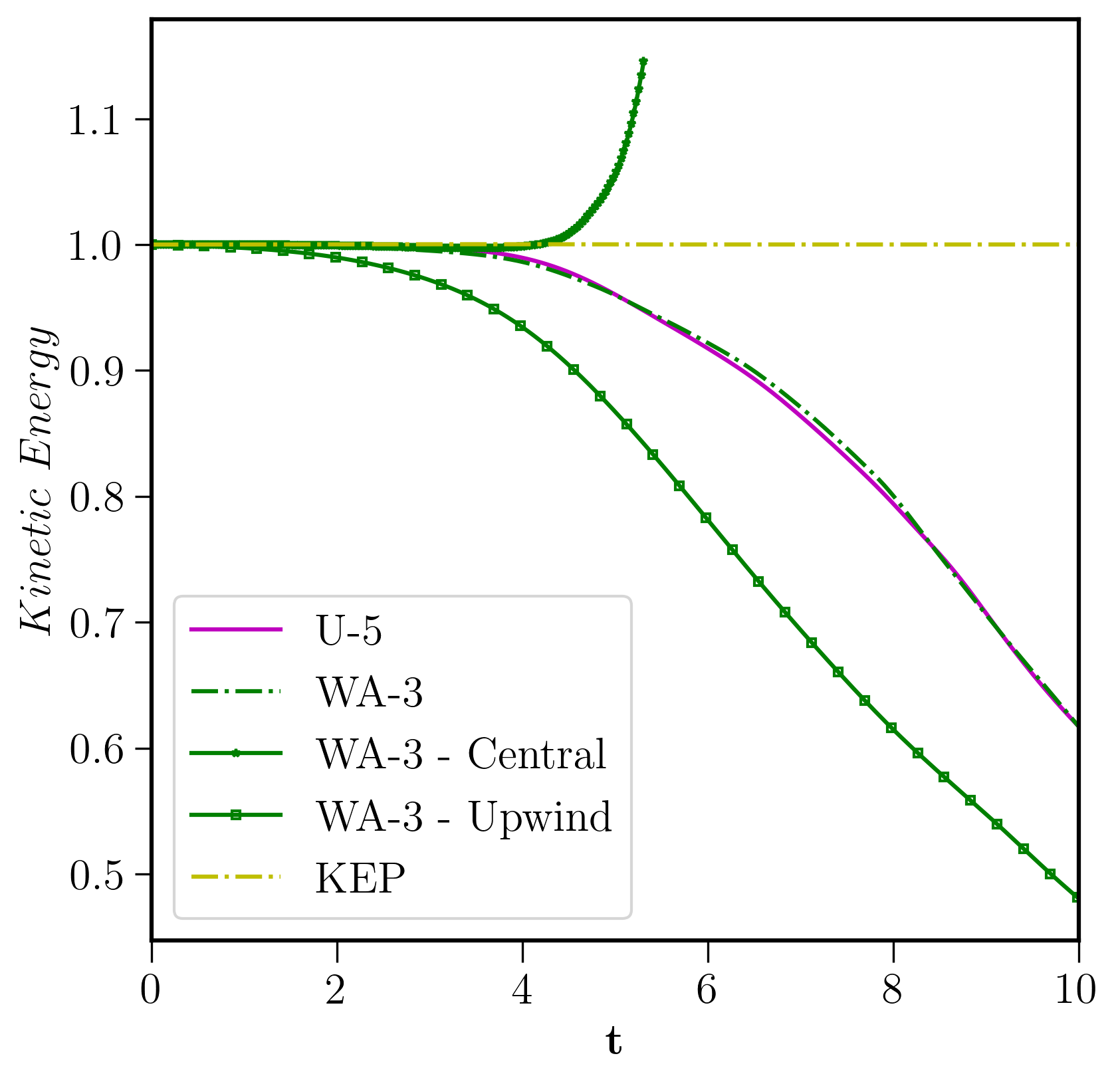}
  \caption{Third order}
  \label{fig:tgv_3}
\end{subfigure}%
\begin{subfigure}{0.4\textwidth}
  \includegraphics[width=\textwidth]{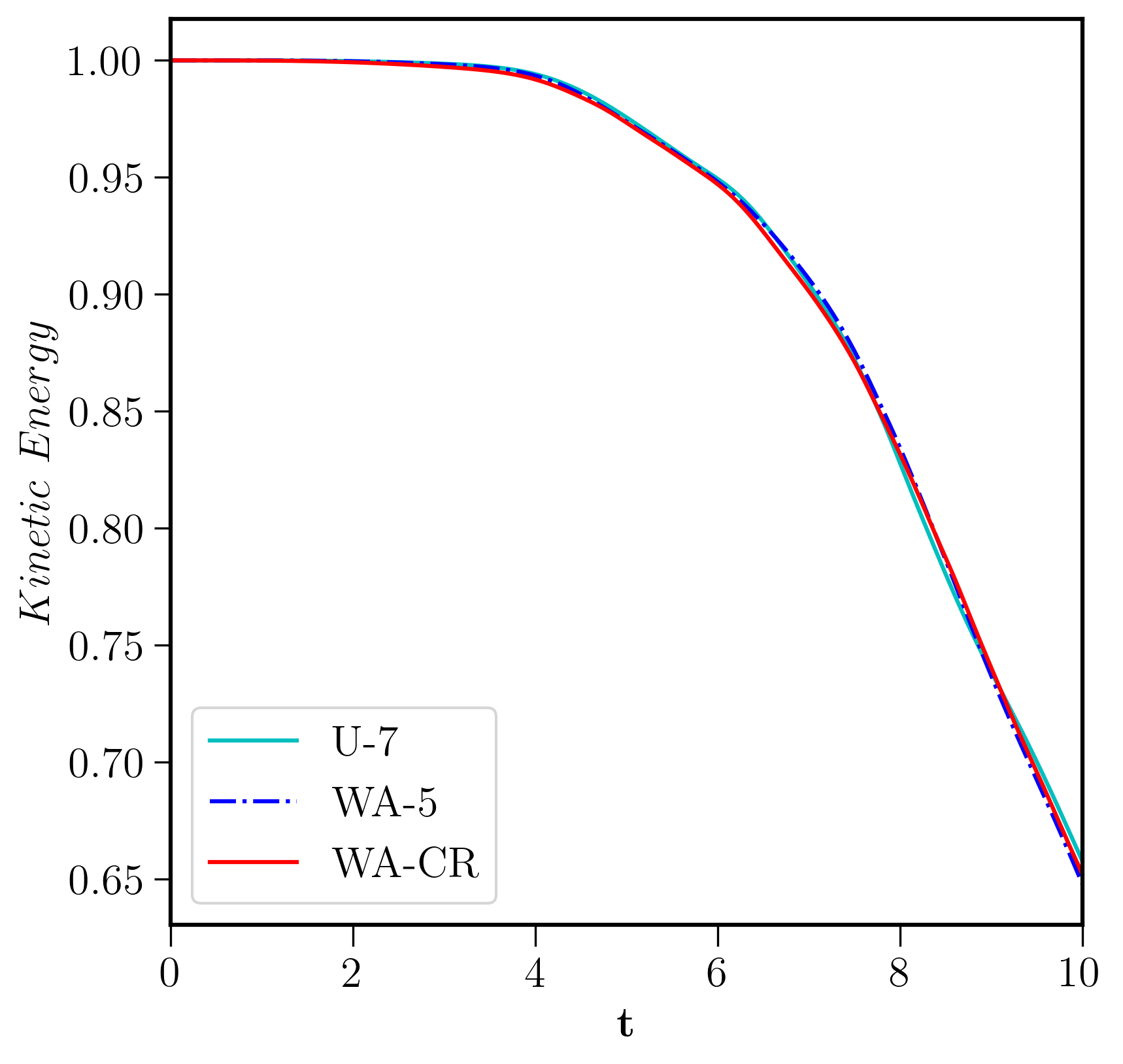}
  \caption{Fifth order}
  \label{fig:tgv_5}
\end{subfigure}
\begin{subfigure}{0.5\textwidth}
  \includegraphics[width=\textwidth]{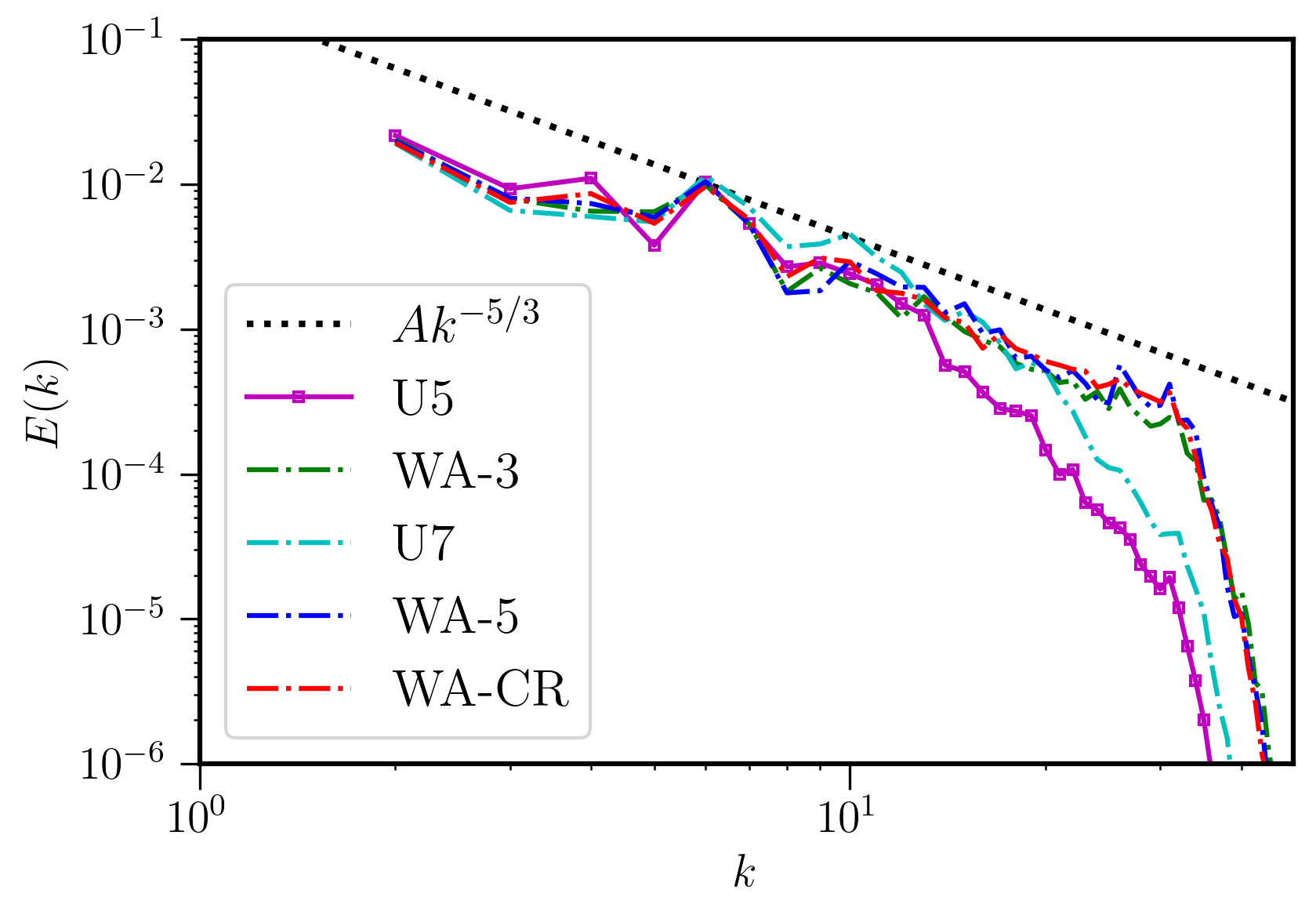}
  \caption{Kinetic energy spectrum at $t=10$}
  \label{fig:tgv_spectrum}
\end{subfigure}
\caption{Subsonic inviscid Taylor-Green vortex ($\mathrm{Ma}=0.1$,
  $p_0=100$, $64^3$ grid, Sec.~\ref{sec:tgv}): time evolution of volume-averaged kinetic
  energy for the (a) third-order and (b) fifth-order schemes, and
  (c) the kinetic energy spectrum at $t=10$.}
\label{fig:tgv}
\end{figure}

The three-dimensional inviscid Taylor-Green vortex (TGV) serves both as
the calibration benchmark for the optimization of $\eta_a$ and as the
primary accuracy assessment. The initial conditions are
\begin{equation}\label{itgv}
\begin{pmatrix}
\rho \\
u \\
v \\
w \\
p \\
\end{pmatrix}
=
\begin{pmatrix}
1 \\
\sin{x} \cos{y} \cos{z} \\
-\cos{x} \sin{y} \cos{z} \\
0 \\
100 + \frac{\left( \cos{(2z)} + 2 \right) \left( \cos{(2x)} + \cos{(2y)} \right) - 2}{16}
\end{pmatrix}.
\end{equation}
on the triply periodic domain $[0,2\pi)^3$ with $\gamma=5/3$. The subsonic variant ($\mathrm{Ma}=0.1$) is used to assess accuracy. All simulations are conducted on a $64^3$ grid until $t=10$. Figure~\ref{fig:tgv_3} shows the kinetic energy decay for the third-order schemes. WA-3 at full upwinding ($\eta_a=1$) is excessively dissipative, decaying significantly faster than the reference U-5. WA-3 at $\eta_a=0.5$ (central) is unstable and diverges at $t\approx 5$, demonstrating the sharp stability cliff identified by the optimizer. WA-3 at $\eta_a^*=0.54$ closely matches the U-5 reference throughout, confirming that the optimized third-order scheme matches the standard linear fifth-order scheme at no additional computational cost. Figure~\ref{fig:tgv_3} also shows the result obtained by the KEP scheme,  where the kinetic energy remains constant, confirming that the scheme  has been correctly implemented.

\textcolor{black}{Three observations from the third-order results are worth noting. First, the instability at $\eta_a=0.5$ is driven by the absence of acoustic upwind bias: even though the entropy wave is treated with the nonlinear MUSCL limiter near contact discontinuities, setting $\eta_a=0.5$ causes the scheme to crash in this smooth acoustic/vortical flow. This confirms that the acoustic bias is the relevant stabilizing mechanism here. Second, the existence of a sharp feasible/infeasible transition near the lower end of the search interval justifies the use of a scalar optimization. Finally, the Ducros sensor detects no shocks throughout this case, as the $\eta_a=0.5$ instability confirms: if the sensor had activated and switched to $\eta_a=1$, the scheme would not have crashed. This validates both the sensor formulation and the $\Omega_d > 0.01$ threshold.}

Figure~\ref{fig:tgv_5} shows the corresponding results for the fifth-order schemes. WA-5 and WA-CR at $\eta_a^*=0.6010$ both overlap the U-7 reference to plotting accuracy. WA-CR matches WA-5 exactly since the flow is shock-free and contact-free throughout: the conservative reconstruction path is always active, and the rank-1 correction is never triggered. This confirms that the conservative variable path introduces no accuracy penalty in smooth flows.

\begin{figure}[H]
\centering
  \includegraphics[width=0.72\textwidth]{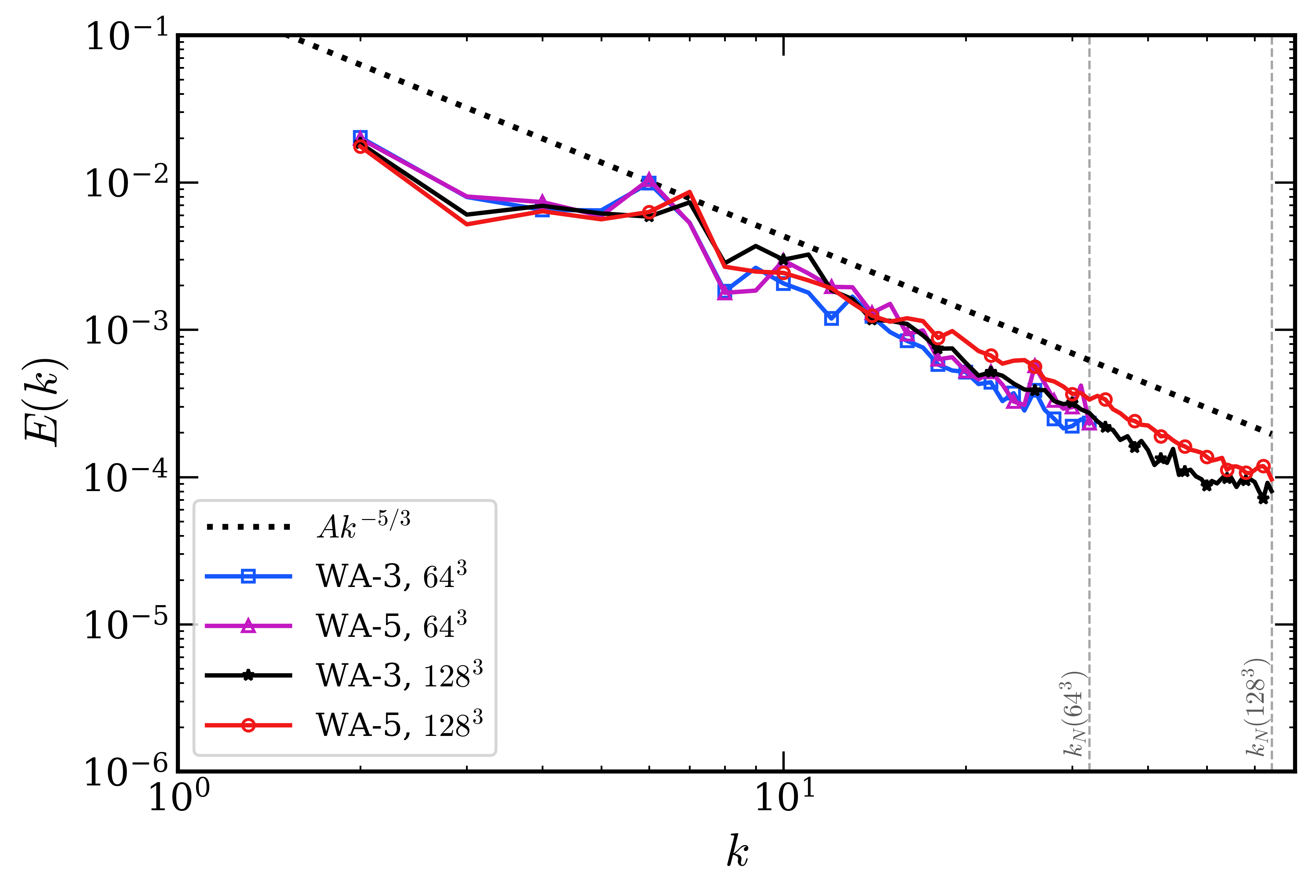}
\caption{\textcolor{black}{Kinetic-energy spectra for optimized WA-3 and WA-5 on $64^3$ and $128^3$ grids using complete Fourier shells. The vertical dashed lines indicate the maximum complete-shell wavenumber for each grid.}}
\label{fig:tgv_spectrum_128}
\end{figure}

Figure~\ref{fig:tgv_spectrum} shows the kinetic energy spectra at $t=10$. All optimized schemes follow the Kolmogorov $k^{-5/3}$ scaling over the resolved inertial range. The spectra of WA-3 and WA-5 both extend further than their respective linear references U-5 and U-7 at high wavenumbers, confirming that the optimized nonlinear $N$th-order schemes resolve small-scale structures at least as well as the standard linear $(N\!+\!2)$th-order schemes they were calibrated against. \textcolor{black}{The close overlap of the optimized WA-3 and WA-5 spectra on the $64^3$ grid should be interpreted as an effective-resolution result for this under-resolved grid and this kinetic-energy calibration, not as a statement that the two reconstructions have the same formal order. A closer inspection of Figure~\ref{fig:tgv_spectrum} still shows that WA-5 is marginally better than WA-3 at high wavenumbers. However, once excess acoustic dissipation is removed, both schemes approach the grid-limited spectral content of the $64^3$ calculation, so the difference between orders is less visible. In addition, the Fourier analysis in Figure~\ref{fig:rank_spectral} shows that the optimized WA-3 is less dissipative than the optimized WA-5, since the third-order scheme admits a lower acoustic bias ($\eta_a^*=0.54$) than the fifth-order scheme ($\eta_a^*=0.6010$). This reduced acoustic dissipation in WA-3 can partly offset the formal-order advantage of WA-5 on the coarse grid. To check that this interpretation is not only a $64^3$ artifact, Figure~\ref{fig:tgv_spectrum_128} compares the optimized WA-3 and WA-5 spectra on both $64^3$ and $128^3$ grids using complete Fourier shells. On the refined grid, WA-5 more clearly retains higher-wavenumber content than WA-3, while the expected grid cutoff shifts from the $64^3$ limit to the $128^3$ limit. The result supports the interpretation that the observed spectra are controlled by both reconstruction order and optimized acoustic dissipation, together with the resolved bandwidth of the grid and the kinetic-energy calibration target.} The KEP spectrum is omitted from Figure~\ref{fig:tgv_spectrum} as it exhibits pronounced energy pile-up near the Nyquist scale, consistent with the known limitation identified by Ghate and Lele~\cite{ghate2023finite}, and its inclusion would obscure the comparison among the wave-appropriate schemes.

\subsubsection{Viscous Taylor-Green vortex}
\label{sec:vtgv}

The three-dimensional viscous Taylor-Green vortex at $Re=1600$ and $\mathrm{Ma}=0.1$ is a standard benchmark for implicit LES schemes, providing a DNS reference~\cite{brachet1983small} against which the dissipation rate $\epsilon = -\mathrm{d}E_k/\mathrm{d}t$ can be compared directly. The initial conditions are identical to Eq.~\eqref{itgv}. Results are presented on both a coarse ($64^3$) and a refined ($96^3$) grid.

\begin{figure}[H]
\centering
\begin{subfigure}{0.45\textwidth}
  \includegraphics[width=\textwidth]{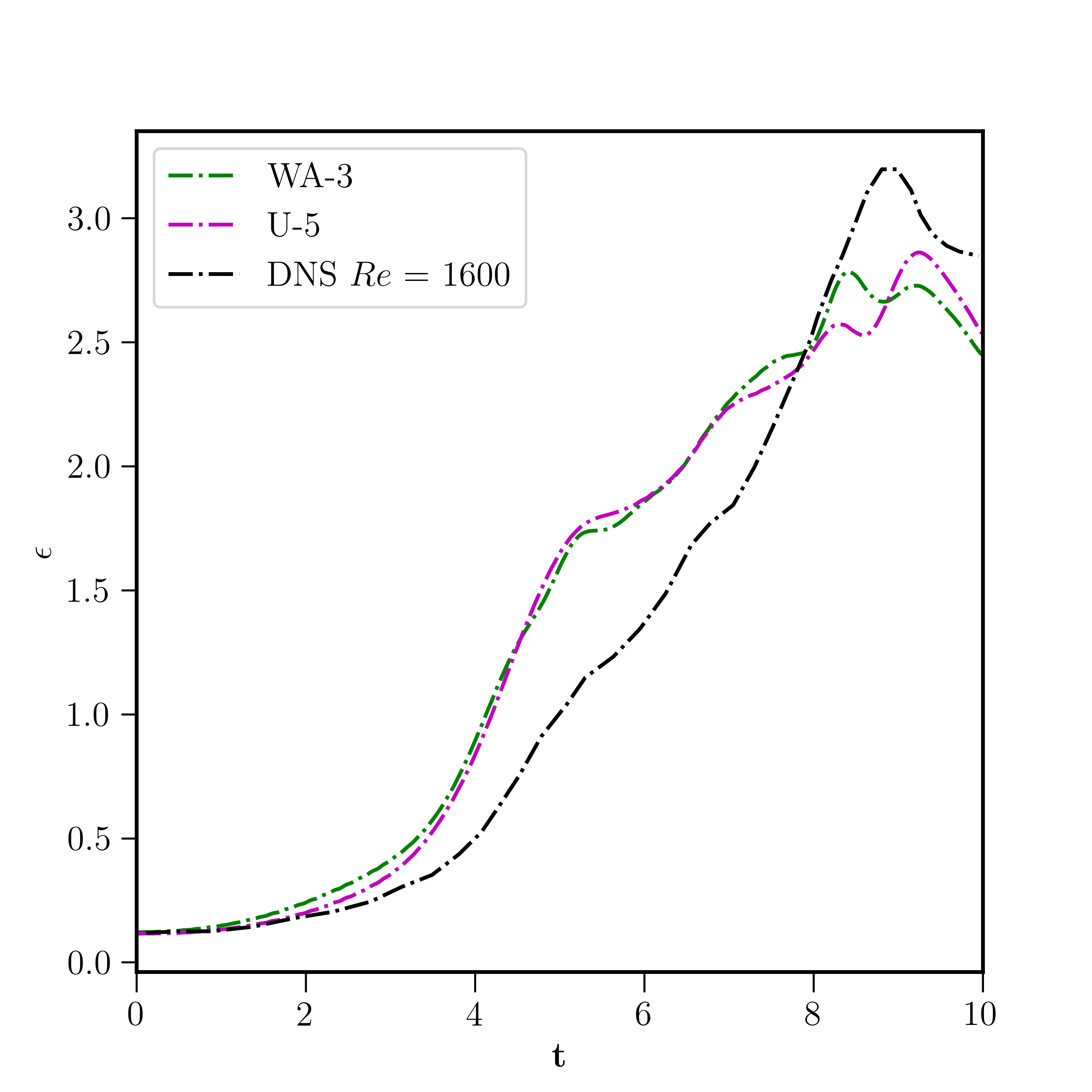}
  \caption{WA-3, $64^3$}
  \label{fig:vtgv_u3_64}
\end{subfigure}\hfill
\begin{subfigure}{0.45\textwidth}
  \includegraphics[width=\textwidth]{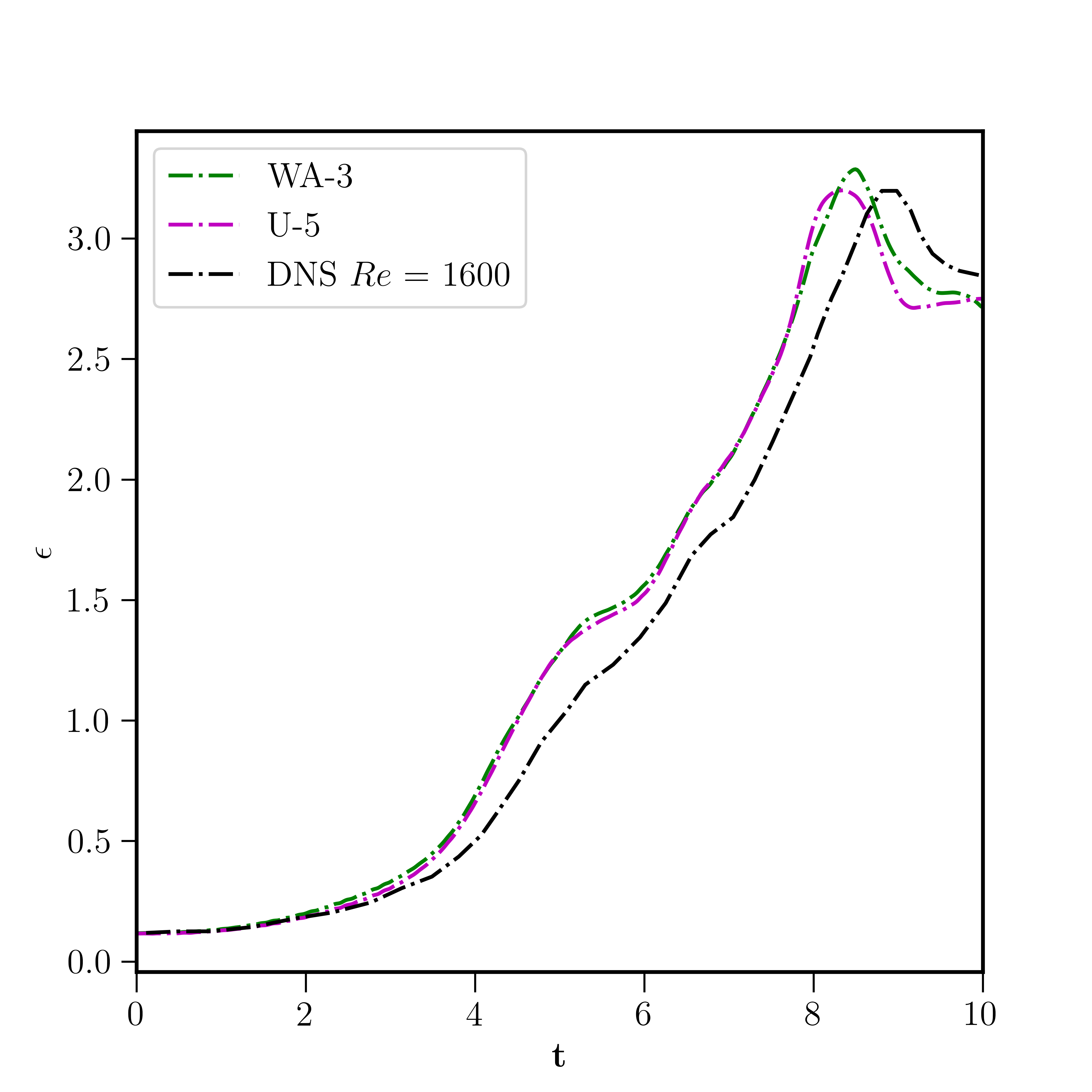}
  \caption{WA-3, $96^3$}
  \label{fig:vtgv_u3_96}
\end{subfigure}
\caption{Viscous Taylor-Green vortex ($Re=1600$, $\mathrm{Ma}=0.1$, Sec.~\ref{sec:vtgv}):
time evolution of the volume-averaged dissipation rate $\epsilon$ on
$64^3$ and $96^3$ grids for WA-3 scheme. DNS reference from~\cite{brachet1983small}.}
\label{fig:vtgv_3}
\end{figure}

Figures~\ref{fig:vtgv_u3_64} and~\ref{fig:vtgv_u3_96} show the dissipation rate for the third-order schemes. On the coarse grid, WA-3 and U-5 both overpredict the dissipation rate and miss the peak, which is a known limitation of third-order accuracy at this resolution. On the refined $96^3$ grid, both schemes improve considerably and approach the DNS profile, with WA-3 and U-5 producing nearly identical results, consistent with the inviscid TGV finding that WA-3 matches its $(N+2)$th-order linear reference.

\begin{figure}[H]
\centering
\begin{subfigure}{0.45\textwidth}
  \includegraphics[width=\textwidth]{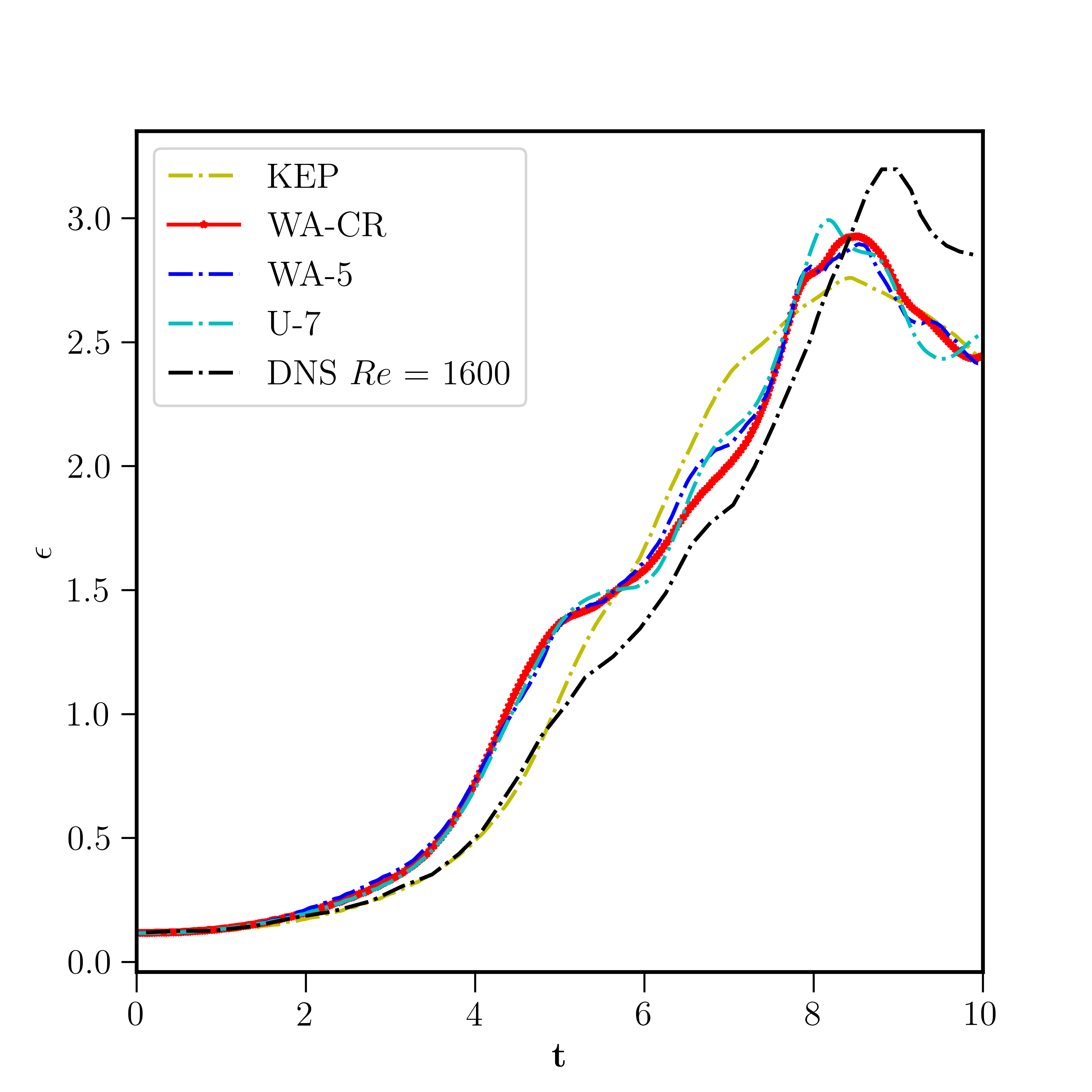}
  \caption{WA-5 and WA-CR, $64^3$}
  \label{fig:vtgv_u5_64}
\end{subfigure}\hfill
\begin{subfigure}{0.45\textwidth}
  \includegraphics[width=\textwidth]{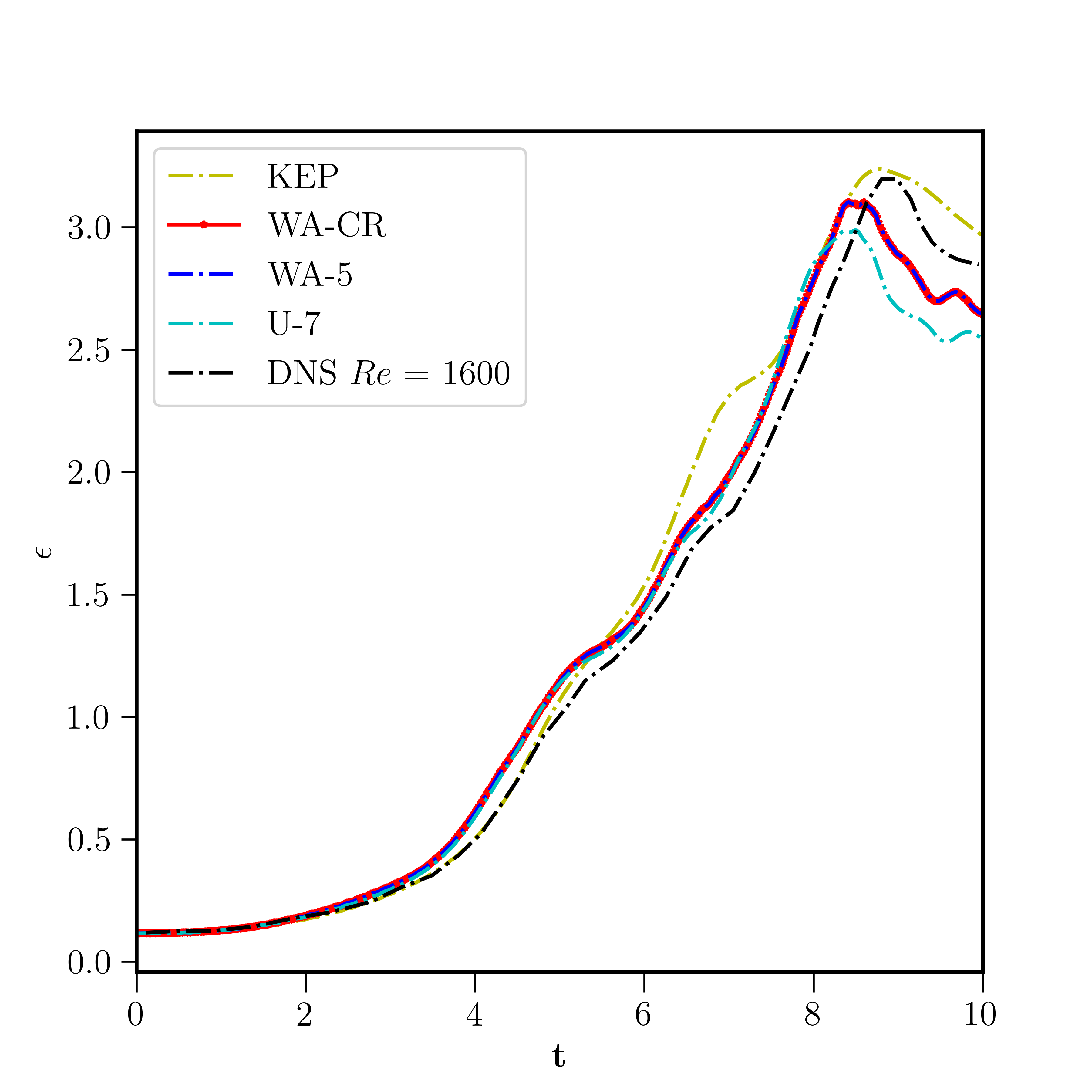}
  \caption{WA-5 and WA-CR, $96^3$}
  \label{fig:vtgv_u5_96}
\end{subfigure}
\caption{Viscous Taylor-Green vortex ($Re=1600$, $\mathrm{Ma}=0.1$, Sec.~\ref{sec:vtgv}):
time evolution of the volume-averaged dissipation rate $\epsilon$ on
$64^3$ and $96^3$ grids for WA-5 scheme. DNS reference from~\cite{brachet1983small}.}
\label{fig:vtgv}
\end{figure}

Figures~\ref{fig:vtgv_u5_64} and~\ref{fig:vtgv_u5_96} show the fifth-order results. On both grids, WA-5 and WA-CR overlap in plotting accuracy and closely follow the DNS profile through the dissipation peak at $t\approx9$. The linear U-7 scheme is slightly more dissipative on the coarse grid but agrees well with DNS on the refined grid. The KEP scheme ~\cite{chandrashekar2013kinetic}, a second-order finite-volume formulation, matches the DNS dissipation rate well until $t\approx5$ but deviates at later times on both grids. The KEP scheme is included in the comparison precisely because it represents the limiting case of zero acoustic dissipation by construction, and its competitive performance here suggests that zero acoustic dissipation is not inherently problematic in smooth shock-free flows. The wave-appropriate schemes are competitive with KEP in this setting and remain applicable to flows with shocks and contact discontinuities, making them more general in scope. The periodic shear layer test in Section~\ref{sec:shear_layer} demonstrates where the fully central scheme approach breaks down.

\begin{figure}[H]
\centering
\begin{subfigure}{0.45\textwidth}
  \includegraphics[width=\textwidth]{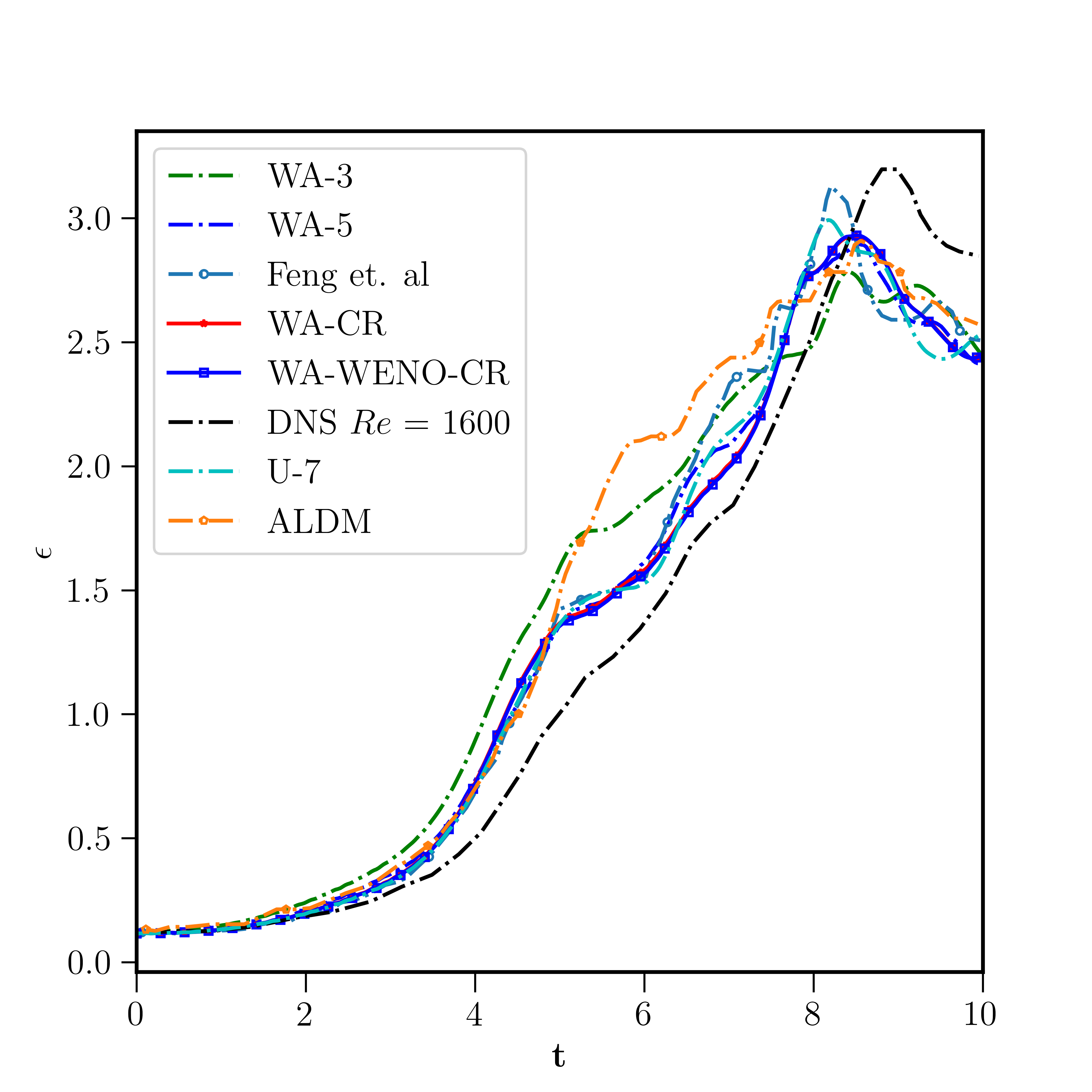}
  \caption{Cross-scheme comparison, $64^3$}
  \label{fig:vtgv_cross}
\end{subfigure}
\begin{subfigure}{0.46\textwidth}
  \includegraphics[width=\textwidth]{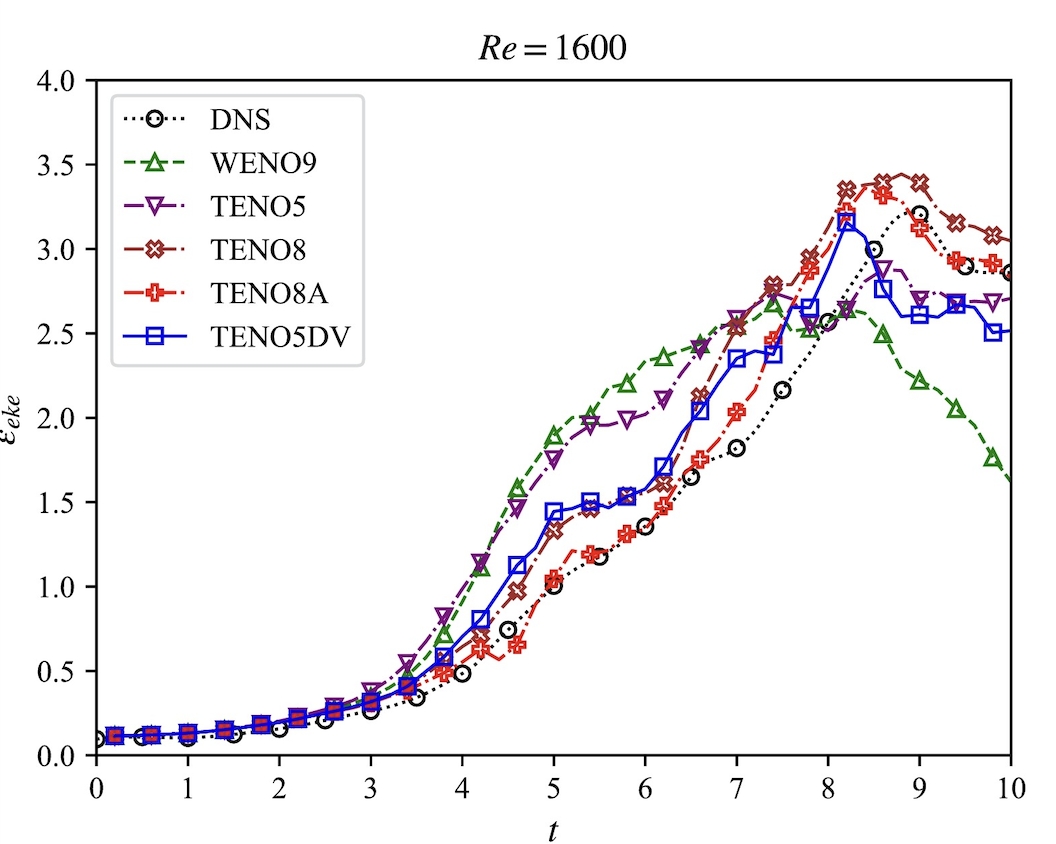}
  \caption{Comparison with Feng et al.~\cite{feng2024general}, $64^3$}
  \label{fig:vtgv_feng}
\end{subfigure}
\begin{subfigure}{0.45\textwidth}
  \includegraphics[width=\textwidth]{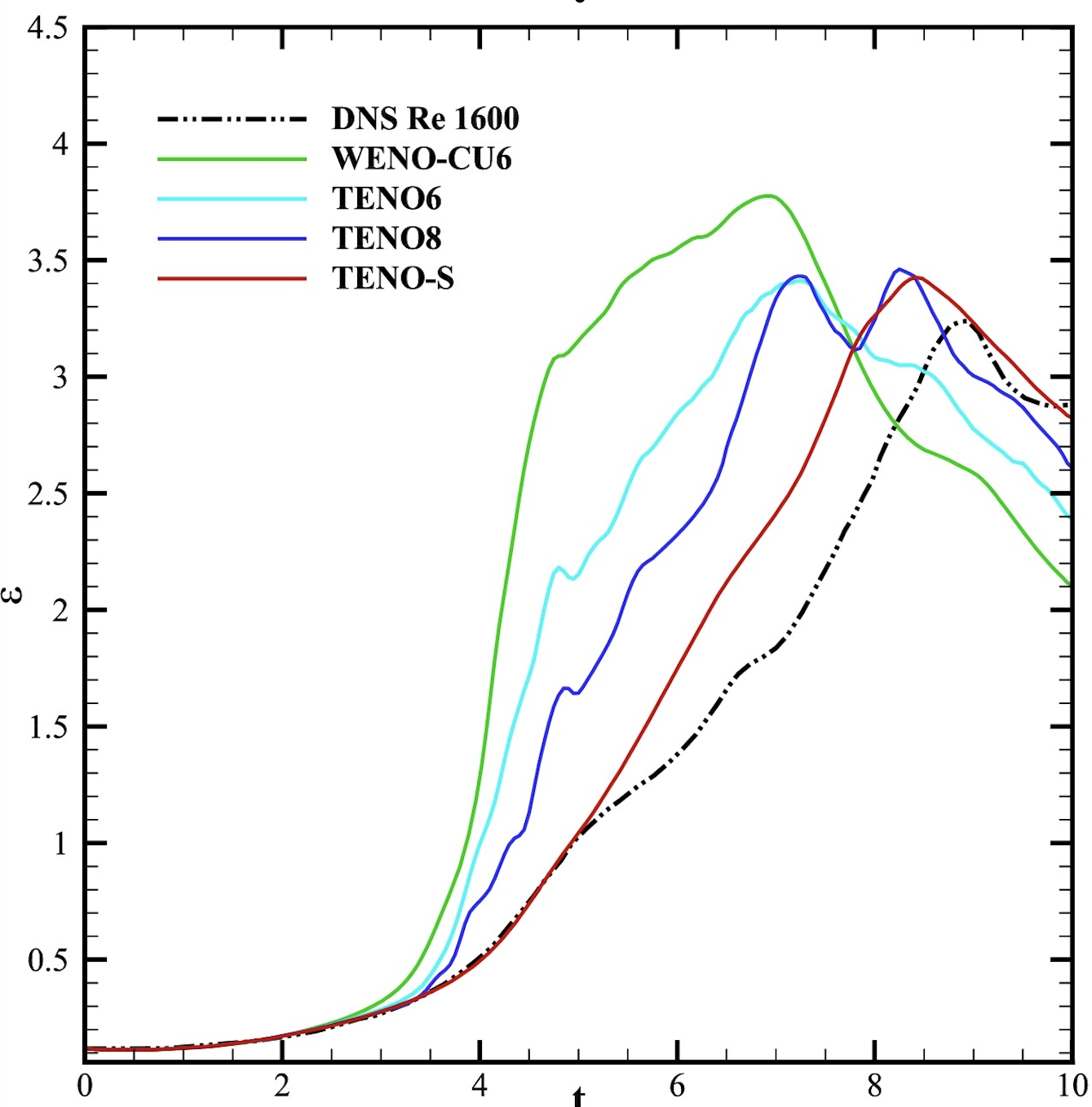}
  \caption{Comparison with Liang and Fu~\cite{liang2024new}, $64^3$}
  \label{fig:vtgv_liang}
\end{subfigure}
\begin{subfigure}{0.45\textwidth}
  \includegraphics[width=\textwidth]{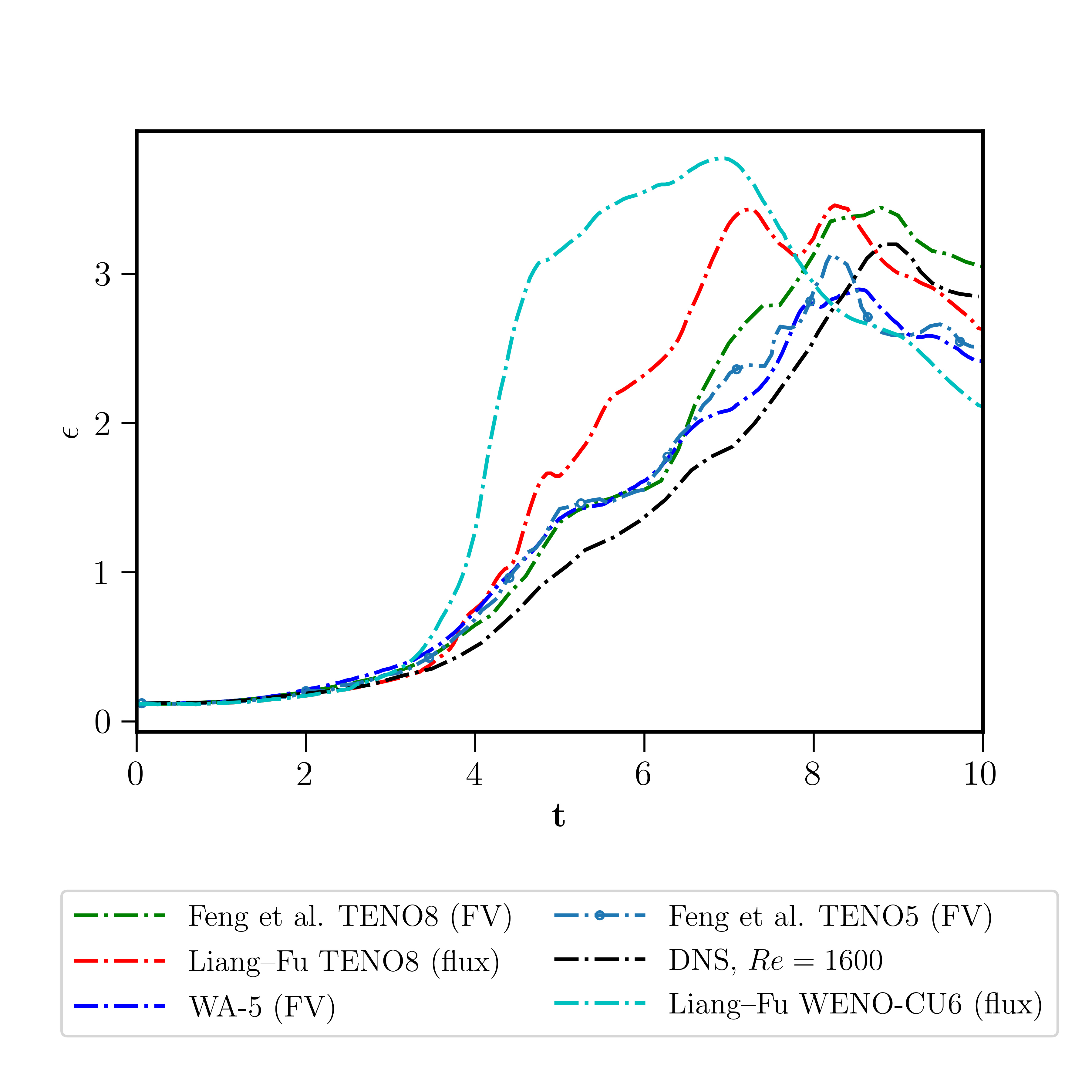}
  \caption{\textcolor{black}{Variable (FV) $\&$ flux-reconstruction comparison}}
  \label{fig:vtgv_combined}
\end{subfigure}
\caption{Viscous Taylor-Green vortex ($Re=1600$, $\mathrm{Ma}=0.1$,
$64^3$ grid, Sec.~\ref{sec:vtgv}): time evolution of volume-averaged kinetic energy
dissipation rate $\epsilon(t)$. WA-3 and WA-5 use the optimized bias
$\eta_a^*$. WA-CR and WA-WENO-CR overlap WA-5 to plotting accuracy,
confirming that the rank-1 correction introduces no accuracy penalty in
smooth flows. \textcolor{black}{Panel (d) overlays the relevant FV variable-reconstruction and flux-formulation literature curves to show that the named reconstruction family alone does not determine the effective dissipation of the complete solver.} DNS reference from Brachet et al.~\cite{brachet1983small}. Figure~\ref{fig:vtgv_liang} is reproduced with permission from Elsevier, License number 6226831189460.}
\label{fig:vtgv_comparison}
\end{figure}

Figure~\ref{fig:vtgv_cross} shows a broader comparison on the $64^3$ grid including WA-WENO-CR, Feng et al.~\cite{feng2024general} (TENO5DV), and ALDM (Adaptive Local Deconvolution Method)~\cite{hickel2006adaptive}. WA-5, WA-CR, and WA-WENO-CR produce nearly identical results and outperform both TENO5DV and ALDM. The agreement between WA-CR and WA-WENO-CR confirms that the rank-1 correction is limiter-agnostic: replacing MP5 with WENO does not affect the dissipation rate for discontinuity-free cases, and both use the same $\eta_a^*=0.6010$.

\textcolor{black}{Figure~\ref{fig:vtgv_feng} is taken from Feng et al.~\cite{feng2024general} and Figure~\ref{fig:vtgv_liang} from Liang and Fu~\cite{liang2024new}. Figure~\ref{fig:vtgv_combined} overlays the relevant curves from these two studies with the present WA-5 result and the DNS reference on the same axes. The comparison is useful because both studies include TENO8-type results, but the corresponding dissipation curves differ substantially. This shows that the label ``TENO8'' does not by itself determine the behavior of the complete solver. A TENO8 flux/finite-difference formulation and a TENO8 finite-volume variable-reconstruction formulation can have different effective dissipation in under-resolved turbulence.}

\textcolor{black}{This distinction is related to the common convention in the shock-capturing literature, where schemes are often named by the reconstruction polynomial, stencil family, or one-dimensional flux formulation. For example, finite-volume ENO-type schemes are frequently referred to as fifth-order or high-order schemes based on their reconstruction procedure~\cite{deng2019fifth}. Similarly, five-point optimized TENO schemes may be described by their stencil class even when the optimized background linear scheme is only third-order ~\cite{huang2023fivepoint}. This convention is useful, but it can obscure the difference between reconstruction order, optimized spectral behavior, and the formal order of the complete multidimensional discretization.}

\textcolor{black}{For this reason, the present comparison distinguishes between finite-volume variable-reconstruction results, denoted here as FV, and high-order flux-formulation results, denoted as flux. Feng et al. use high-order TENO reconstruction in a finite-volume framework without matching high-order Gaussian quadrature for flux integration, whereas Liang and Fu employ high-order flux/finite-difference formulations. In Figure~\ref{fig:vtgv_combined}, the FV variable-reconstruction results are closer to the DNS dissipation curve than the high-order flux-formulation results on this coarse grid. This indicates that nominal order alone is not sufficient: the effective dissipation of the complete scheme, including its spectral behavior and implementation form, controls the under-resolved Taylor--Green vortex result. The present WA schemes perform favorably in this comparison because they impose an additional physical constraint that is absent from purely stencil- or spectral-optimization viewpoints: dissipation is assigned according to the characteristic wave family, with non-acoustic waves kept minimally dissipative in smooth regions and the acoustic bias $\eta_a^*$ optimized for stability and accuracy.}

In implicit large-eddy simulation (ILES), the numerical truncation error acts as the subgrid-scale model~\cite{grinstein2007implicit}. In the wave-appropriate framework, as the results suggest, $\eta_a$ is the only tunable dissipation mechanism in smooth regions. Non-acoustic waves are centralized. The Ducros sensor and MP limiting (or WENO) contribute only near discontinuities. The optimized value $\eta_a^*$ marks the least dissipative implicit subgrid model the scheme can support without instability.

\subsubsection{Supersonic viscous Taylor-Green vortex ($\mathrm{Ma}=1.25$)}
\label{sec:stgv}

The supersonic viscous TGV at $\mathrm{Ma}=1.25$ and $Re=1600$ tests the scheme in a compressible turbulent regime. Here, acoustic effects are significant. Unlike the subsonic case, the volume-averaged kinetic energy initially increases above its initial value. This increase is due to the conversion of acoustic energy before viscous dissipation takes over \cite{lusher2021assessment}. Simulations use $64^3$ and $128^3$ grids and are compared against the DNS reference \cite{lusher2021assessment}.

\begin{figure}[H]
\centering
\begin{subfigure}{0.45\textwidth}
  \includegraphics[width=\textwidth]{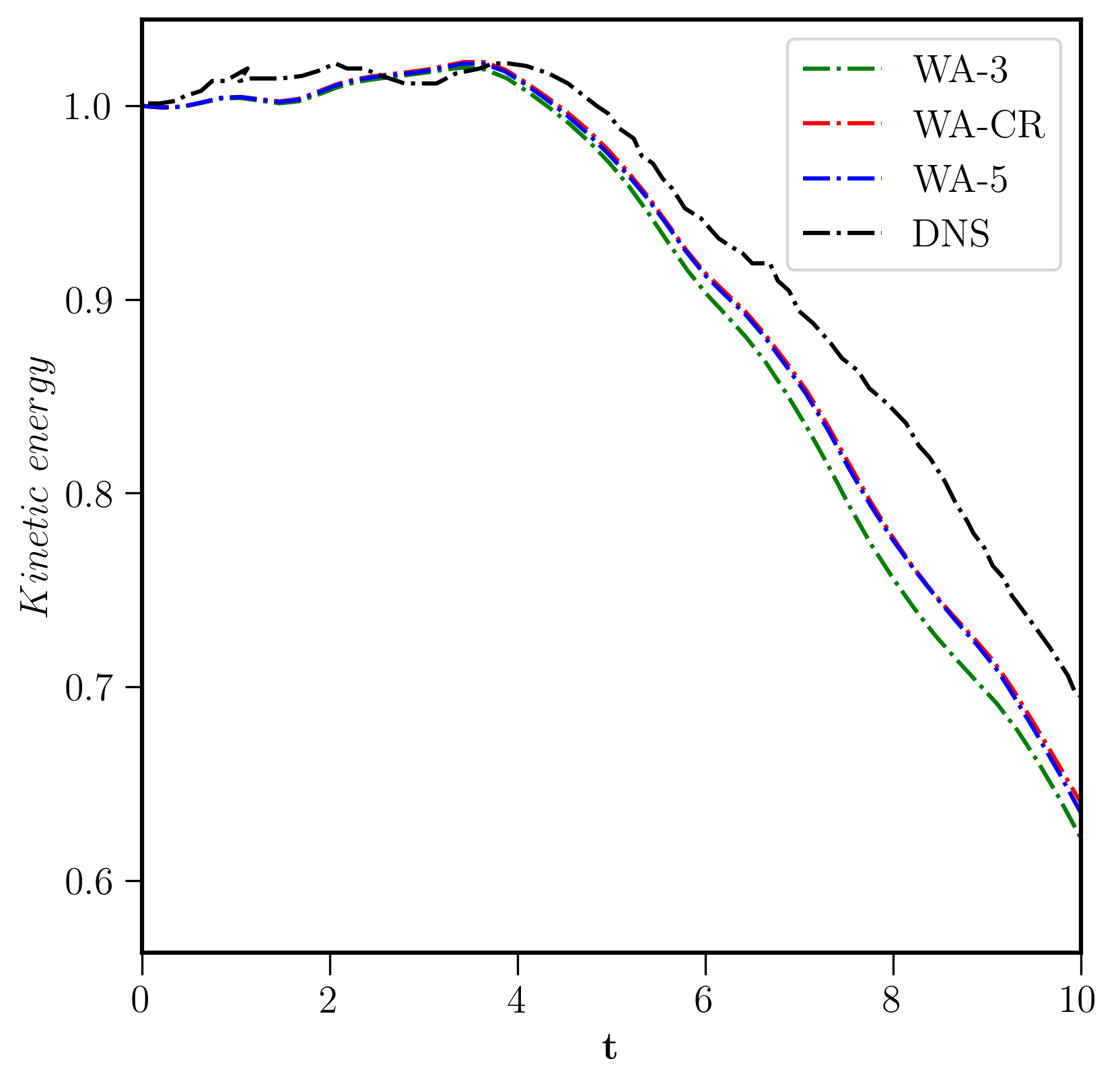}
  \caption{$64^3$}
  \label{fig:stgv_64}
\end{subfigure}\hfill
\begin{subfigure}{0.45\textwidth}
  \includegraphics[width=\textwidth]{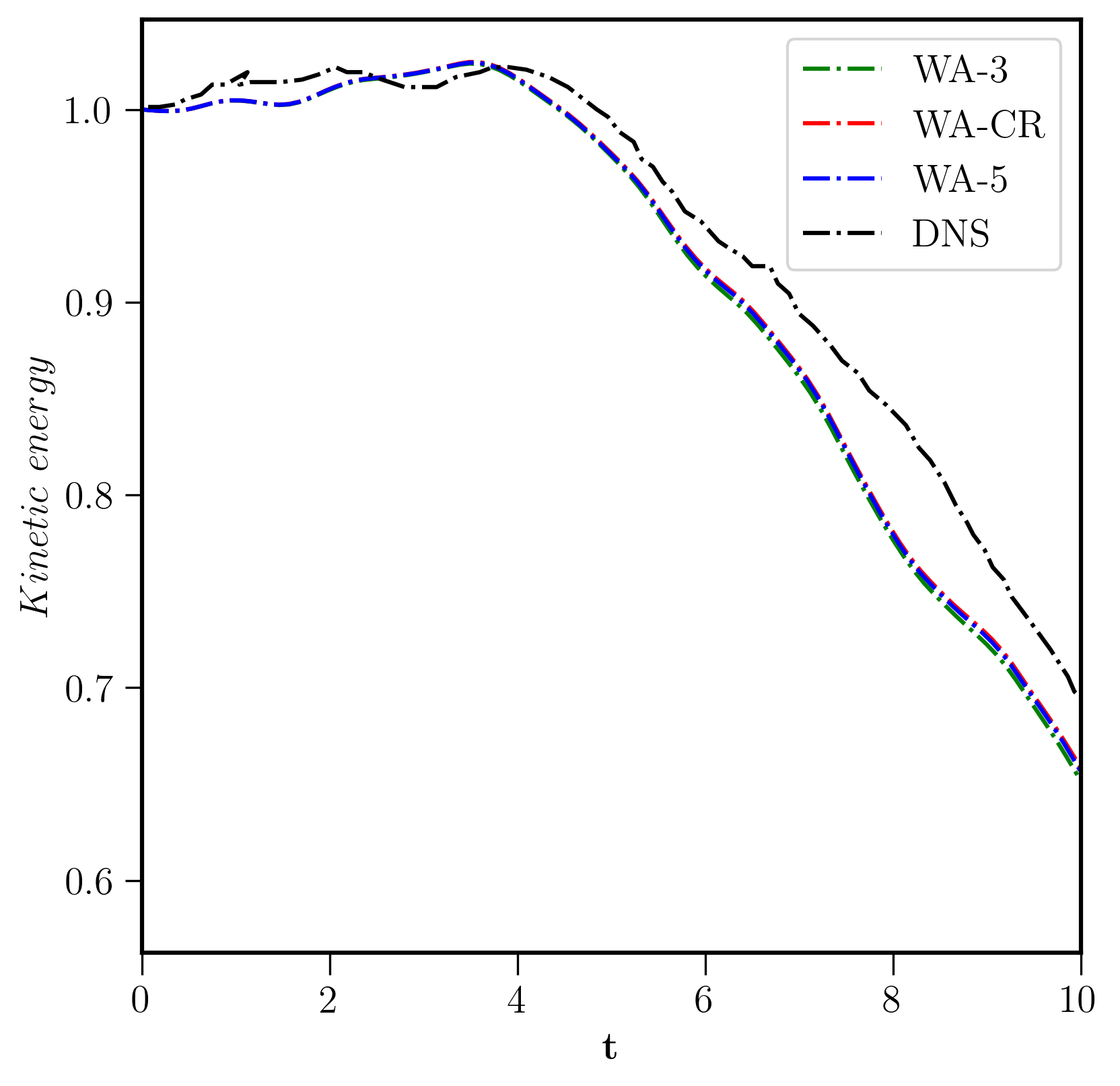}
  \caption{$128^3$}
  \label{fig:stgv_128}
\end{subfigure}
\caption{Supersonic viscous Taylor-Green vortex ($\mathrm{Ma}=1.25$,
$Re=1600$, Sec.~\ref{sec:stgv}): time evolution of volume-averaged kinetic energy at $64^3$
and $128^3$. The initial increase above unity reflects acoustic-to-kinetic
energy conversion at supersonic Mach number. WA-CR overlaps WA-5 on both
grids. All schemes converge toward the DNS reference with mesh
refinement.}
\label{fig:stgv}
\end{figure}

Figure~\ref{fig:stgv} shows the kinetic energy evolution at both resolutions. On the $64^3$ grid, WA-5 and WA-CR overlap to plotting accuracy. WA-3 decays slightly faster at late times due to higher background dissipation. On the $128^3$ grid, all three schemes converge and give nearly identical results. Both the initial kinetic energy overshoot and the subsequent decay closely match the DNS profile. The agreement between WA-CR and WA-5 on both grids confirms that the conservative reconstruction path introduces no accuracy penalty in compressible turbulent flows.

\subsection{Double periodic shear layer}
\label{sec:shear_layer}

The double periodic shear layer serves as the counterpart to the inviscid Taylor-Green vortex in understanding the effects of acoustic wave dissipation. The inviscid TGV demonstrated that the KEP scheme with zero dissipation can perform adequately in smooth homogeneous turbulence; the shear layer demonstrates that the same approach is unstable when vorticity gradients are sharp and sustained. Schranner et al.~\cite{schranner2013physically} noted that central discretization requires artificial viscosity for stability while upwind discretization is excessively dissipative, identifying the challenge of devising a scheme that simultaneously provides high wavenumber resolution and sufficient dissipation to prevent vortical instability. The wave-appropriate framework resolves this by restricting upwinding to acoustic characteristic waves, leaving the vortical waves discretized centrally. The initial conditions~\cite{minion1997performance} for this case over $[0,1]^2$ are

\begin{subequations}\label{eq:dsl_ic}
\begin{align}
\rho &= 1,\quad p = \frac{1}{\gamma\,\mathrm{Ma}^2},\quad
u = \begin{cases}
      \tanh[\theta(y-0.25)], & y\leq0.5, \\
      \tanh[\theta(0.75-y)], & y>0.5,
    \end{cases} \\
v &= 0.05\sin[2\pi(x+0.25)],
\end{align}
\end{subequations}
with $\mathrm{Ma}=0.1$, $\gamma=1.4$, and shear layer width $\theta=80$. The flow is inviscid. A reference solution is computed on a $1600\times1600$ grid, and test simulations use a $320\times320$ grid. Since this test has no shocks or contact discontinuities, the Ducros sensor is never triggered, and all numerical dissipation is attributable solely to the acoustic upwind bias $\eta_a$, making it a clean diagnostic for the effect of $\eta_a$ on vortical structures.  Importantly, $\eta_a^*$ was determined exclusively from the Taylor-Green vortex suite and was not tuned for the shear layer in any way. The fact that it prevents spurious vortex generation on this qualitatively different flow suggests that $\eta_a^*$ is intrinsic to the scheme.

\begin{figure}[H]
\centering
\begin{subfigure}{0.35\textwidth}
  \includegraphics[width=\textwidth]{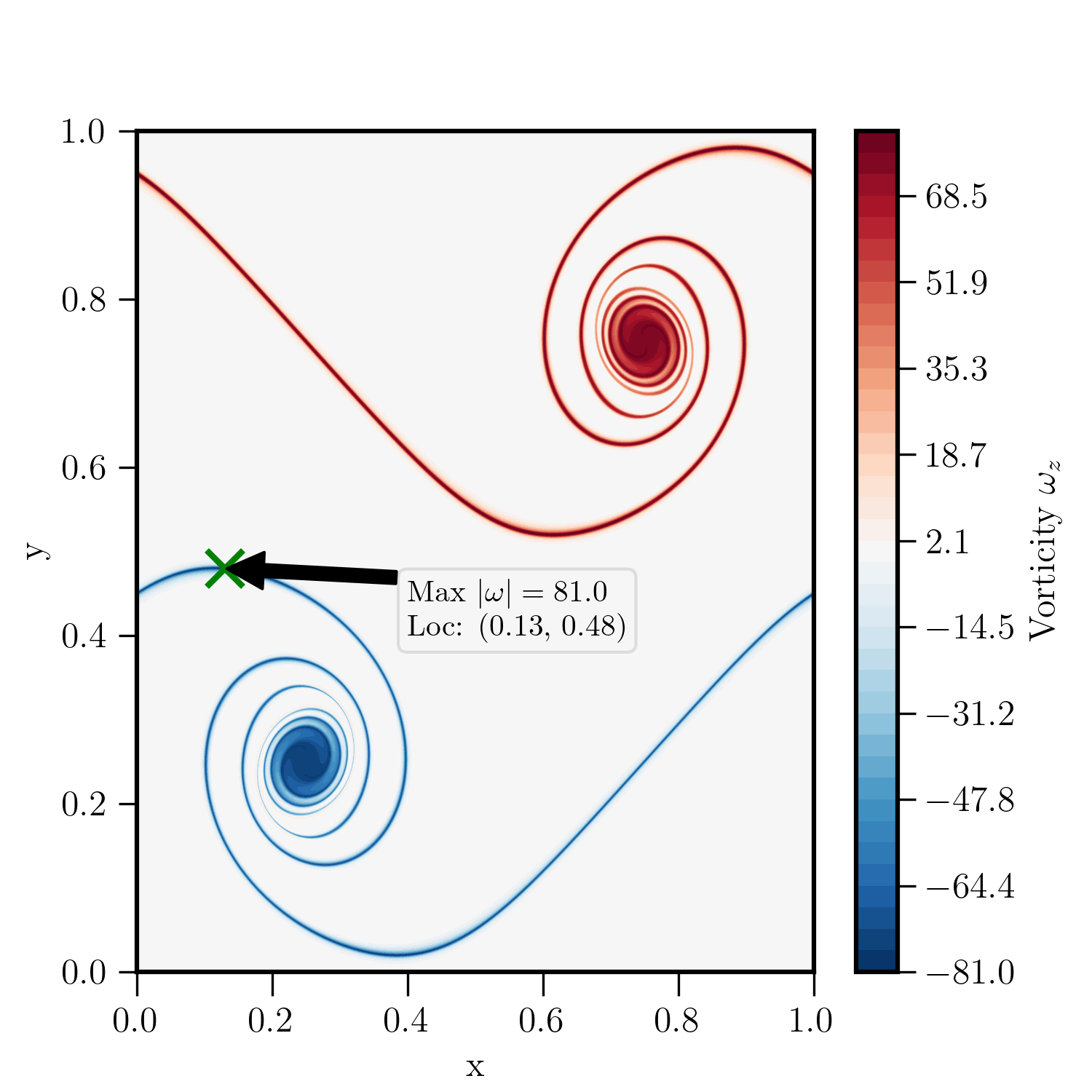}
  \caption{Reference ($1600^2$)}
  \label{fig:dsl_ref_a}
\end{subfigure}%
\begin{subfigure}{0.35\textwidth}
  \includegraphics[width=\textwidth]{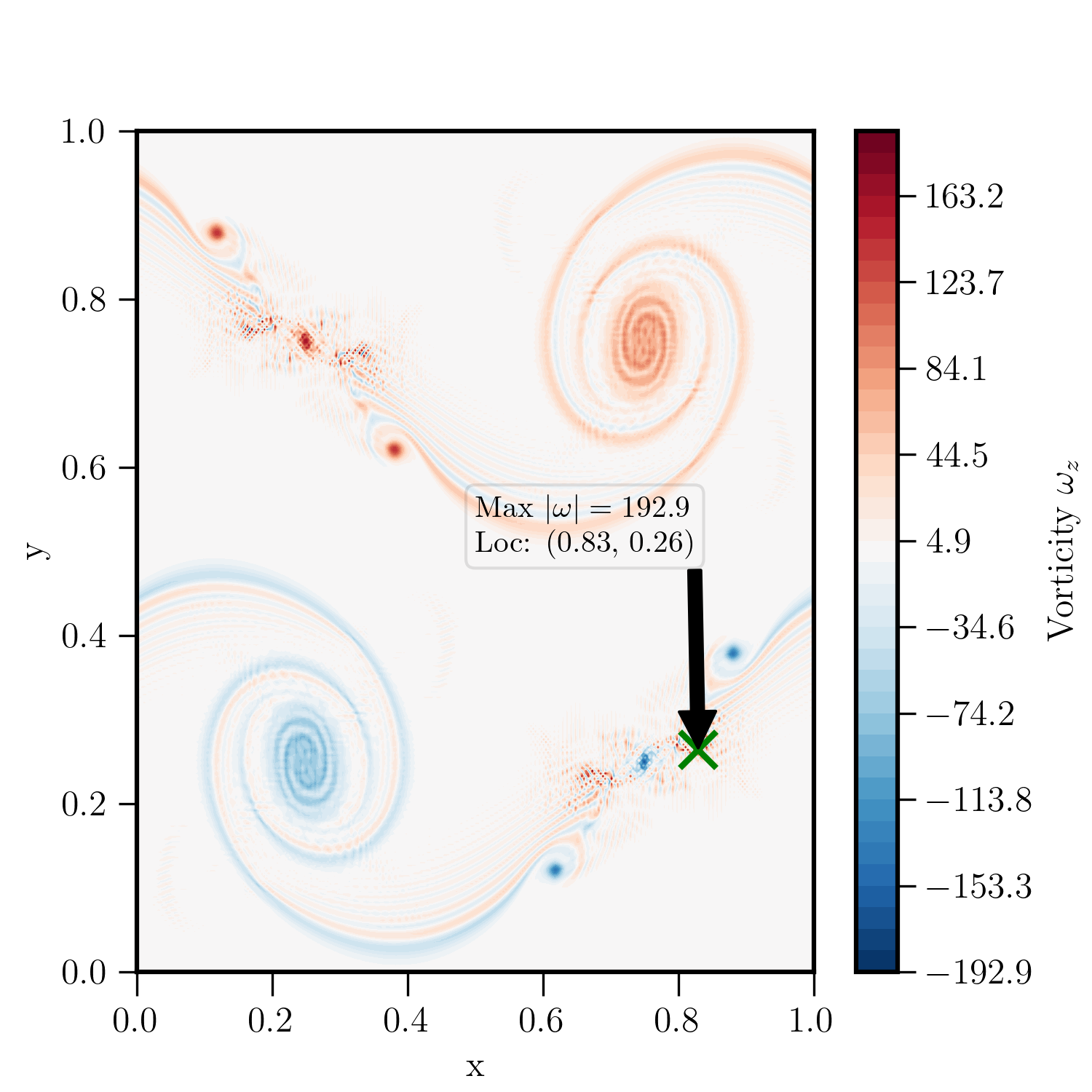}
  \caption{KEP}
  \label{fig:dsl_ref_b}
\end{subfigure}
\begin{subfigure}{0.35\textwidth}
  \includegraphics[width=\textwidth]{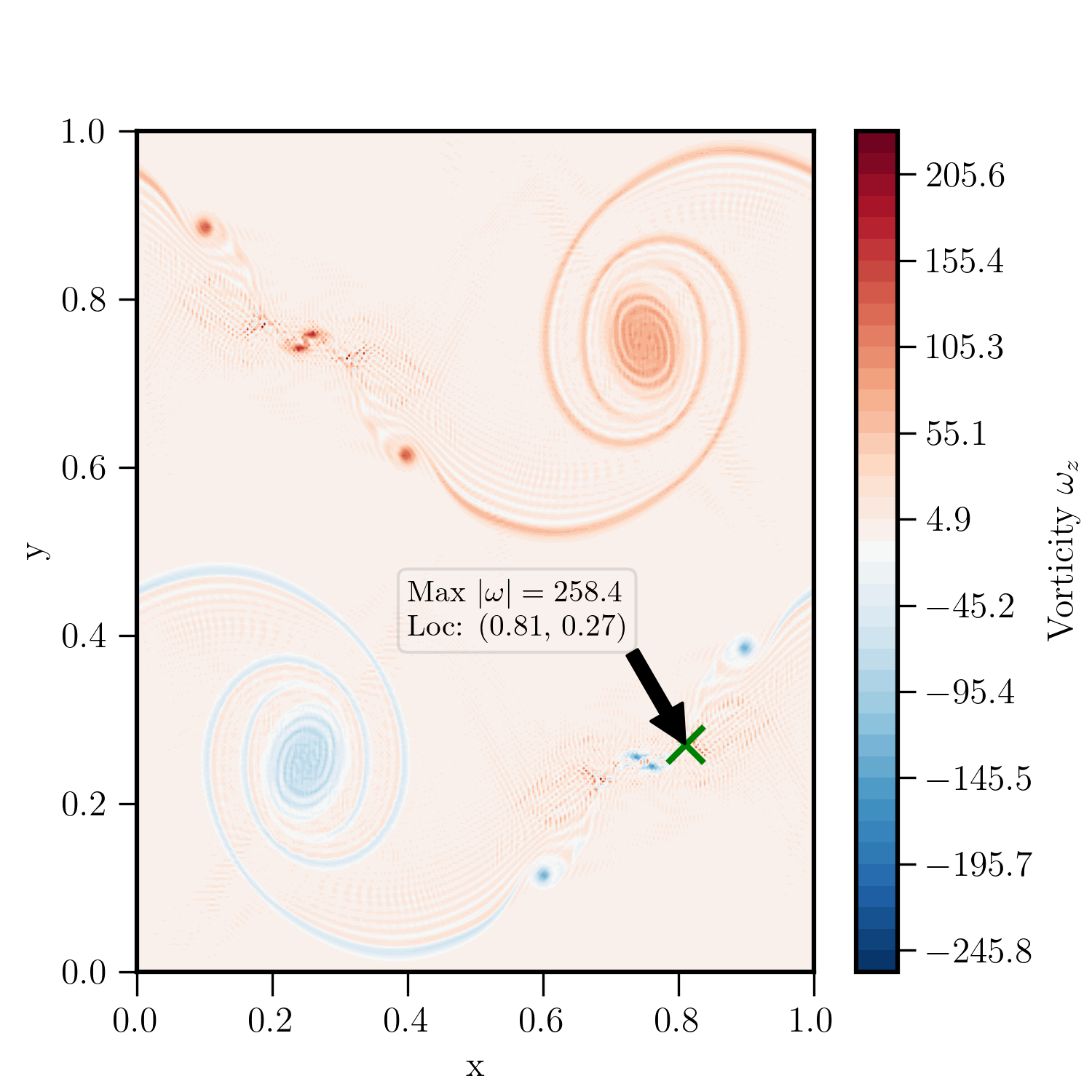}
  \caption{WA-3, $\eta_a=0.5$}
  \label{fig:dsl_ref_c}
\end{subfigure}
\begin{subfigure}{0.35\textwidth}
  \includegraphics[width=\textwidth]{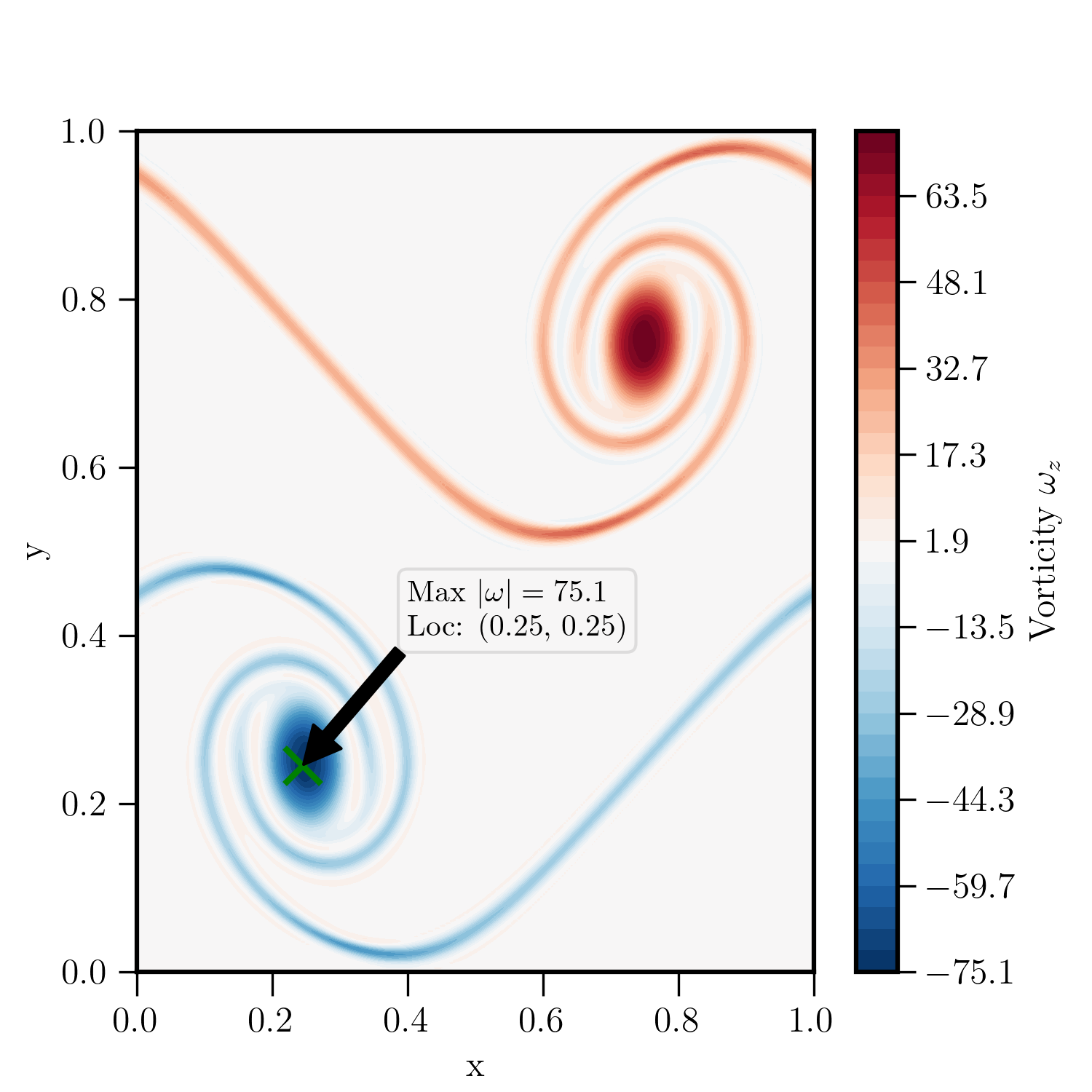}
  \caption{WA-3, $\eta_a=1$}
  \label{fig:dsl_ref_d}
\end{subfigure}
\caption{Double periodic shear layer ($\theta=80$, inviscid, $t=1$,
$320\times320$ grid, Sec.~\ref{sec:shear_layer}): $z$-vorticity contours for the reference solution
and three baseline variants. The KEP scheme and WA-3 at $\eta_a=0.5$ are both unstable with 
oscillations; WA-3 at full upwinding ($\eta_a=1$) is stable but 
overly dissipative.}
\label{fig:dsl_ref}
\end{figure}

Figure~\ref{fig:dsl_ref} presents the reference $z$-vorticity and compares three baseline variants on the coarse grid to clarify the specific behaviors of each scheme. The $z$-vorticity field is calculated using spectral derivatives via the fast Fourier transform, eliminating numerical differentiation error from the vorticity calculation regardless of the flow solver's order. When using the KEP scheme, which conserves energy exactly, spurious braid vortices do appear, underscoring the need for a controlled dissipation to suppress vortices and that energy conservation alone is insufficient. \textcolor{black}{WA-3 at $\eta_a=0.5$ (fully central acoustics) also produces oscillations characteristic of those discussed in Section~\ref{sec:intro}.} In contrast, WA-3 at $\eta_a=1.0$ is overly dissipative but effectively eliminates both spurious vortices and oscillations.

\begin{figure}[H]
\centering
\begin{subfigure}{0.35\textwidth}
  \includegraphics[width=\textwidth]{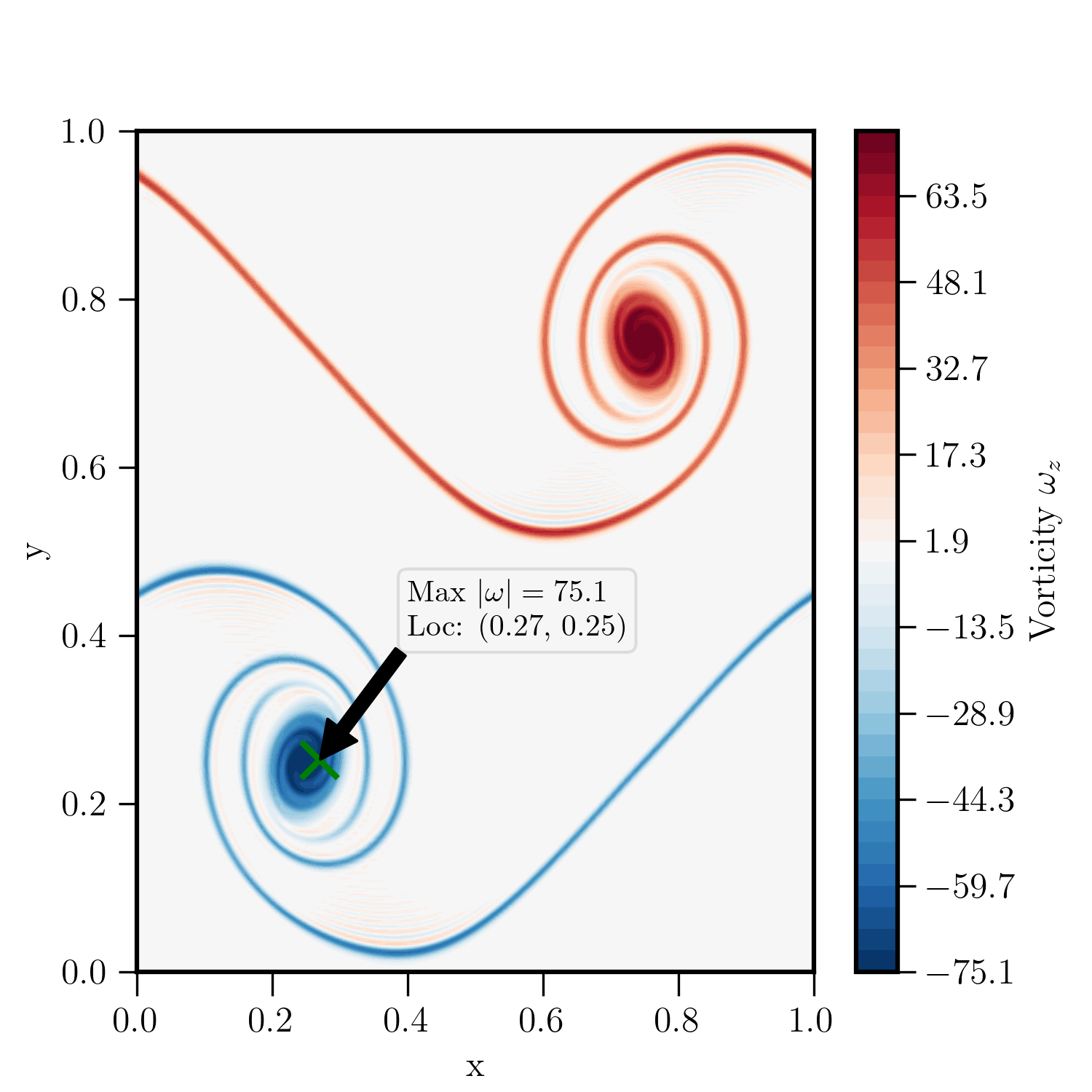}
  \caption{WA-3, $\eta_a^*=0.54$}
  \label{fig:dsl_a}
\end{subfigure}%
\begin{subfigure}{0.35\textwidth}
  \includegraphics[width=\textwidth]{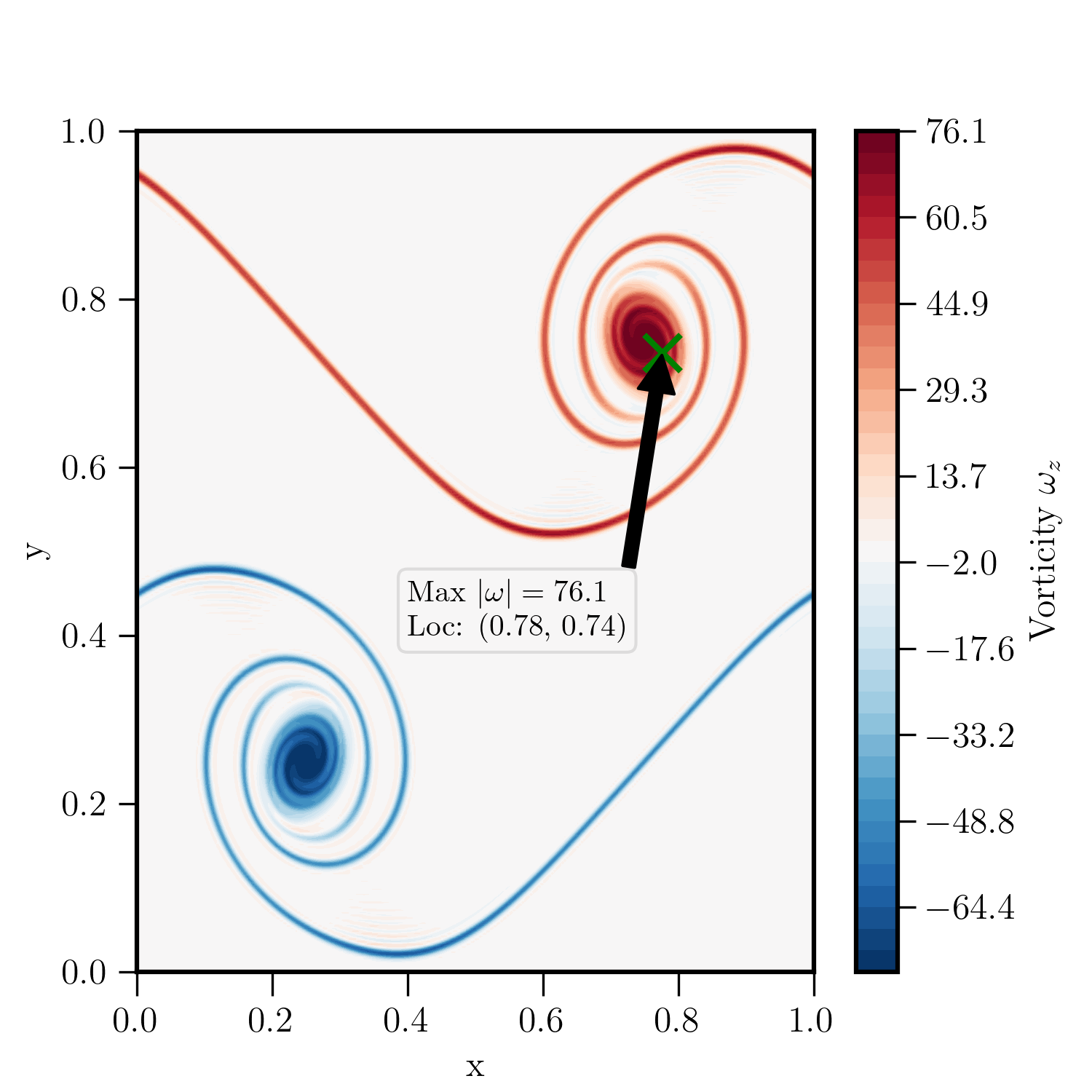}
  \caption{WA-5, $\eta_a^*=0.6010$}
  \label{fig:dsl_b}
\end{subfigure}
\begin{subfigure}{0.35\textwidth}
  \includegraphics[width=\textwidth]{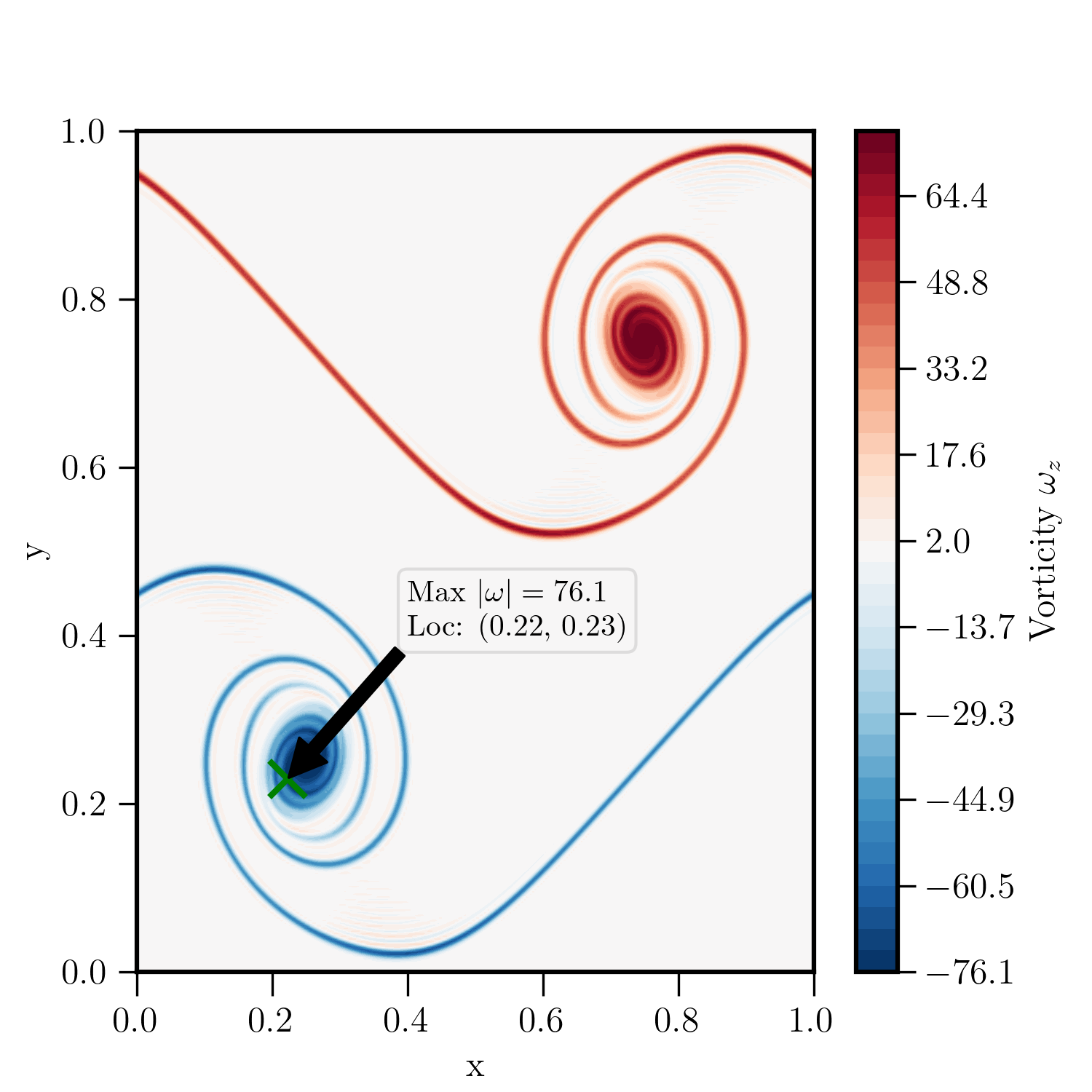}
  \caption{WA-CR}
  \label{fig:dsl_c}
\end{subfigure}%
\begin{subfigure}{0.35\textwidth}
  \includegraphics[width=\textwidth]{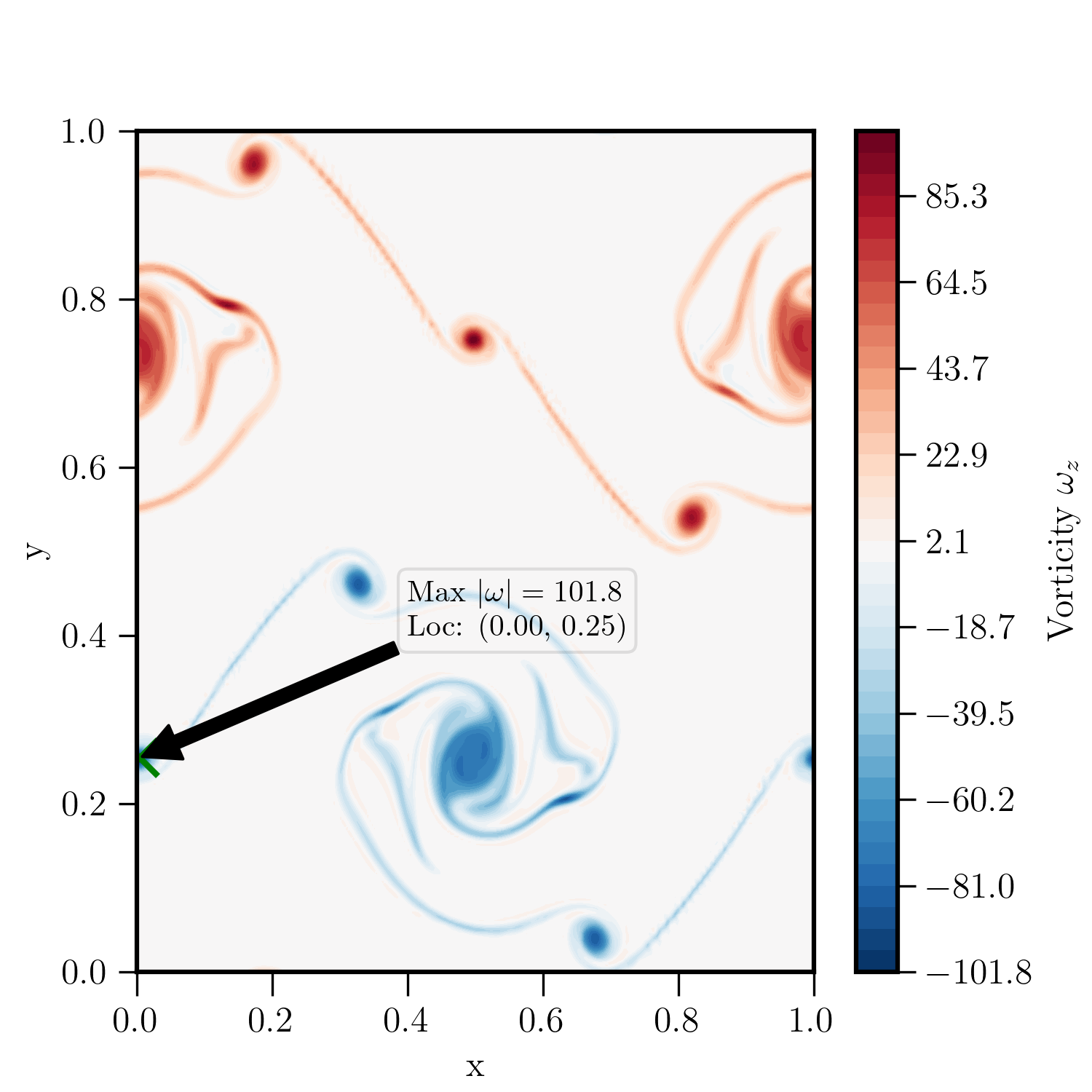}
  \caption{TENO5}
  \label{fig:dsl_d}
\end{subfigure}%
\begin{subfigure}{0.35\textwidth}
  \includegraphics[width=\textwidth]{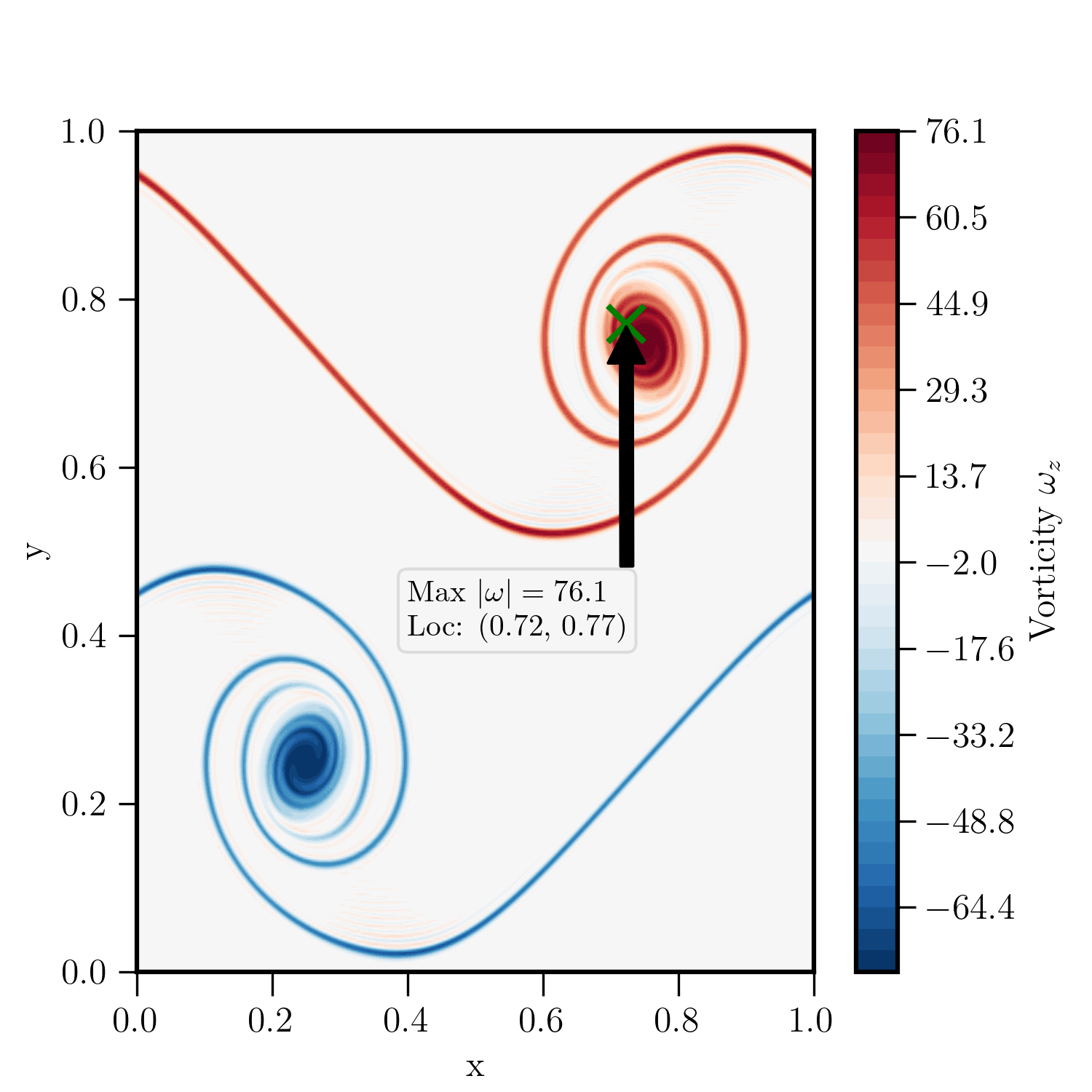}
  \caption{WA-WENO-CR}
  \label{fig:dsl_e}
\end{subfigure}
\caption{Double periodic shear layer ($\theta=80$, inviscid, $t=1$,
$320 \times 320$ grid, Sec.~\ref{sec:shear_layer}): $z$-vorticity contours for the optimized schemes.
WA-3 at $\eta_a^*=0.54$ and WA-5 at $\eta_a^*=0.6010$ both reproduce
two clean primary vortices. WA-CR and WA-WENO-CR match WA-5. TENO5
produces spurious braid vortices on this grid.}
\label{fig:dsl}
\end{figure}

Figure~\ref{fig:dsl} shows the results for the optimized schemes. WA-3 at $\eta_a^*=0.54$ and WA-5 at $\eta_a^*=0.6010$ both reproduce the two clean primary vortices without spurious structures on the $320\times320$ grid. WA-CR matches WA-5 to plotting accuracy, as expected for a shock-free contact-free flow where the conservative path is always active and the rank-1 correction has no effect. WA-WENO-CR also produces clean results, consistent with the limiter-agnostic property of the rank-1 correction. TENO5 produces spurious braid vortices on the $320\times320$ grid, requiring a finer grid to obtain an acceptable result.

\begin{figure}[H]
\centering
\begin{subfigure}{0.35\textwidth}
  \includegraphics[width=\textwidth]{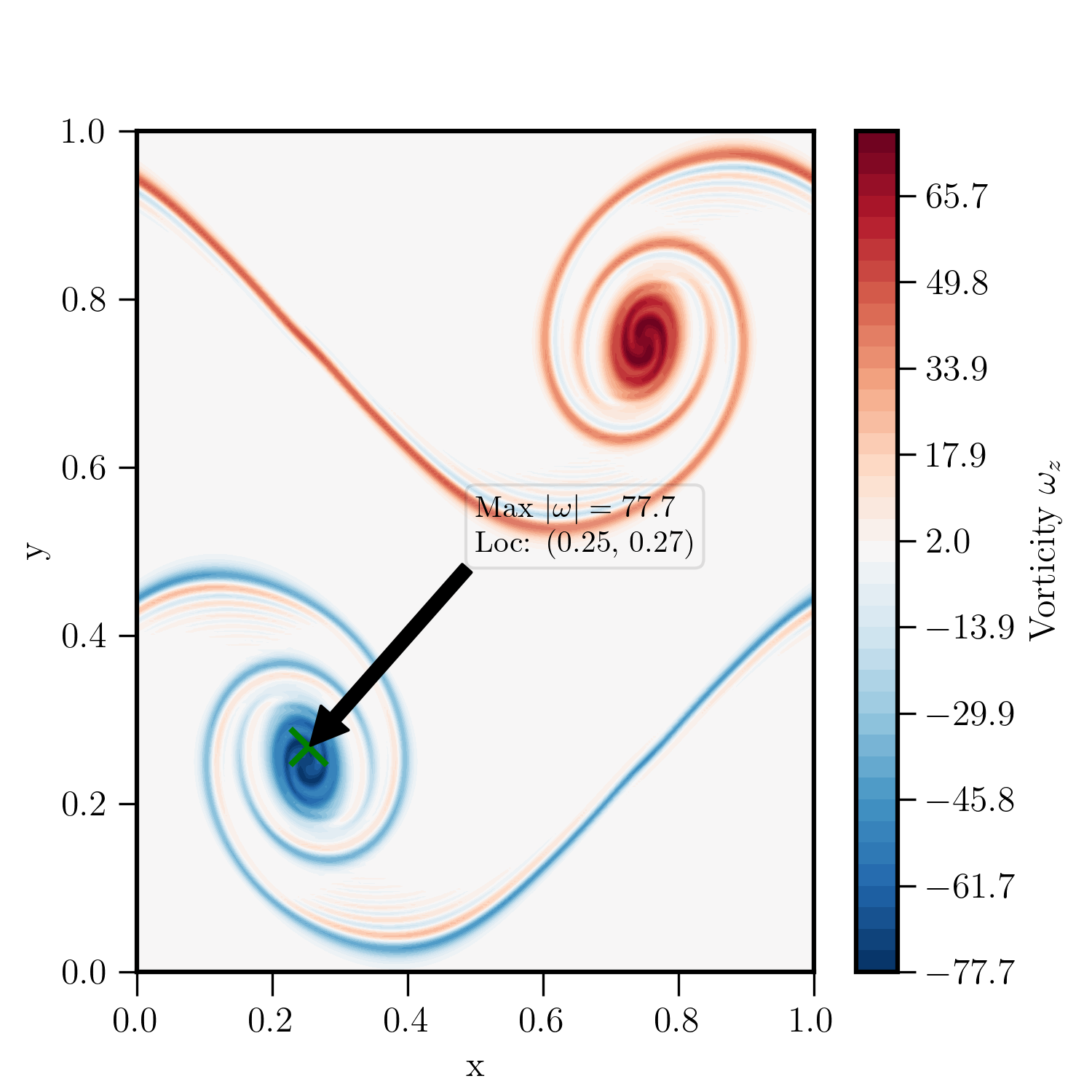}
  \caption{WA-KEP, $\eta_a =0.56$}
  \label{fig:dsl_akep}
\end{subfigure}%
\begin{subfigure}{0.35\textwidth}
  \includegraphics[width=\textwidth]{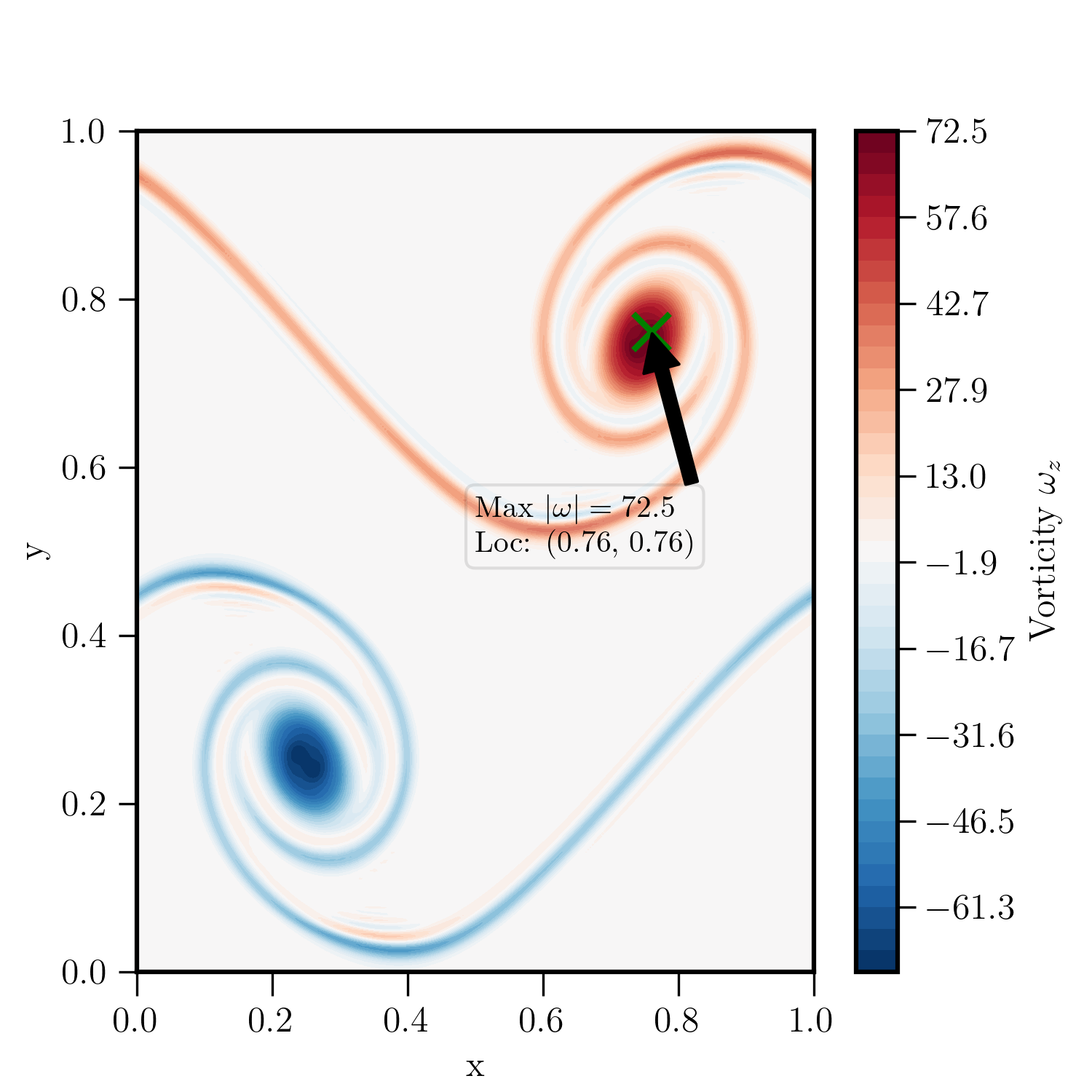}
  \caption{WA-KEP, $\eta_a =1.0$}
  \label{fig:dsl_bkep}
\end{subfigure}
\begin{subfigure}{0.95\textwidth}
  \includegraphics[width=\textwidth]{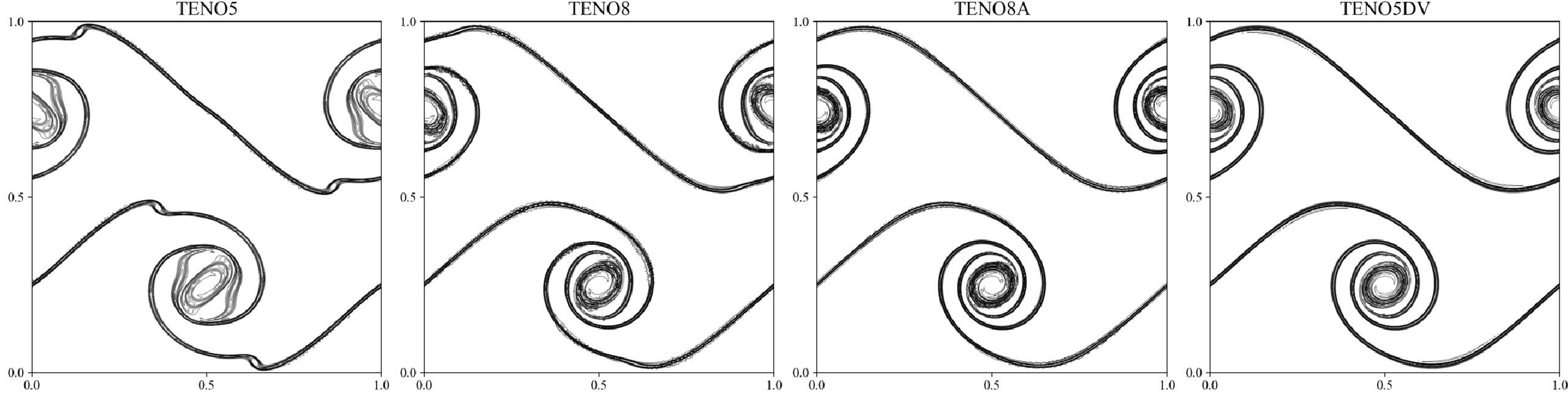}
  \caption{Figure is taken from Reference \cite{feng2024general}, where the simulations are computed on a grid size of $512^2$.}
  \label{fig:dsl_feng}
\end{subfigure}
\caption{Double periodic shear layer ($\theta=80$, inviscid, $t=1$,
$320 \times 320$ grid, Sec.~\ref{sec:shear_layer}): $z$-vorticity contours for the WA-KEP approach and that of Feng et al. \cite{feng2024general}.}
\label{fig:dsl_kep}
\end{figure}

The WA-KEP scheme, formulated in Section~\ref{sec:WA-KEP}, is tested here  on the double periodic shear layer. The unmodified KEP scheme produces  spurious braid vortices ($\omega_{z,\max} \approx 192.9$ versus a reference  of 81) because there is no dissipation for any of the fluxes or waves. Figure~\ref{fig:dsl_kep} shows  the $z$-vorticity contours at $t=1$ on the $320\times320$ grid for  $\eta_a = 0.56$ and $\eta_a = 1.0$. Introducing dissipation ($\eta_a = 0.56$) exclusively  through the normal momentum flux recovers the two clean primary vortices; the maximum $z$-vorticity is 77.7, within 4\%  of the reference. \textcolor{black}{Small residual oscillations may still be visible locally because this value was not obtained from a separate WA-KEP optimization. Increasing $\eta_a$ damps these oscillations further, but also increases acoustic dissipation.} At $\eta_a = 1.0$, the maximum vorticity is 72.5, and the shear layers are  visibly diffused due to excessive dissipation. Figure~\ref{fig:dsl_feng},  adapted from~\cite{feng2024general}, shows their results on a $512\times512$  grid, which is 2.5 times finer than the grid used here. Despite the higher  resolution and the optimized TENO8 scheme, spurious vortices are still  visible. The present schemes eliminate these structures on the coarser  $320\times320$ grid, demonstrating that physics-based wave-appropriate  dissipation is more effective than increasing resolution or reconstruction order alone.

The value $\eta_a = 0.56$ is slightly above the WA-3 threshold of  $\eta_a^* = 0.54$. This is expected: WA-3 applies either a third-order  upwind or fourth-order central scheme to the flow variables, whereas KEP  is purely second-order central apart from the normal momentum flux. In  WA-KEP, the acoustic bias is the sole source of dissipation in the entire  scheme, so it must compensate for the complete absence of dissipation on  all other wave families. The upward shift from 0.54 to 0.56 is consistent  with this reasoning.

\begin{itemize}
\item First, the acoustic dissipation enters through a completely different mechanism than in WA-3 or WA-5. In the wave-appropriate reconstruction framework, the bias is applied during the reconstruction step by blending left- and right-biased polynomials in characteristic space. In WA-KEP, the reconstruction is first-order piecewise constant, and the bias is instead applied through the Rusanov dissipative flux acting on the normal momentum alone. Despite this difference in mechanism, the stabilizing effect is identical: a controlled acoustic upwind bias suppresses the vortical instability.

\item Second, \textcolor{black}{the value $\eta_a \approx 0.56$ stabilizes the dominant shear-layer instability in a scheme with zero background dissipation on any wave family, while $\eta_a^* = 0.54$ and $\eta_a^* = 0.6010$ are the optimized values for the reconstruction-based WA-3 and WA-5 schemes. The comparison suggests that a small acoustic upwind bias above the central value of 0.5 is important across fundamentally different discretization strategies, but the WA-KEP value itself should not be interpreted as a separately optimized universal threshold.}
\end{itemize}

Table~\ref{tab:dsl_vorticity} reports the maximum $z$-vorticity at $t=1$ on the $320\times320$ grid. The reference value computed on the $1600\times1600$ grid is $\omega_{z,\max}=81$. Physically acceptable results should reproduce this value within a reasonable tolerance; values significantly above the reference indicate spurious vorticity amplification due to numerical oscillations rather than physical roll-up.

\begin{table}[H]
\centering
\caption{Maximum $z$-vorticity at $t=1$ for the double periodic shear
layer ($\theta=80$, $320\times320$ grid). Reference value:
$\omega_{z,\max}=81$ ($1600\times1600$ grid).}
\label{tab:dsl_vorticity}
\begin{tabular}{lc}
\hline
Scheme & $\omega_{z,\max}$ \\
\hline
Reference ($1600^2$) & 81 \\
WA-3                 & 75.1 \\
WA-5                 & 76.1 \\
WA-CR                & 76.1 \\
TENO5                & $\approx 101.8$ \\
KEP                  & $\approx 192.9$ \\
WA-3, $\eta_a=0.5$  & $\approx 258.4$ \\
WA-KEP, $\eta_a=0.56$  & 77.7 \\
\hline
\end{tabular}
\end{table}

WA-3, WA-5, and WA-CR all produce values within 7\% of the reference, confirming that the optimized acoustic bias yields the correct dissipation level for this flow. The cases of insufficient dissipation, KEP and WA-3 at $\eta_a=0.5$, produce values far above the reference due to unrestrained growth of numerical oscillations at the shear interface. The elevated value for TENO5 ($\approx 101.8$) has a different origin: rather than a global lack of upwinding, the TENO smoothness indicators may misidentify the sharp vorticity gradients at the shear interface as near-discontinuous, triggering low-order stencil selection and generating spurious small-scale oscillations along the interface. These oscillations introduce spurious vorticity on top of the physical roll-up, elevating the maximum above the reference level. \textcolor{black}{The wave-appropriate schemes avoid this by design: the shear/vortical waves are reconstructed with the central scheme in smooth regions, and the Ducros sensor correctly identifies the shear layer as shock-free, so no shock-triggered upwind fallback is applied.} WA-KEP at $\eta_a = 0.56$ also produces a maximum vorticity within 4\% of  the fine-grid reference, consistent with the wave-appropriate schemes.

\subsection{Rayleigh-Taylor instability}
\label{sec:rt}

The Rayleigh-Taylor instability develops when a heavy fluid is accelerated into a lighter one, producing a hierarchy of mushroom-cap structures whose fine-scale detail is highly sensitive to numerical dissipation. It is a particularly challenging test for the rank-1 entropy wave correction because the interface between the two fluids is a pure contact discontinuity: density jumps sharply while velocity and pressure remain continuous, which is precisely the configuration that a conservative reconstruction without entropy wave correction would mishandle. The Ducros sensor remains inactive throughout, since there are no shocks (the pressure-based Ducros sensor cannot activate at a contact discontinuity because pressure remains continuous there~\cite{chamarthi2023wave}), so all interface treatment is performed solely by the rank-1 correction. The initial conditions on $[0,\tfrac{1}{4}]\times[0,1]$ are
\begin{equation}\label{eq:rt_ic}
(\rho,u,v,p)=\begin{cases}
(2,\;0,\;-0.025c\cos(8\pi x),\;2y+1), & 0\leq y<0.5,\\[3pt]
(1,\;0,\;-0.025c\cos(8\pi x),\;y+1.5), & y\geq0.5,
\end{cases}
\end{equation}
where $c=\sqrt{\gamma p/\rho}$ and $\gamma=1.4$, with appropriate source terms added to the momentum and energy equations. Boundary conditions are $(\rho,p,u,v)=(1,2.5,0,0)$ at the top and $(2,1,0,0)$ at the bottom. Figure~\ref{fig:rt} shows density contours at $t=1.95$ on the $128\times512$ and $512\times2048$ grids. Several observations follow from the results:

\begin{figure}[H]
\centering\offinterlineskip
\begin{subfigure}{0.18\textwidth}
  \includegraphics[height=0.5\textheight]{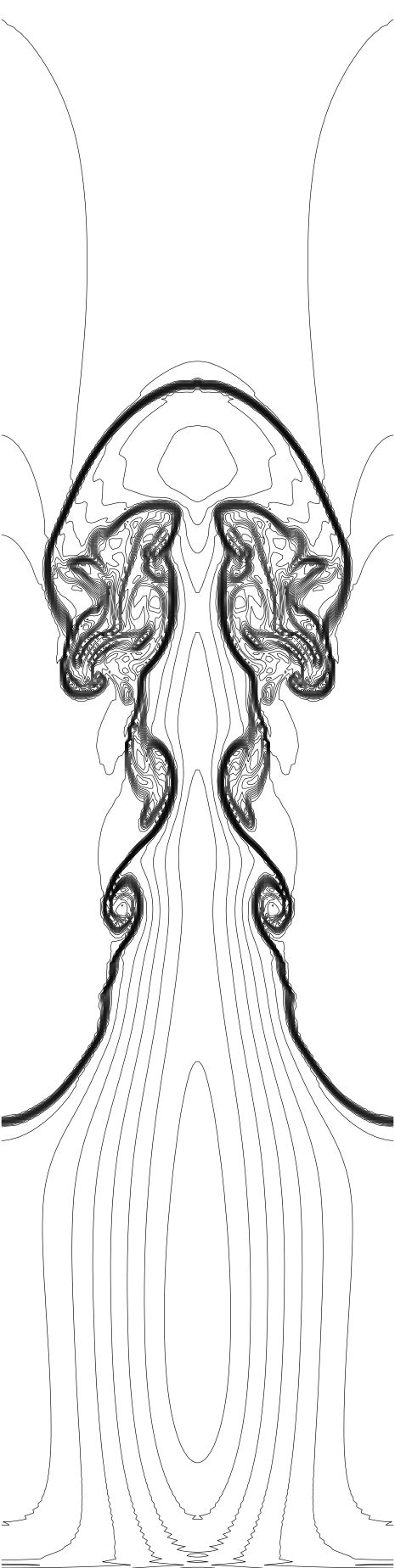}
  \caption{WA-5\\$128\times512$}
  \label{fig:rt_a}
\end{subfigure}%
\begin{subfigure}{0.18\textwidth}
  \includegraphics[height=0.5\textheight]{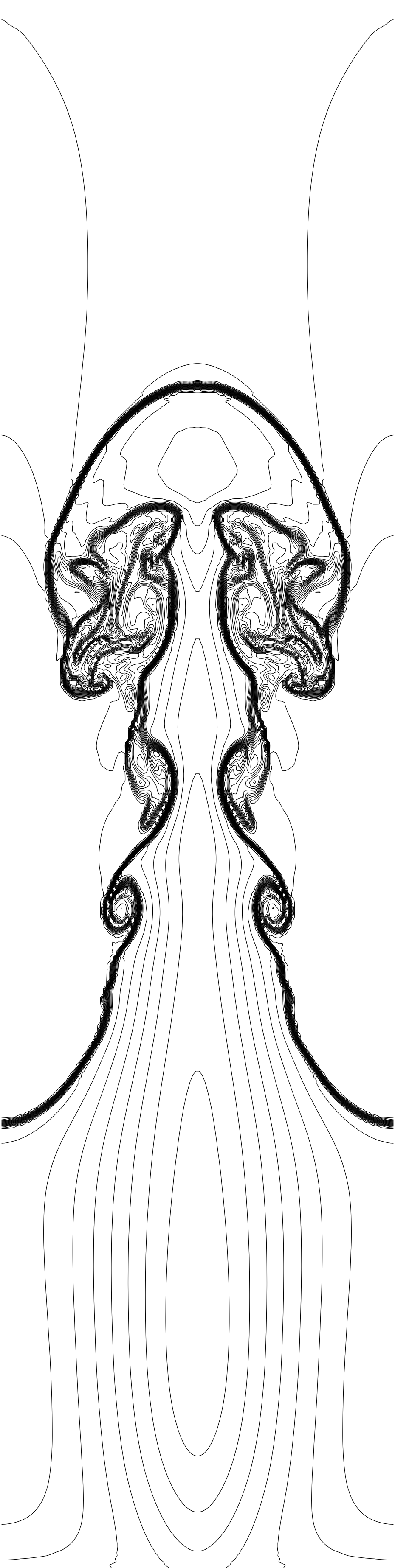}
  \caption{WA-CR\\$128\times512$}
  \label{fig:rt_b}
\end{subfigure}%
\begin{subfigure}{0.18\textwidth}
  \includegraphics[height=0.5\textheight]{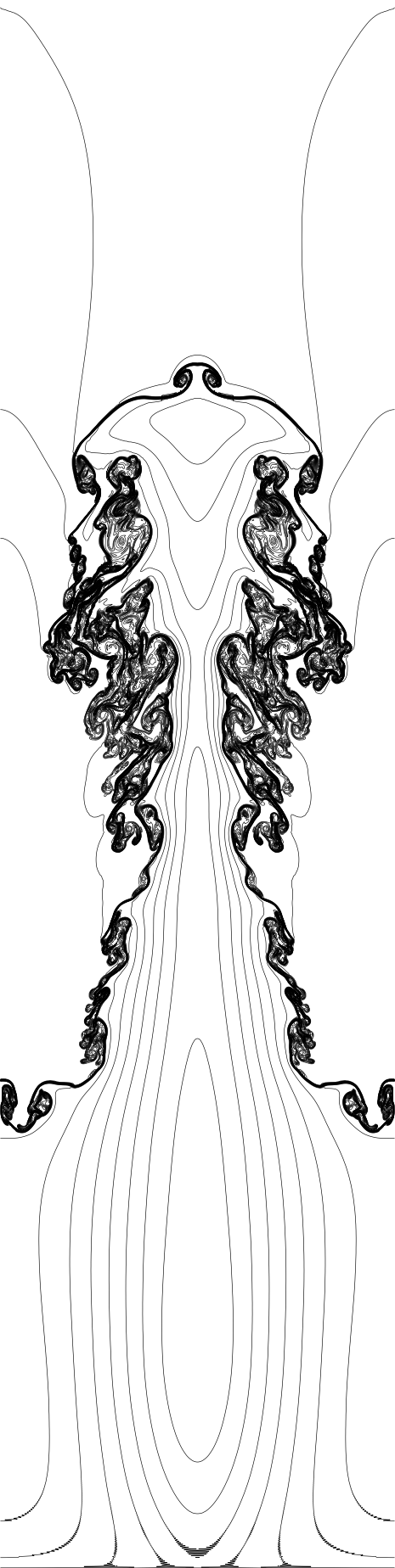}
  \caption{WA-5\\$512\times2048$}
  \label{fig:rt_c}
\end{subfigure}%
\begin{subfigure}{0.18\textwidth}
  \includegraphics[height=0.5\textheight]{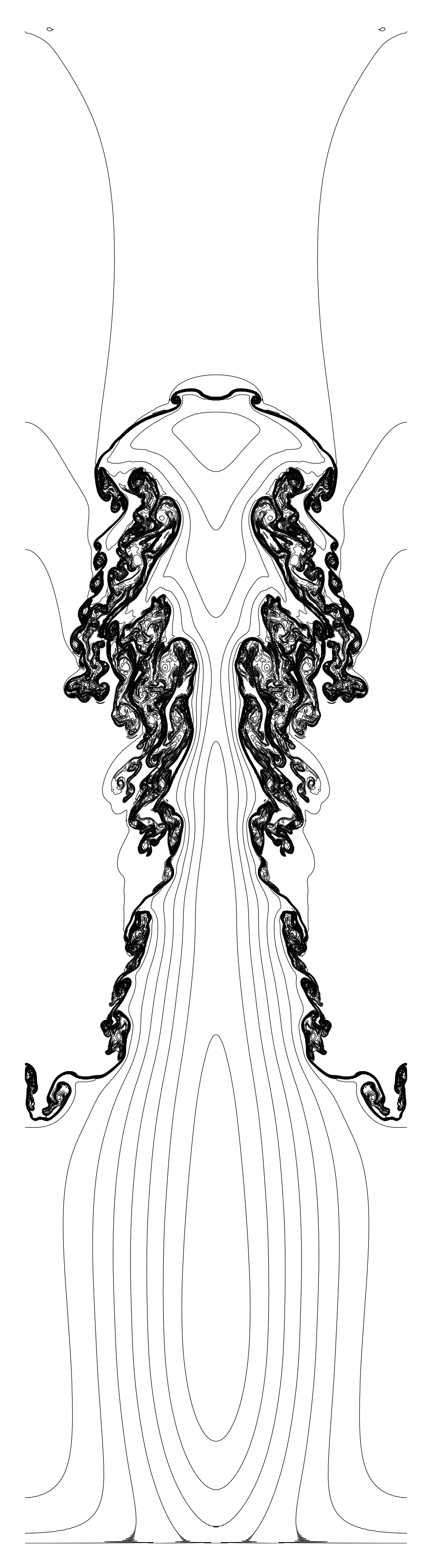}
  \caption{WA-CR\\$512\times2048$}
  \label{fig:rt_d}
\end{subfigure}%
\begin{subfigure}{0.18\textwidth}
  \includegraphics[height=0.5\textheight]{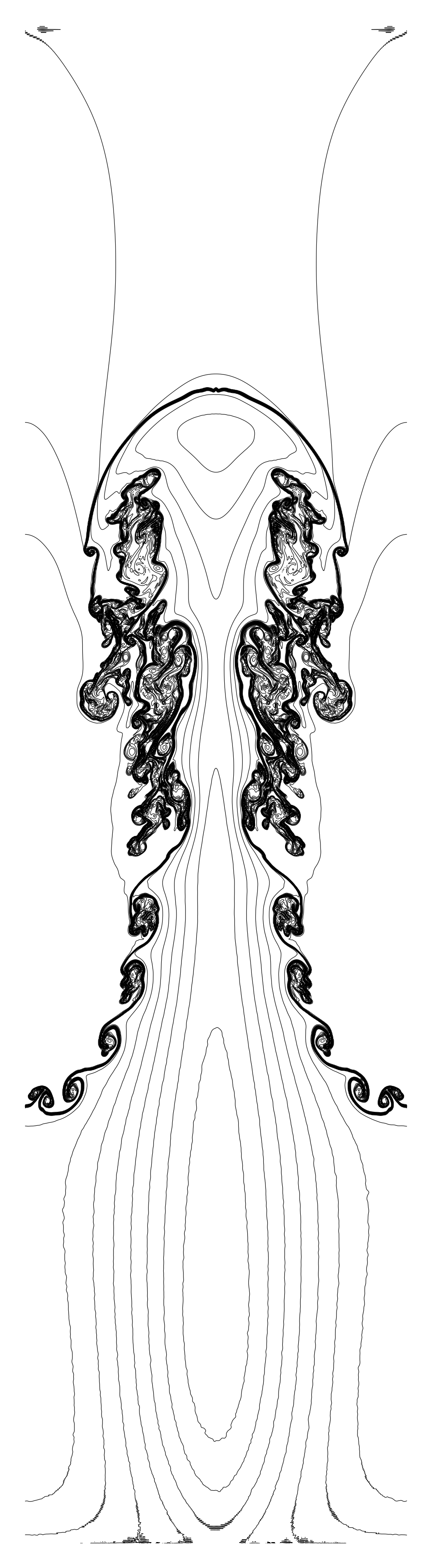}
  \caption{WA-WENO-CR\\$512\times2048$}
  \label{fig:rt_e}
\end{subfigure}
\caption{Rayleigh-Taylor instability ($t=1.95$, Sec.~\ref{sec:rt}): density contours at
$128\times512$ (a,b) and $512\times2048$ (c,d,e). WA-CR matches WA-5
at both resolutions. WA-WENO-CR matches WA-CR on the fine grid.}
\label{fig:rt}
\end{figure}
On the coarse $128\times512$ grid, WA-5 and WA-CR produce nearly identical density contours. The mushroom cap structure, rolled-up vortex sheets, and interface symmetry are preserved equally well. This result confirms that the rank-1 correction replicates the accuracy of the full characteristic decomposition at the material interface—but at lower cost. On the fine $512\times2048$ grid, both WA-5 and WA-CR fully resolve secondary instabilities and fine-scale filaments along the stem. Spurious density oscillations do not appear at the interface. This indicates that the rank-1 entropy wave correction works as intended. Purely conservative reconstructions without this correction would introduce density overshoots at the contact; these are not present. WA-WENO-CR on the fine grid also shows no oscillations. This confirms the rank-1 correction is independent of the limiter. The quality of the interface representation depends on the MP5 or WENO limiter applied to the scalar entropy wave stencil.

\begin{figure}[H]
\centering
\begin{subfigure}[b]{0.5\textwidth}
  \centering
  \includegraphics[height=0.38\textheight]{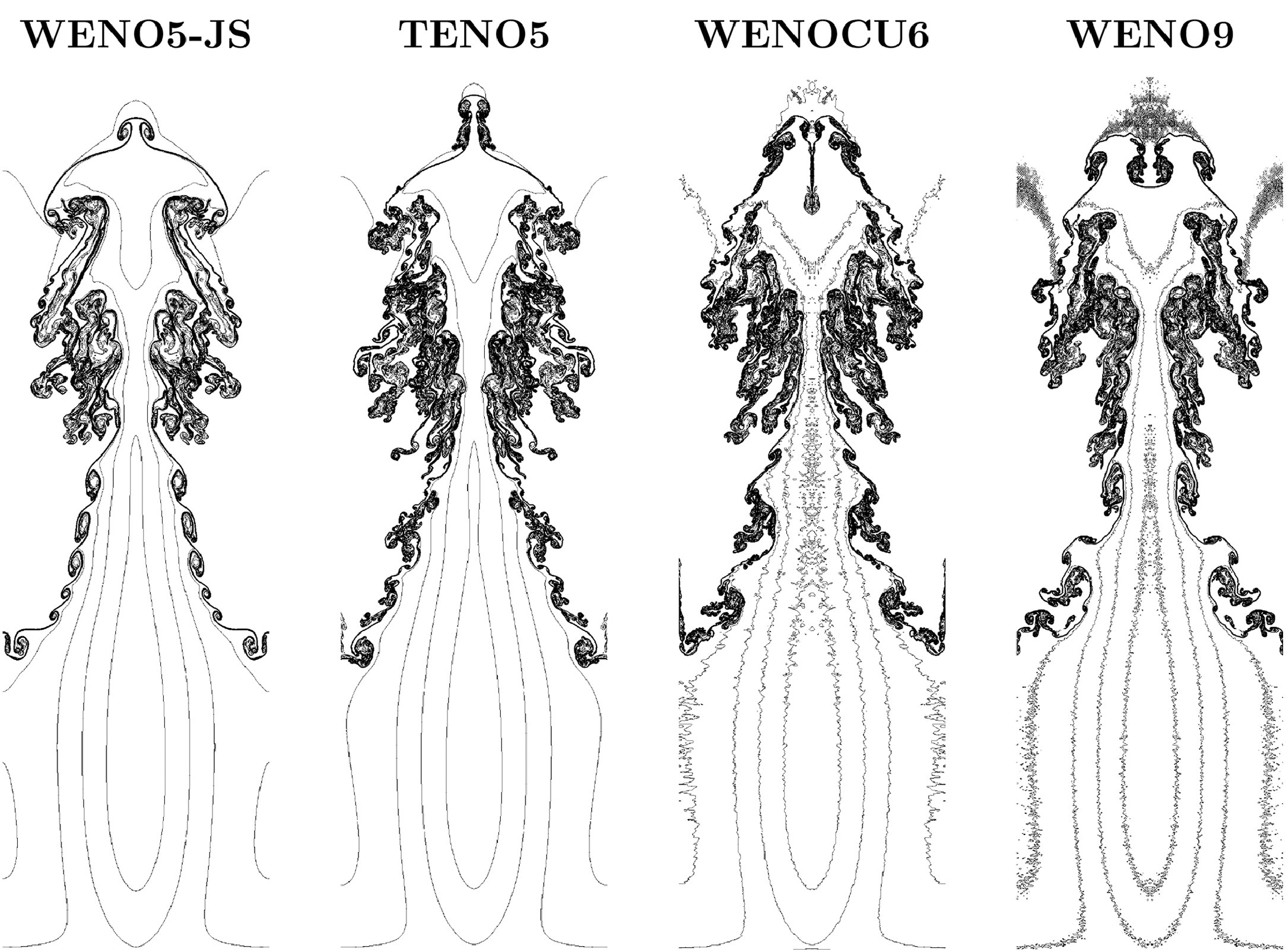}
  \caption{Ref.~\cite{fleischmann2019numerical}:
           WENO5-JS, TENO5, WENOCU6, WENO9.}
  \label{fig:rt_adams}
\end{subfigure}
\hfill
\begin{subfigure}[b]{0.4\textwidth}
  \centering
  \includegraphics[height=0.38\textheight]{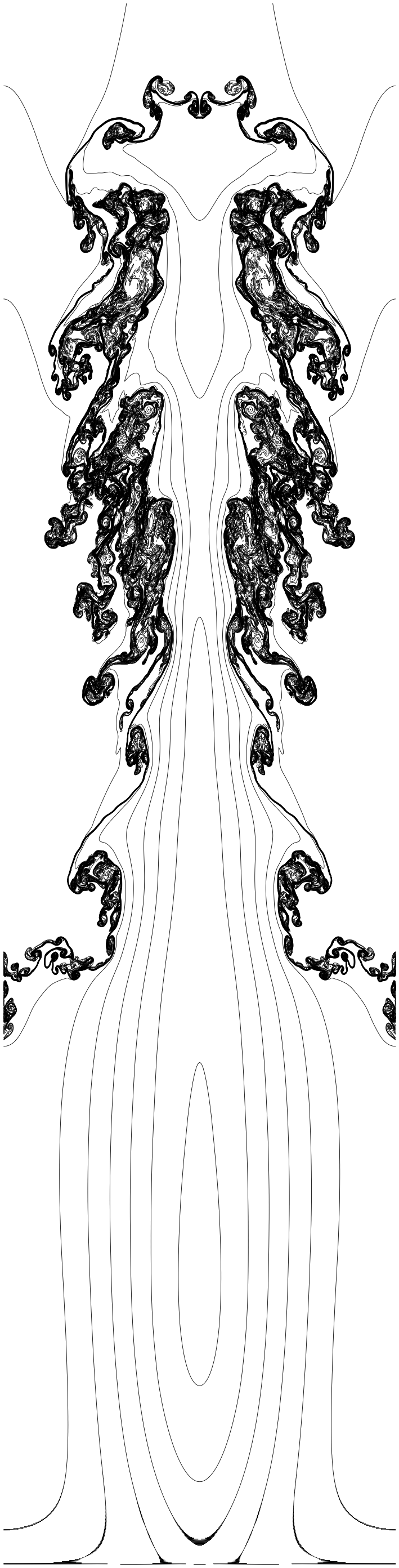}
  \caption{Present work: WA-CR.}
  \label{fig:rt_sainath}
\end{subfigure}
\caption{Rayleigh-Taylor instability ($t = 1.95$, Sec.~\ref{sec:rt}):
density contours. Left: reference results from Fleischmann et
al.~\cite{fleischmann2019numerical} on the same configuration.
Right: WA-CR at $1024\times4096$ resolution. The present scheme
resolves finer secondary instabilities and filaments along the stem
while remaining free of spurious density oscillations at the
interface.}
\label{fig:rt_compare}
\end{figure}

Figure~\ref{fig:rt_compare} shows density contours at $t = 1.95$, comparing the present WA-CR scheme at $1024 \times 4096$ resolution to reference results from Fleischmann et al.~\cite{fleischmann2019numerical} on the same configuration using WENO5-JS, TENO5, WENOCU6, and WENO9. The WA-CR solution preserves the instability's symmetry along the vertical centerline throughout the simulation. The mushroom cap geometry, rolled-up vortex sheets, and secondary instabilities all appear symmetrically, with no signs of symmetry-breaking due to numerical noise. No spurious density oscillations occur at the material interface, confirming that the rank-1 entropy wave correction properly limits the contact discontinuity without affecting the smooth flow. At $1024 \times 4096$ resolution, finer secondary instabilities and filaments are fully resolved, with detail comparable to the reference schemes from Fleischmann et al.

\subsection{Explosion problem}
\label{sec:explosion}
\begin{figure}[H]
\centering
\begin{subfigure}{0.4\textwidth}
  \includegraphics[width=\textwidth]{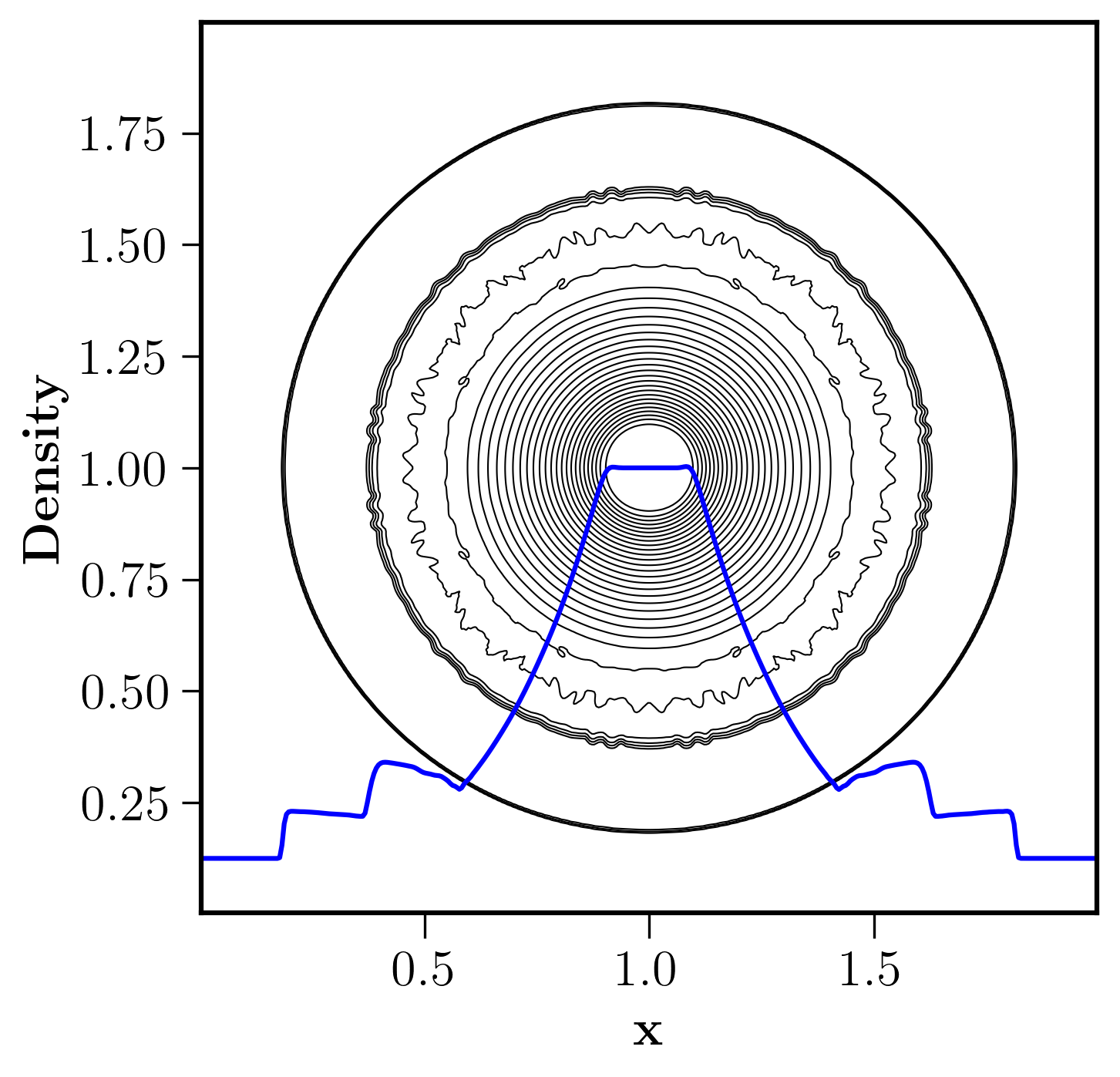}
  \caption{WA-3}
  \label{fig:exp_a}
\end{subfigure}%
\begin{subfigure}{0.4\textwidth}
  \includegraphics[width=\textwidth]{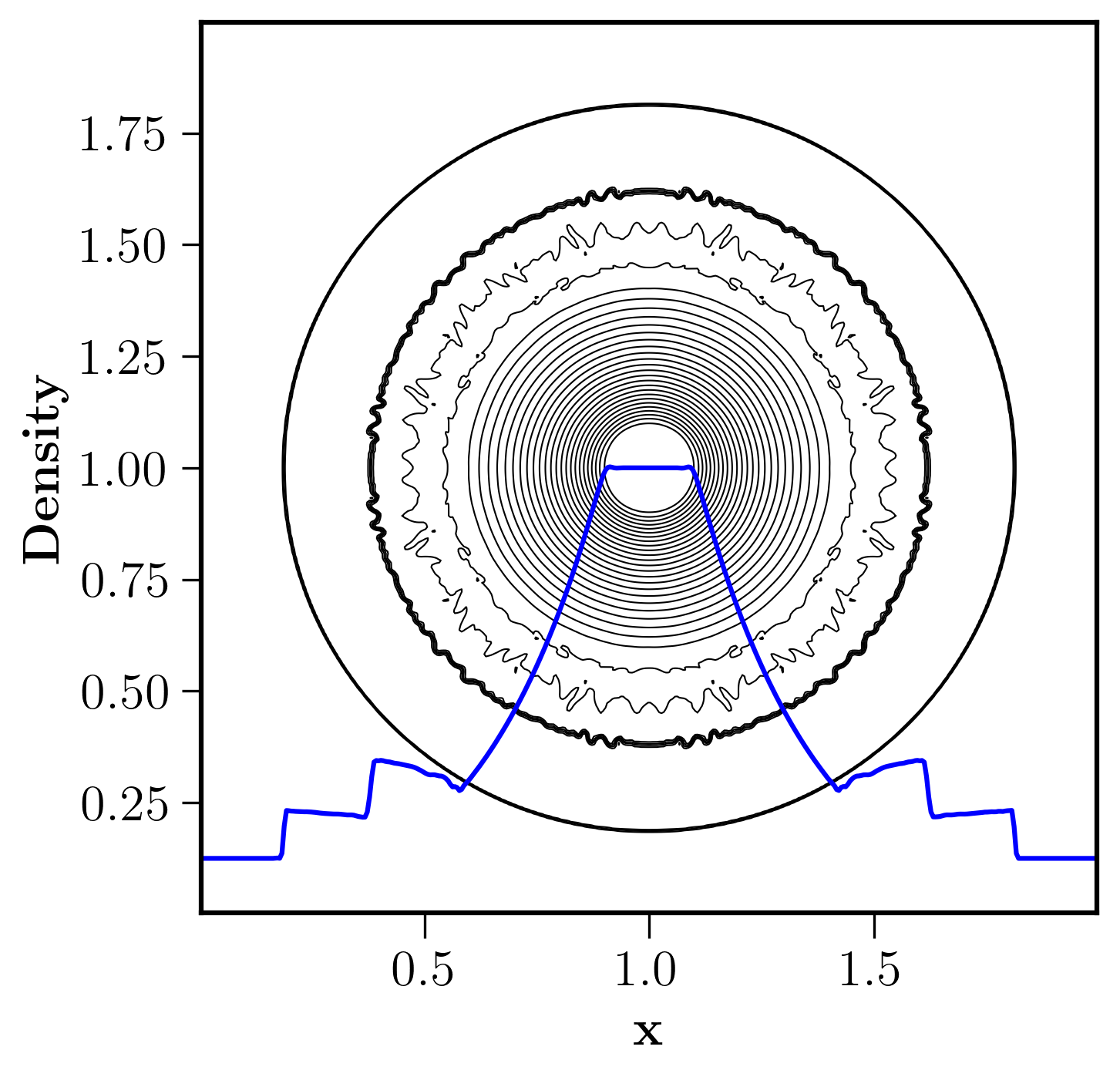}
  \caption{WA-5}
  \label{fig:exp_b}
\end{subfigure}
\begin{subfigure}{0.4\textwidth}
  \includegraphics[width=\textwidth]{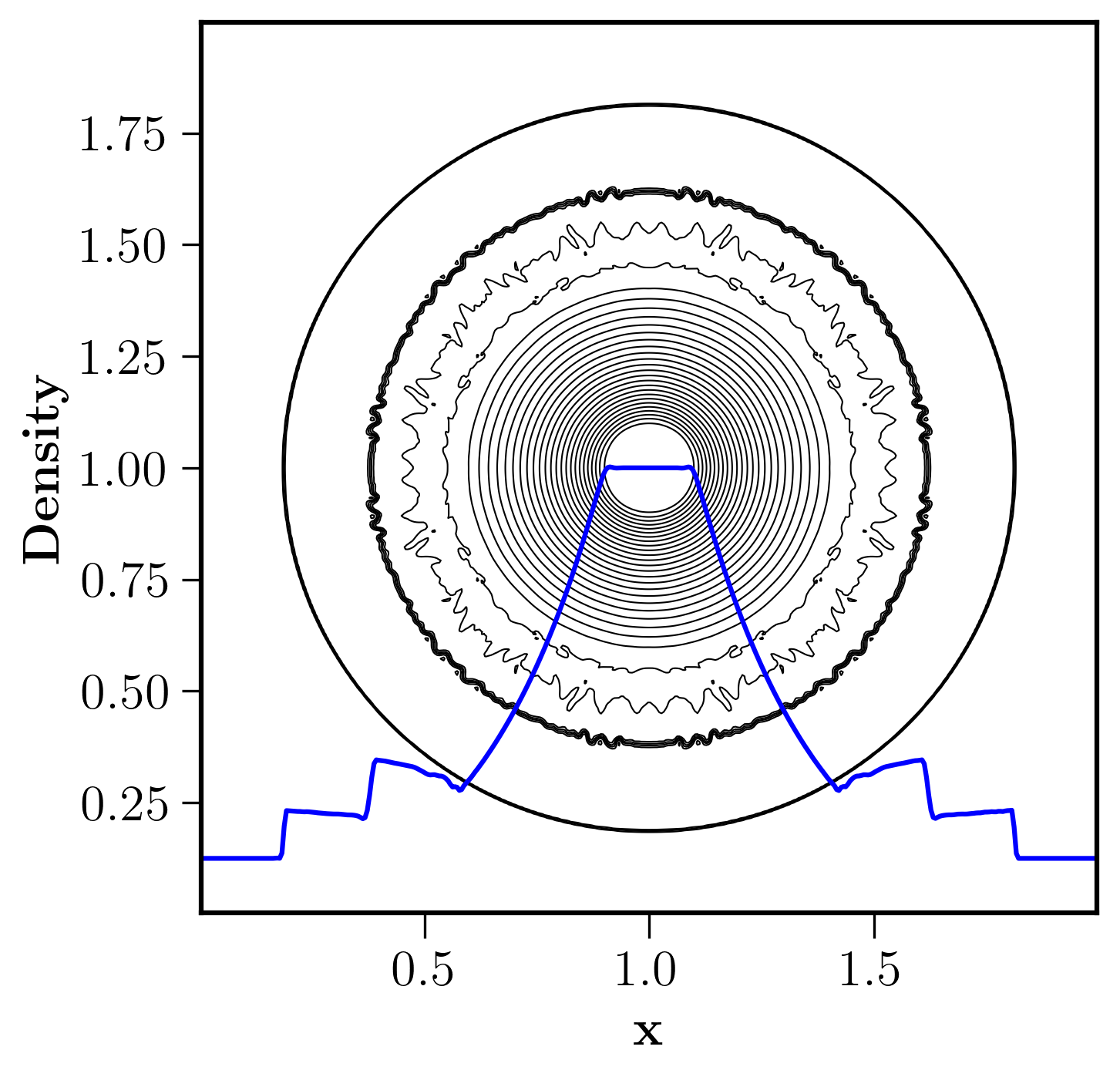}
  \caption{WA-CR}
  \label{fig:exp_c}
\end{subfigure}
\begin{subfigure}{0.4\textwidth}
  \includegraphics[width=\textwidth]{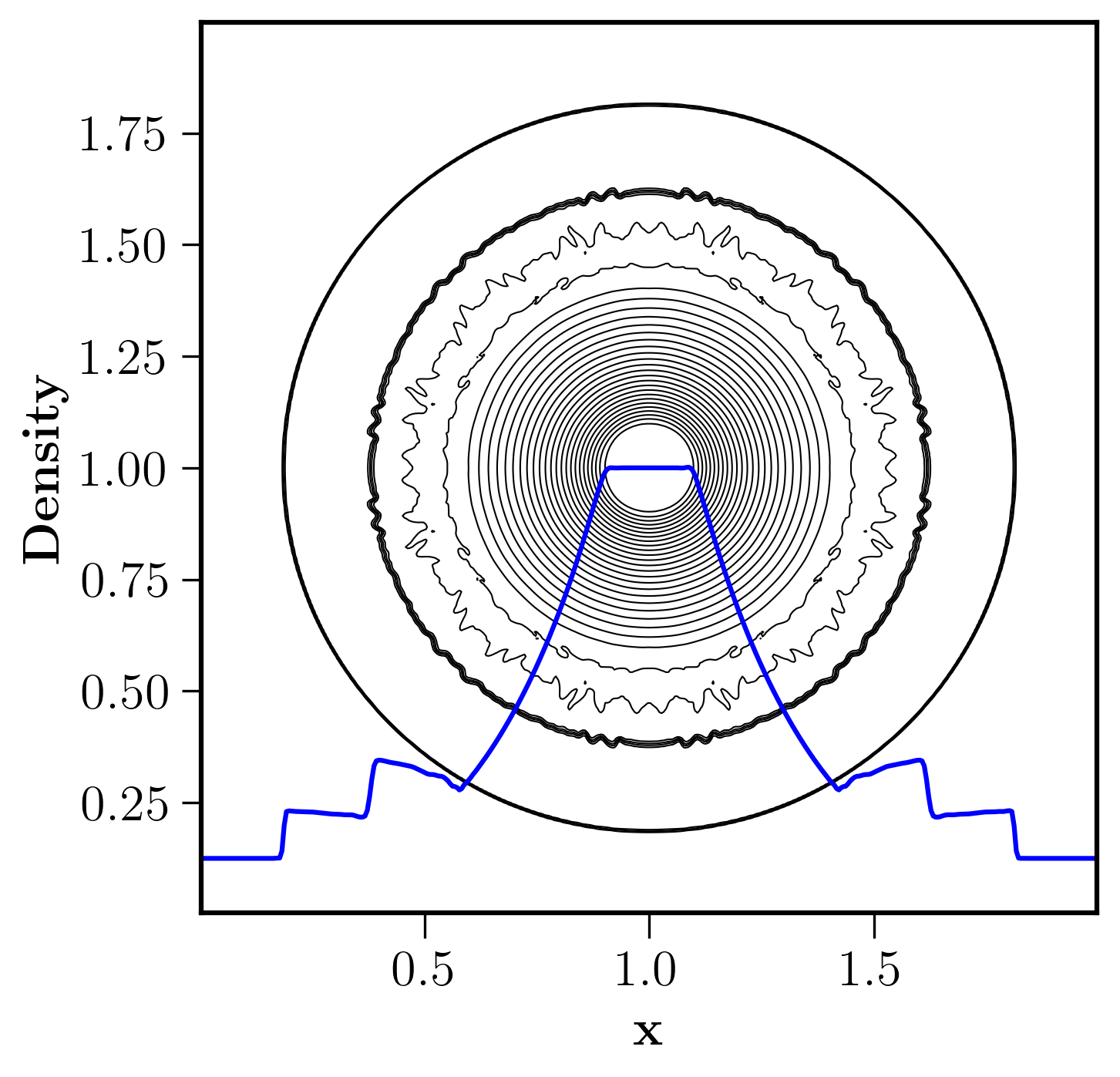}
  \caption{WA-WENO-CR}
  \label{fig:exp_d}
\end{subfigure}
\caption{Explosion problem ($400\times400$ grid, $t=0.25$, Sec.~\ref{sec:explosion}): density
contours with the cross-sectional density profile along $y=0$ shown
in blue. The shock and contact discontinuity are captured cleanly by all
schemes.}
\label{fig:exp}
\end{figure}

In this example, the initial condition consists of two constant states  for the flow variables, a circular region of radius $r=0.4$ centered  at $(1,1)$ and the region outside of it as mentioned in  Toro~\cite{toro2009riemann}. The initial conditions are given as:
\begin{equation}
(\rho, u, v, p)=\left\{\begin{array}{ll}
(1, 0, 0, 1), & \text{if } (x-1)^2+(y-1)^2 < r^2, \\
(0.125, 0, 0, 0.1), & \text{otherwise}.
\end{array}\right.
\end{equation}
In the present case, numerical simulations are carried out over a square domain of size $[0,2]\times [0,2]$ until a final time $t=0.25$ on a uniform grid of resolution $400\times 400$. Figure~\ref{fig:exp} shows density contours and the cross-sectional density profile along $y=0$ for all four schemes. Each scheme maintains circular symmetry and resolves the shock and contact discontinuity cleanly without oscillations. The cross-sectional profile shows sharp capture of the shock and contact in all cases. WA-5, WA-CR, and WA-WENO-CR yield nearly identical results, while WA-3 offers slightly less resolution at the contact, as expected from its lower order.

\subsection{2-D shock-entropy wave interaction}
\label{sec:shock_entropy}

The two-dimensional shock-entropy wave interaction of Acker et al.~\cite{acker2016improved} tests the ability of a scheme to resolve fine-scale entropy waves. This occurs in the presence of a strong moving shock. The initial conditions are
\begin{equation}\label{eq:shock_entropy_ic}
(\rho,u,p)=
\begin{cases}
(3.857143,\;2.629369,\;10.3333), & x<-4,\\
(1+0.2\sin(10x\cos\theta+10y\sin\theta),\;0,\;1), & \text{otherwise},
\end{cases}
\end{equation}
with $\theta=\pi/6$ over the domain $[-5,5]\times[-1,1]$. The initial sine waves make an angle of $\theta$ with the $x$-axis. Following Deng et al.~\cite{deng2019fifth}, a higher frequency is used for the initial sine waves compared to the original formulation to provide a more demanding test of small-scale resolution. A mesh of $400\times80$ ($\Delta x=\Delta y=1/40$) is used.

\begin{figure}[H]
\centering
\begin{subfigure}{0.5\textwidth}
  \includegraphics[width=\textwidth]{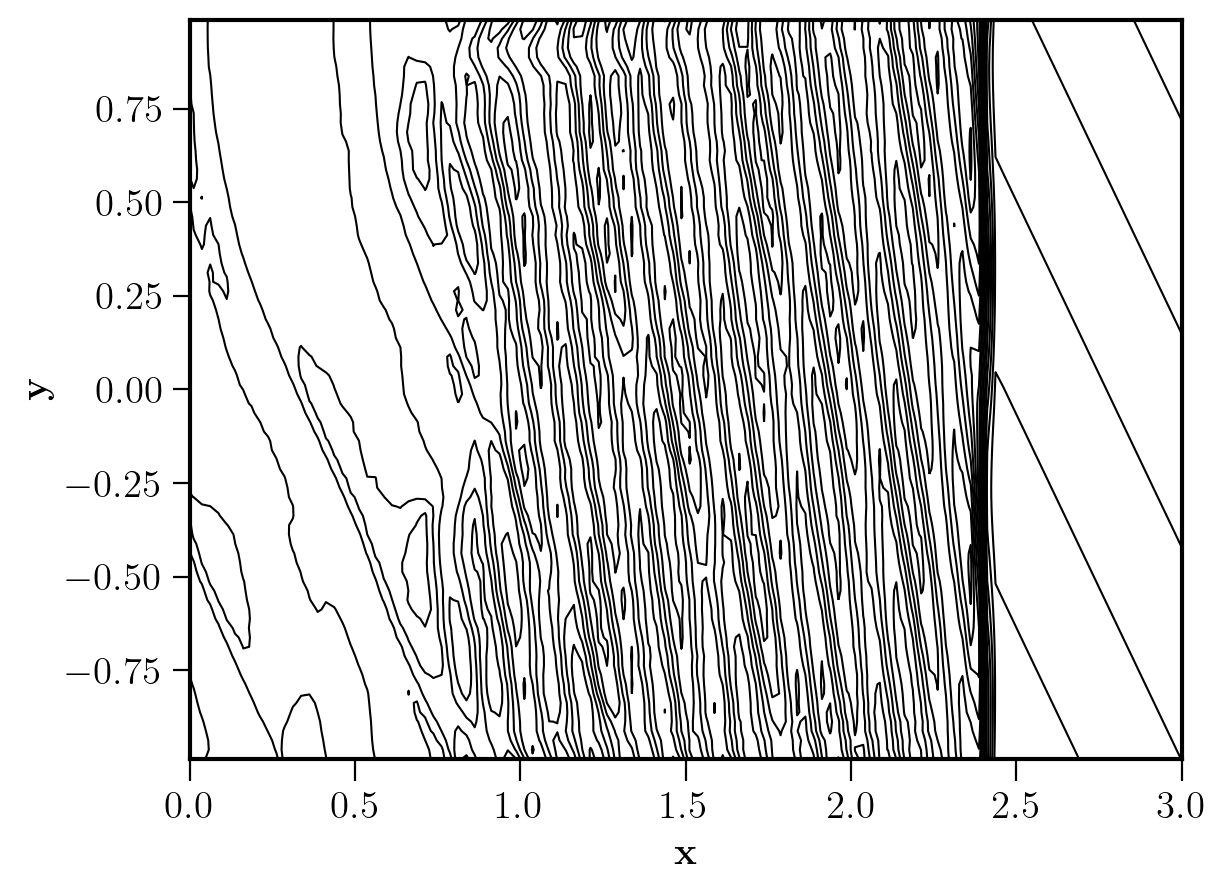}
  \caption{Density contours}
  \label{fig:se_contours}
\end{subfigure}%
\begin{subfigure}{0.5\textwidth}
  \includegraphics[width=\textwidth]{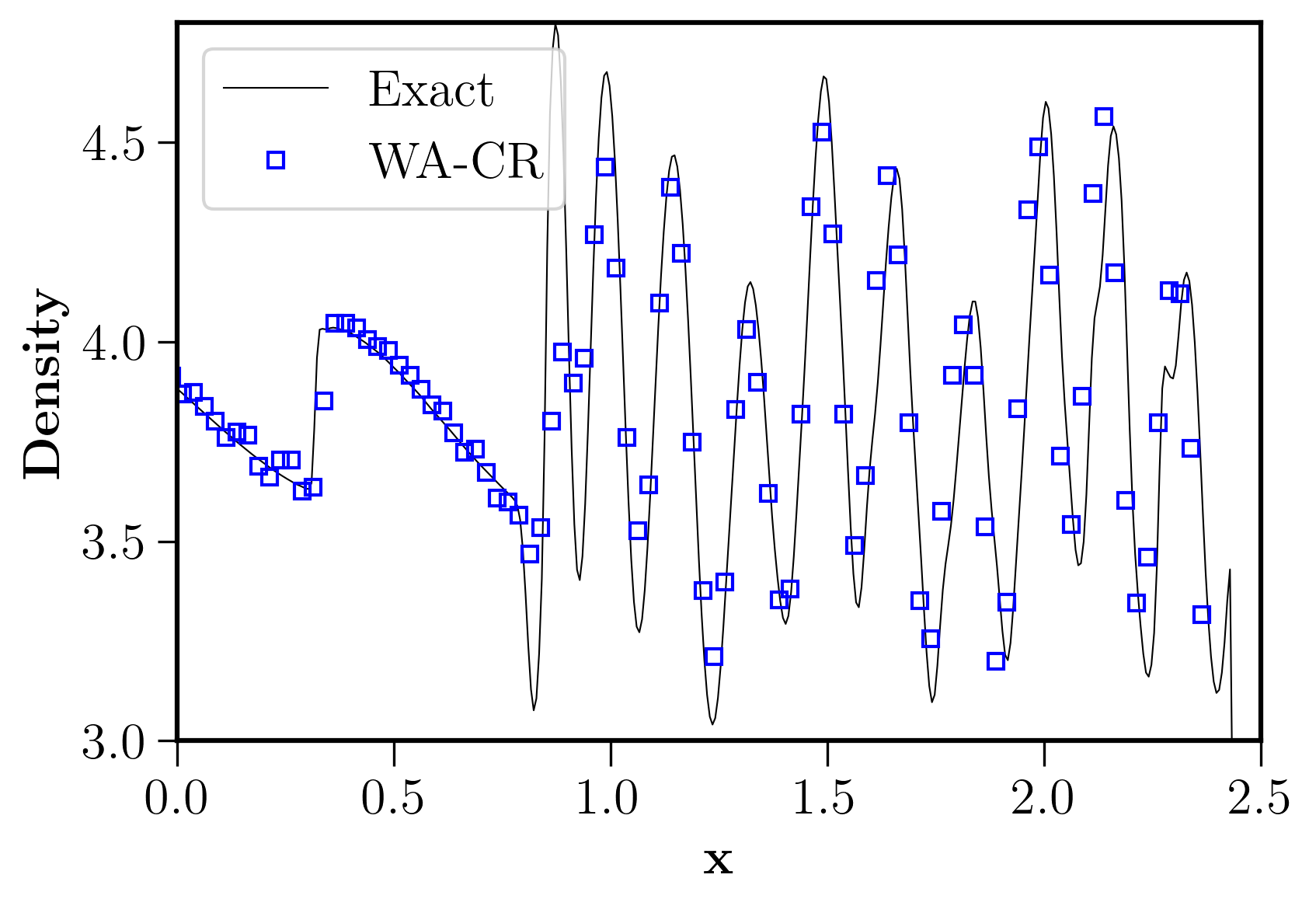}
  \caption{Local density profile}
  \label{fig:se_profile}
\end{subfigure}
\caption{2-D shock-entropy wave interaction ($t=1.8$, $400\times80$
grid, Sec.~\ref{sec:shock_entropy}): density contours and local density profile along $y=0$ for
WA-CR. The exact solution is shown for reference in (b).}
\label{fig:shock_entropy}
\end{figure}

Figure~\ref{fig:shock_entropy} shows the density contours and local density profile at $t=1.8$. The shock is captured cleanly on the right side of the domain. Behind the shock, the entropy wave structures are well resolved by WA-CR, with the local density profile closely following the exact solution across the full range of oscillations. This case exercises both paths of the WA-CR algorithm simultaneously: the Ducros sensor activates the full characteristic reconstruction in the vicinity of the shock, while the conservative reconstruction with rank-1 entropy correction handles the smooth entropy wave region. The clean resolution of the entropy wave structures confirms that the rank-1 correction applies the appropriate limiting to the entropy characteristic without contaminating the smooth flow away from the shock.

\subsection{Double Mach reflection}
\label{sec:dmr}

\begin{figure}[H]
\centering
\begin{subfigure}{0.45\textwidth}
  \includegraphics[width=\textwidth]{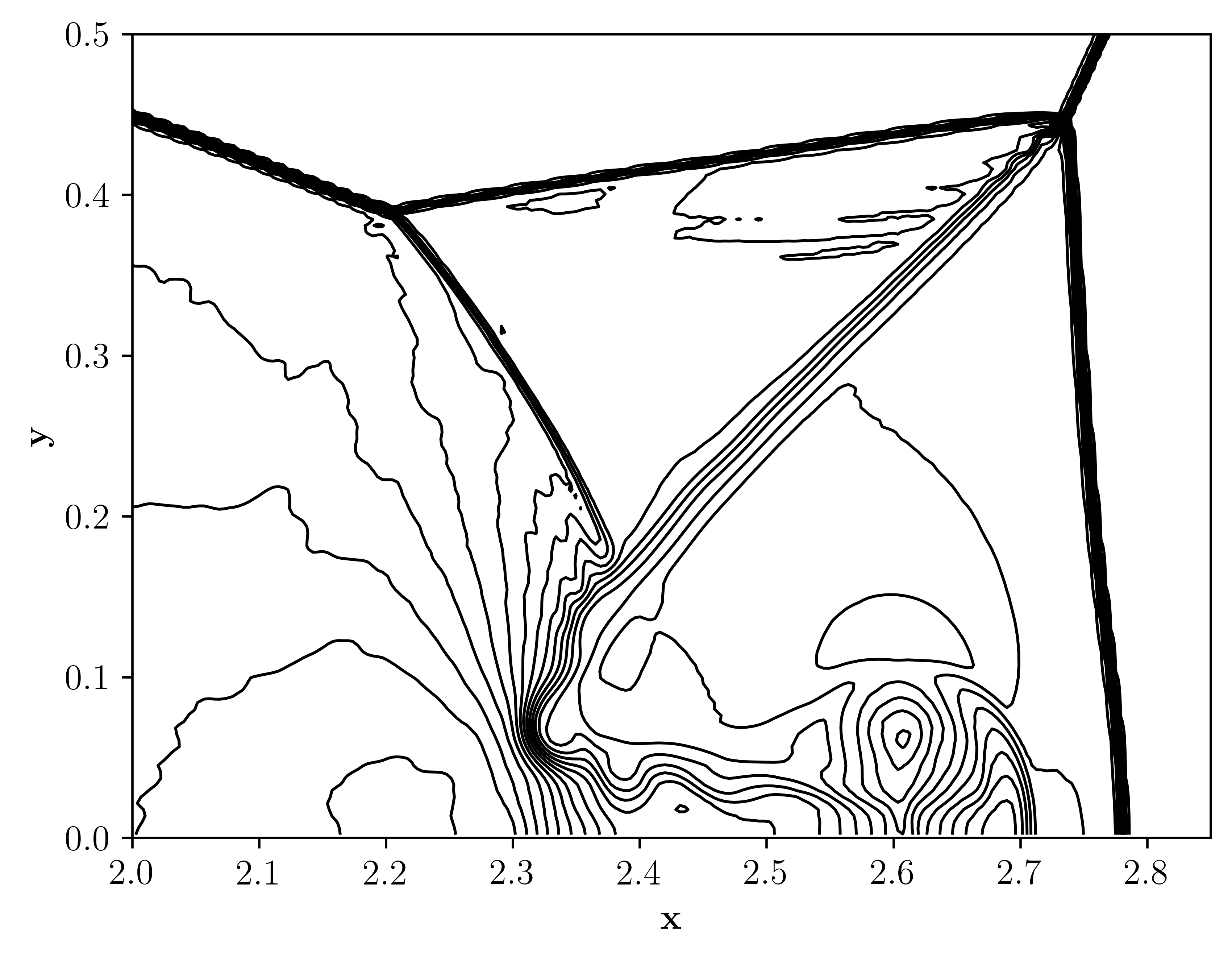}
  \caption{WA-3}
  \label{fig:dmr_a}
\end{subfigure}%
\begin{subfigure}{0.45\textwidth}
  \includegraphics[width=\textwidth]{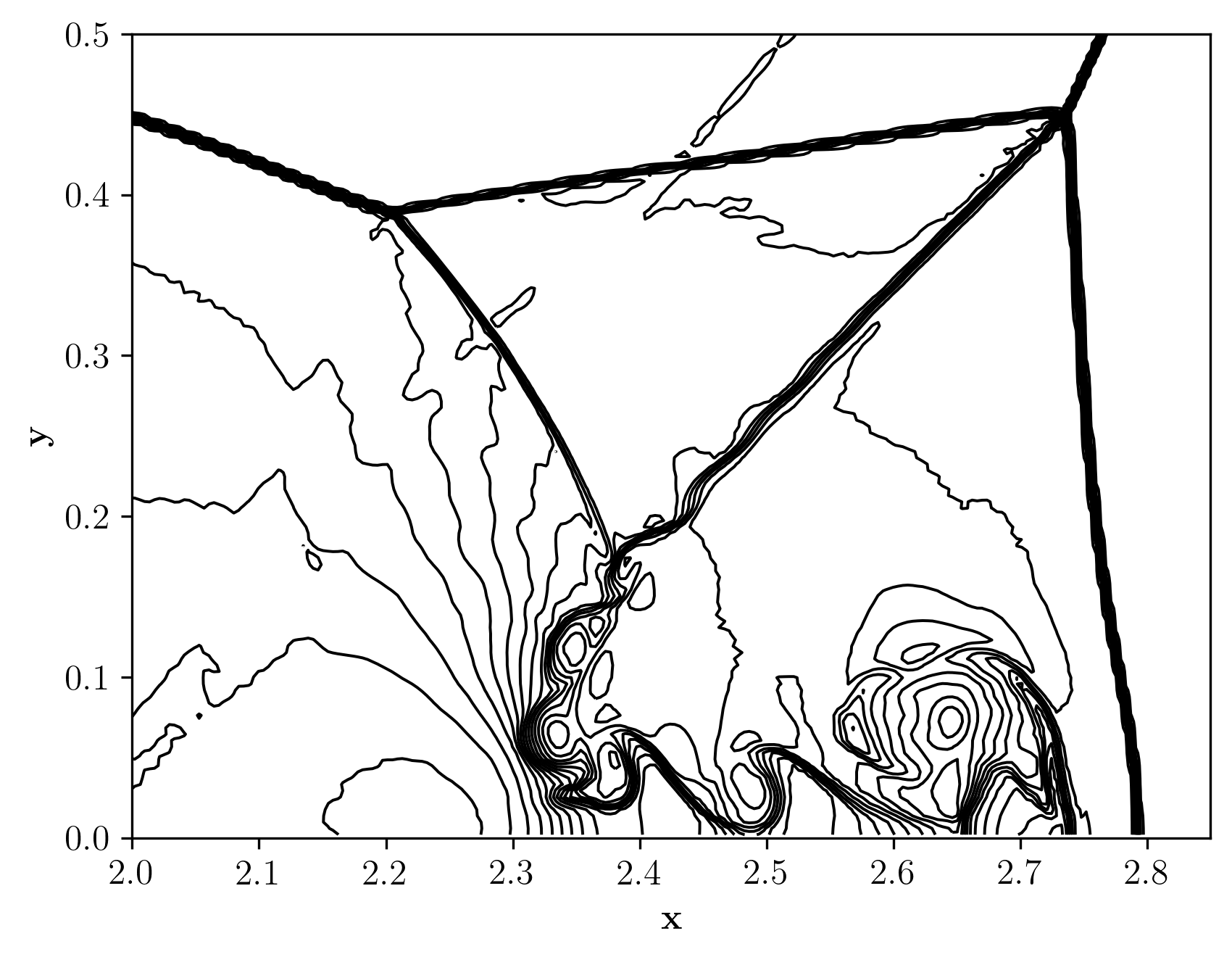}
  \caption{WA-5}
  \label{fig:dmr_b}
\end{subfigure}
\begin{subfigure}{0.45\textwidth}
  \includegraphics[width=\textwidth]{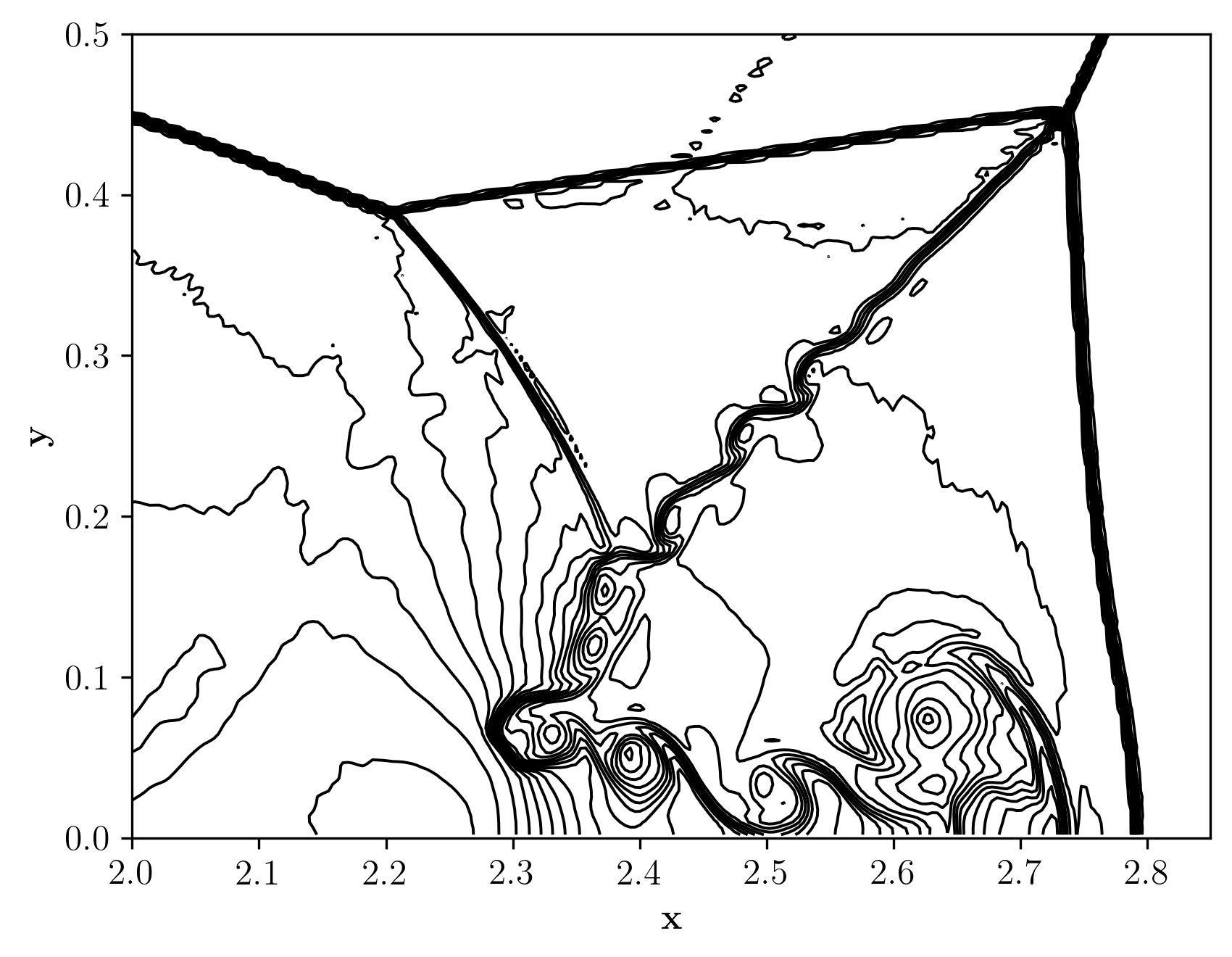}
  \caption{WA-CR}
  \label{fig:dmr_c}
\end{subfigure}
\begin{subfigure}{0.45\textwidth}
  \includegraphics[width=\textwidth]{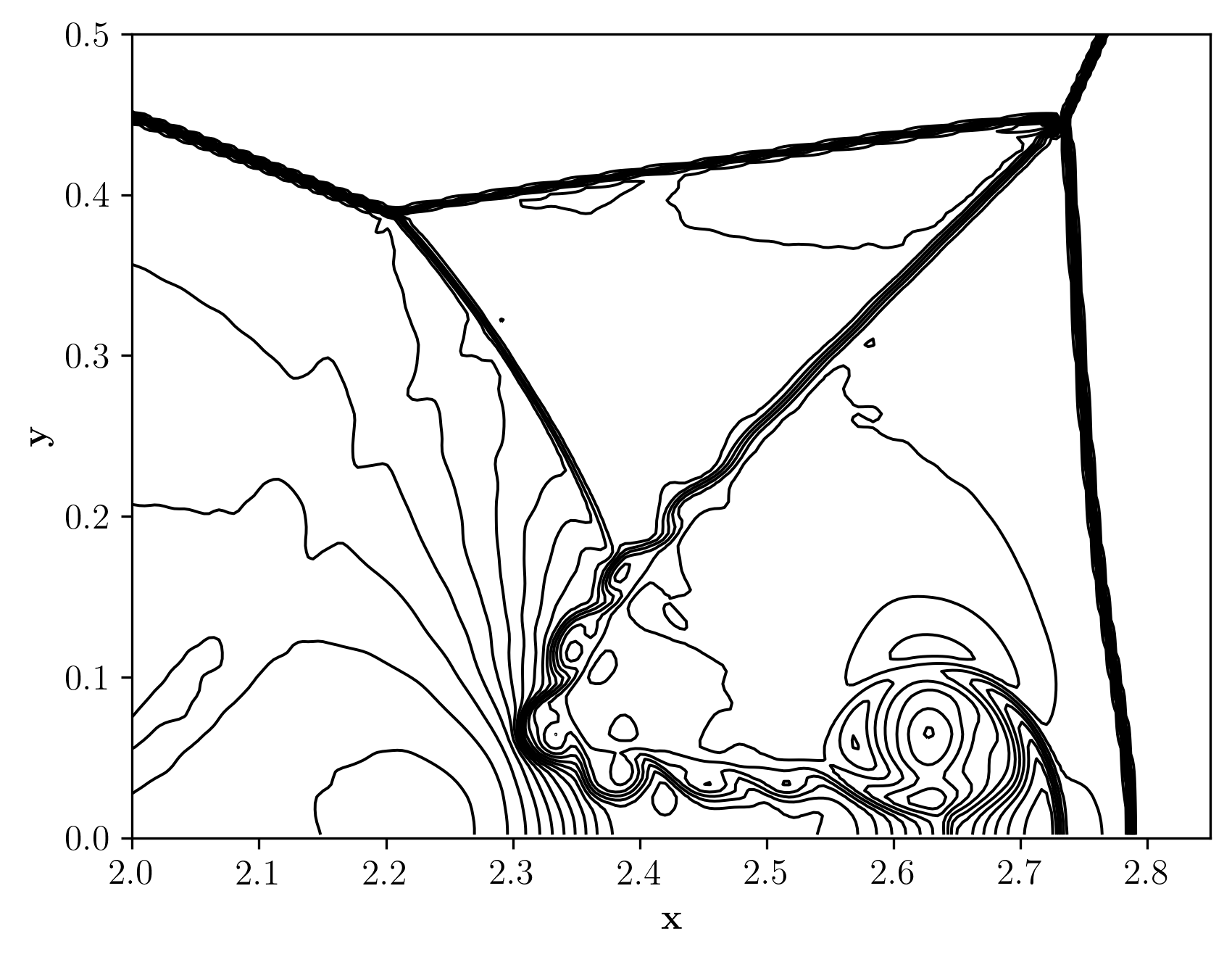}
  \caption{WA-WENO-CR}
  \label{fig:dmr_d}
\end{subfigure}
\caption{Double Mach reflection ($\mathrm{Ma}=10$, $768\times256$
grid, $t=0.3$, Sec.~\ref{sec:dmr}): density contours zoomed into the Mach stem region.
WA-5, WA-CR, and WA-WENO-CR all resolve the slip-layer roll-up
structures. WA-3 resolves fewer vortices due to its higher background
dissipation.}
\label{fig:dmr}
\end{figure}

The double Mach reflection~\cite{woodward1984numerical} involves a Mach~10 shock impinging on a $30^\circ$ wedge, testing both shock-capturing fidelity and the resolution of slip-layer vortices formed behind the Mach stem. For this scenario, the domain $[0,3]\times[0,1]$ is discretized on a $768\times256$ grid, and the simulation runs to $t=0.3$. The initial conditions for this test case are as follows:
\begin{equation}\label{eq:dmr_ic}
(\rho,u,v,p)=\begin{cases}
(1.4,\;0,\;0,\;1),              & y<1.732(x-0.1667),\\
(8,\;7.145,\;{-4.125},\;116.8), & \text{otherwise}.
\end{cases}
\end{equation}
To prevent the carbuncle instability, the HLL Riemann solver is used. Reflecting wall conditions are applied on the bottom boundary for $x>0.1667$, while post-shock conditions hold for $x\leq0.1667$. The top boundary follows the exact moving-shock solution. Throughout the shock system, the Ducros sensor activates, ensuring the full characteristic reconstruction path is used in shocked regions. Conversely, the conservative path with rank-1 correction is active only in the smooth post-shock regions away from the Mach stem. As illustrated in Figure~\ref{fig:dmr}, which shows density contours zoomed into the Mach stem region, WA-3 resolves fewer slip-layer vortices than WA-5. This is consistent with lower formal order of WA-3 and the increased dissipation of the MUSCL limiter applied to the entropy wave. Both WA-5 and WA-WENO-CR produce sharp, well-resolved roll-up structures. Notably, WA-CR matches WA-5 in the shock region, as expected, since the Ducros sensor activates the full characteristic path there. However, the cost reduction of WA-CR relative to WA-5 is realized in the smooth regions with improvements in the slip lines.

\subsection{Two-dimensional Riemann problem}
\label{sec:riemann}

\begin{figure}[H]
\centering
\begin{subfigure}{0.45\textwidth}
  \includegraphics[width=\textwidth]{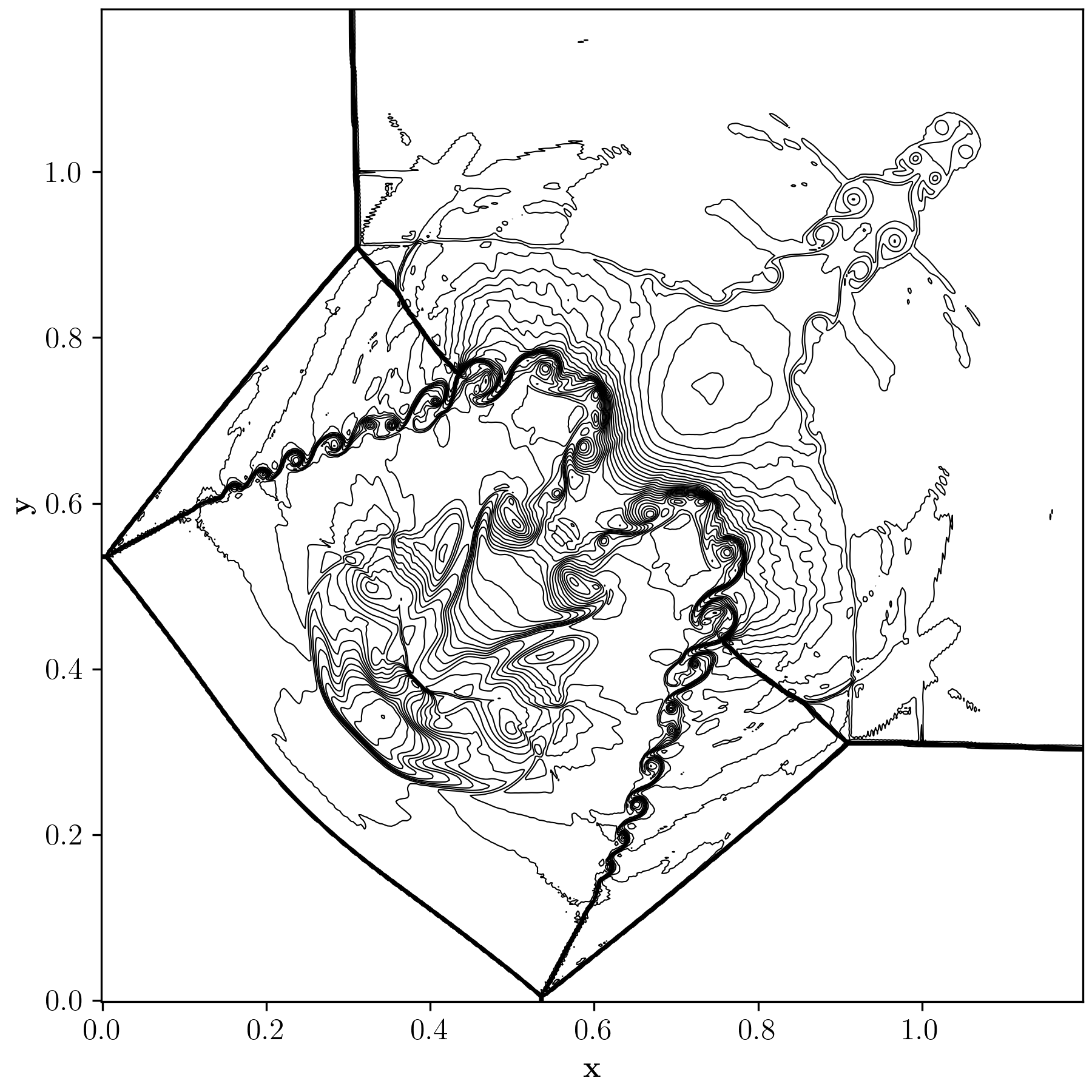}
  \caption{WA-3}
  \label{fig:rp_a}
\end{subfigure}%
\begin{subfigure}{0.45\textwidth}
  \includegraphics[width=\textwidth]{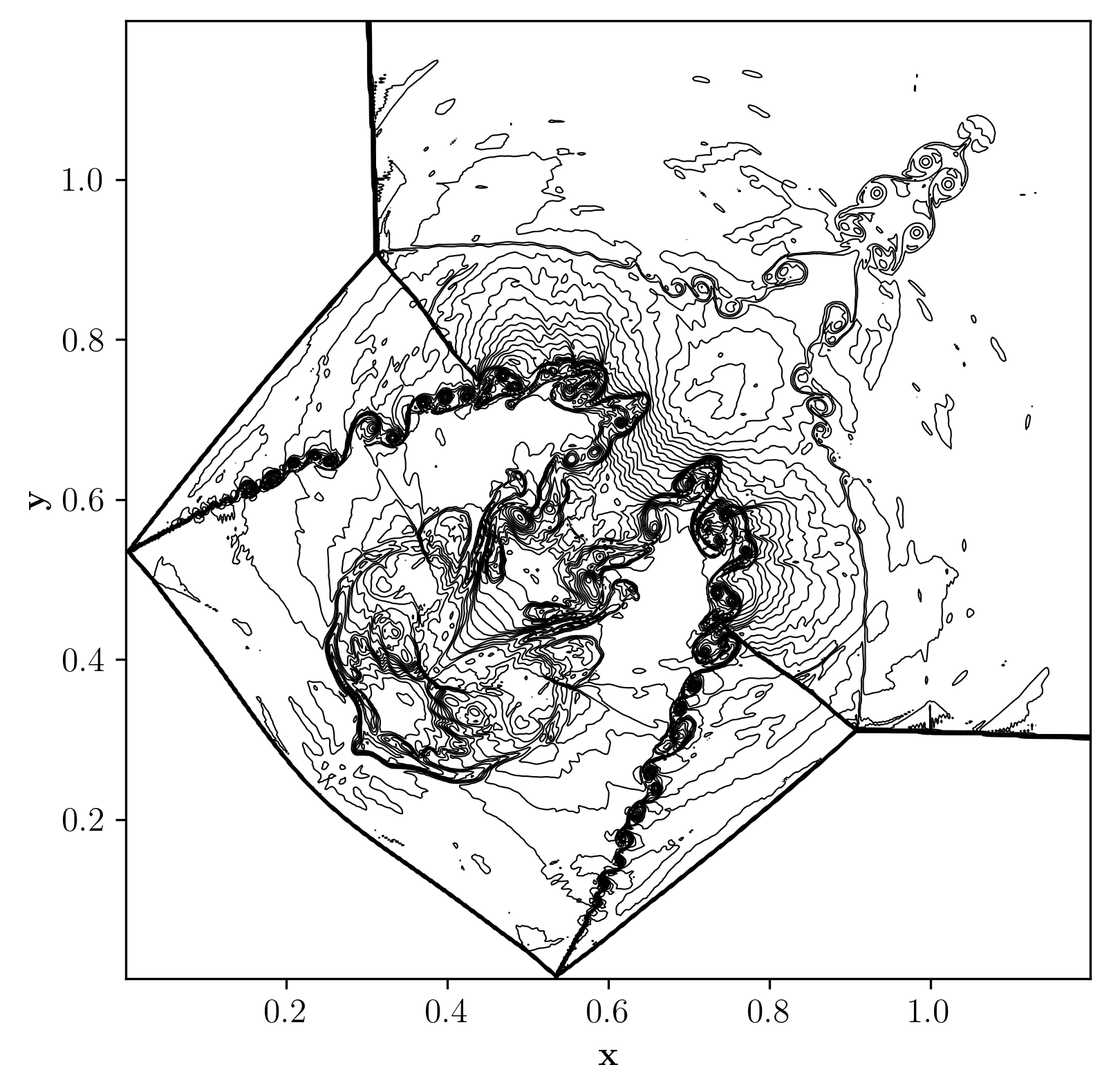}
  \caption{WA-5}
  \label{fig:rp_b}
\end{subfigure}
\begin{subfigure}{0.32\textwidth}
  \includegraphics[width=\textwidth]{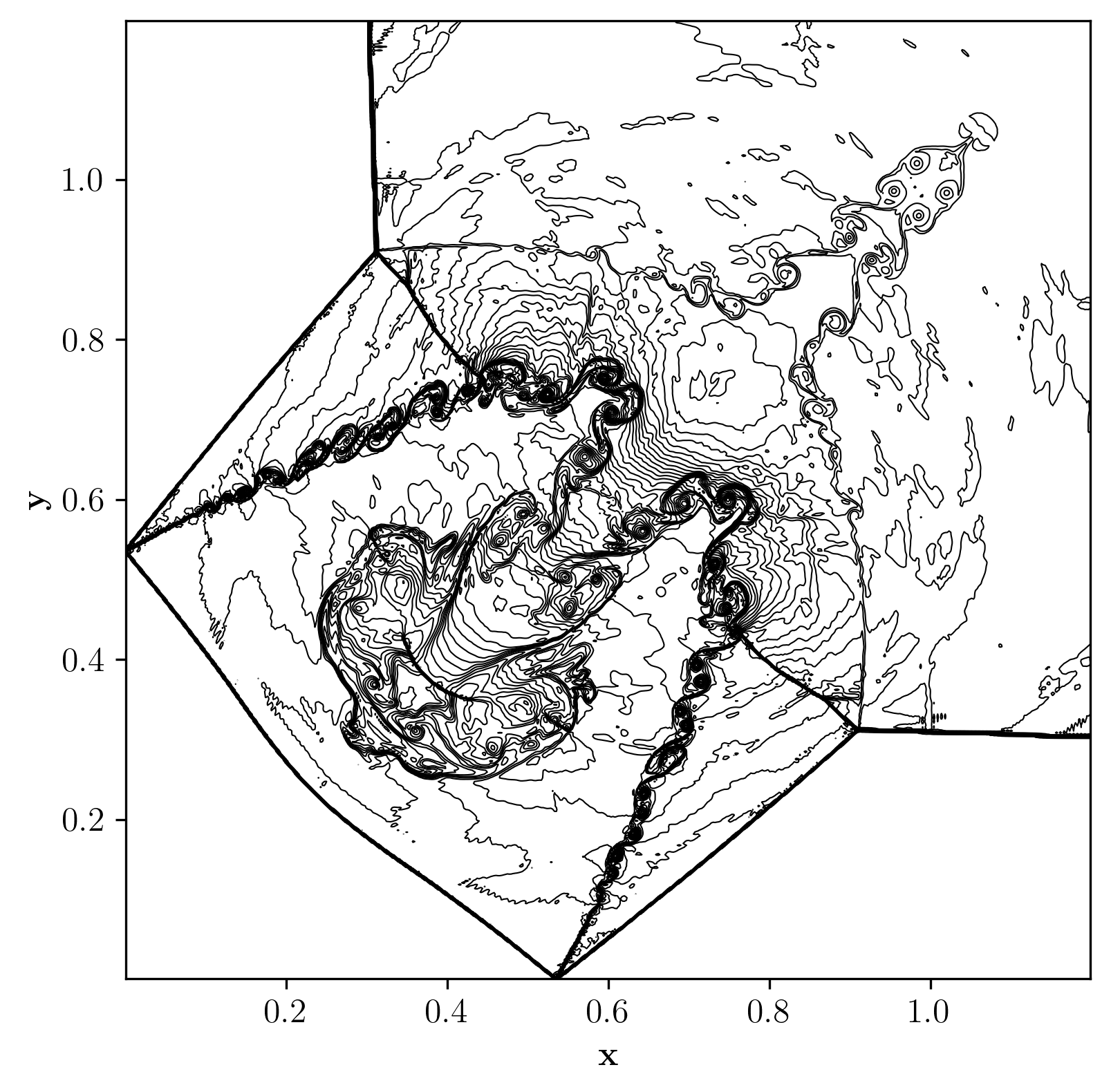}
  \caption{WA-CR}
  \label{fig:rp_c}
\end{subfigure}%
\begin{subfigure}{0.32\textwidth}
  \includegraphics[width=\textwidth]{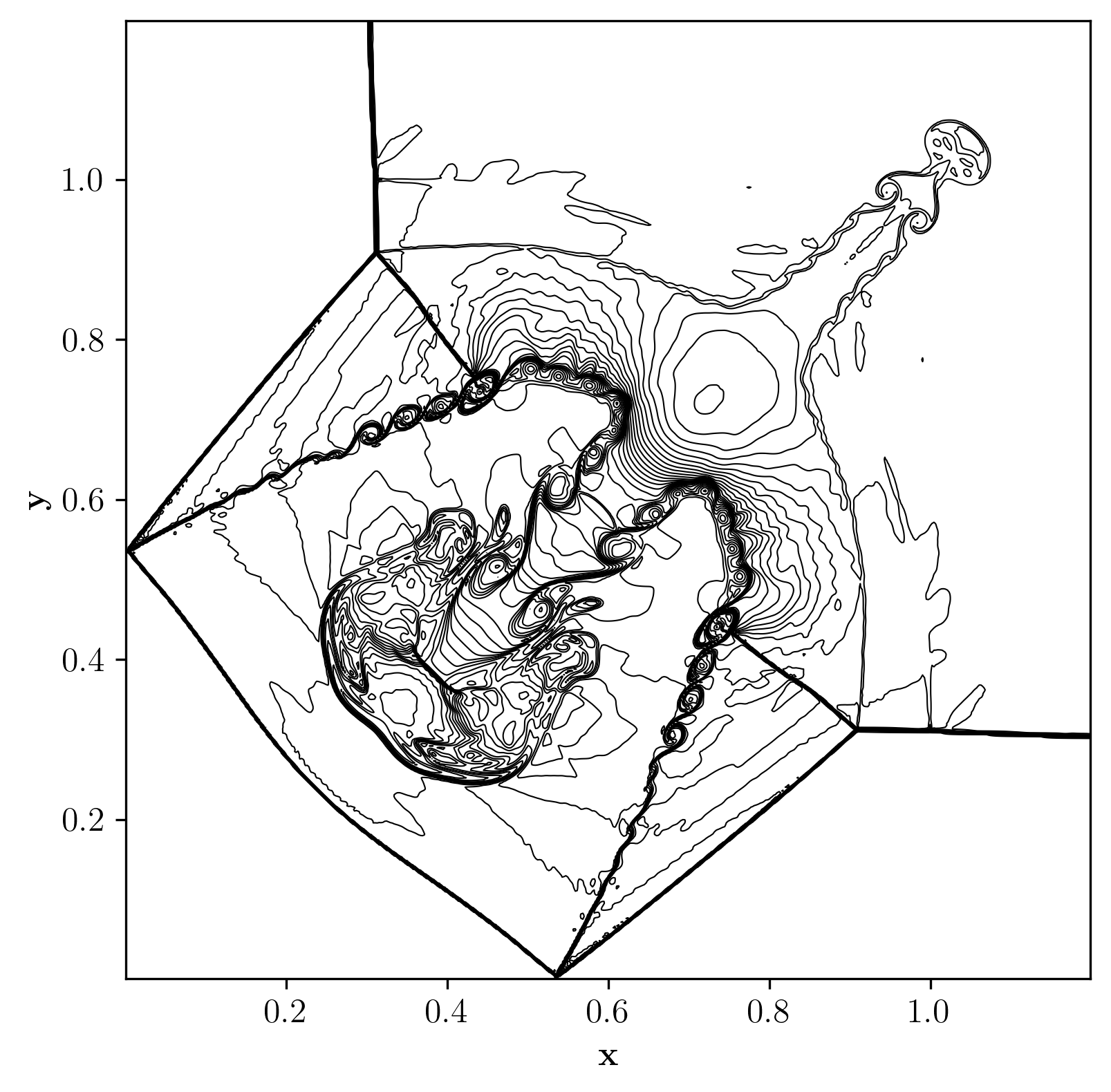}
  \caption{WENO5}
  \label{fig:rp_d}
\end{subfigure}%
\begin{subfigure}{0.32\textwidth}
  \includegraphics[width=\textwidth]{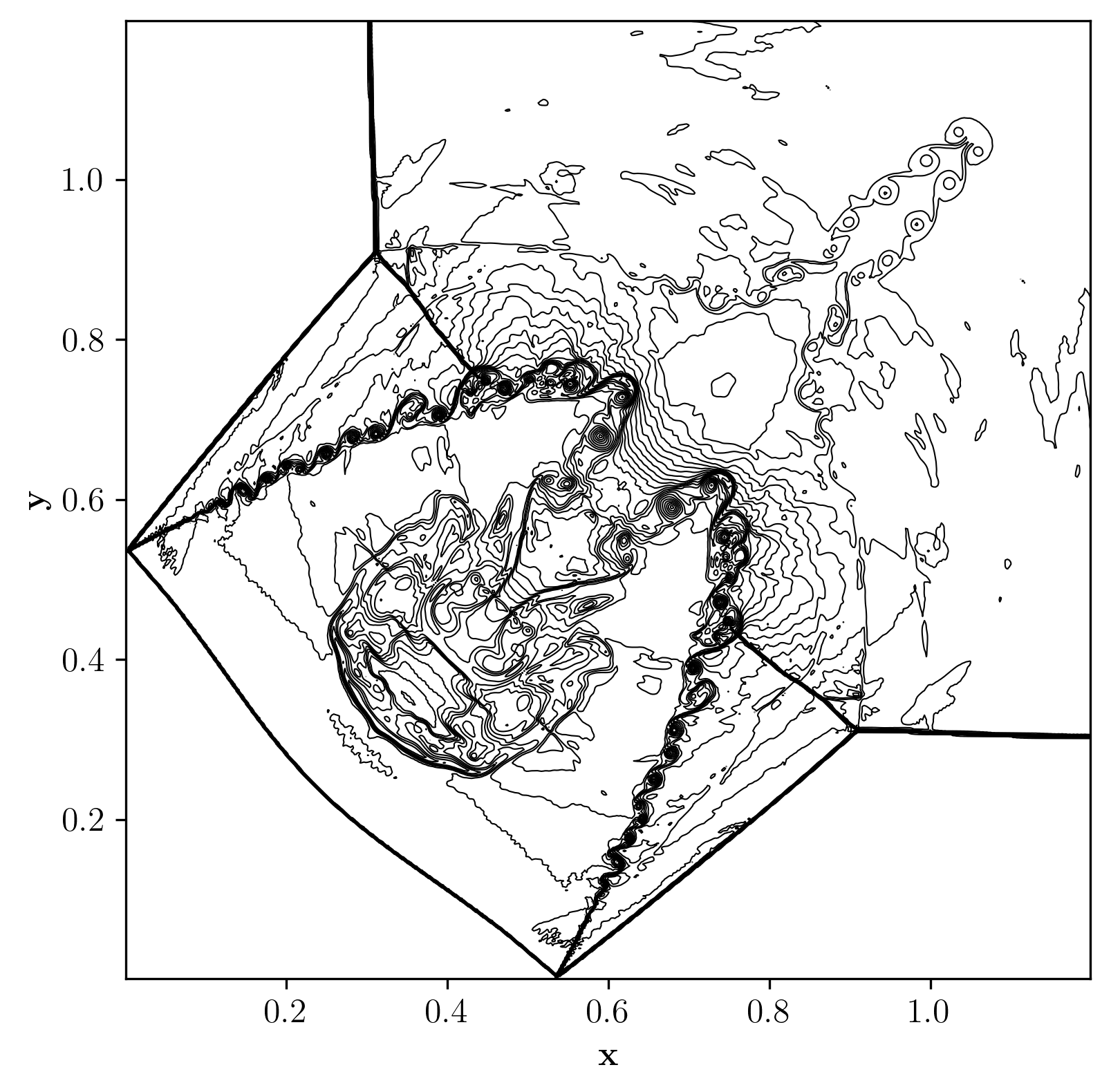}
  \caption{WA-WENO-CR}
  \label{fig:rp_e}
\end{subfigure}
\caption{Two-dimensional Riemann problem, configuration~3
($512\times512$, $t=1.1$, \ref{sec:riemann}): density contours. Base WENO5 smears the slip-line vortices compared to that of WA-WENO-CR.}
\label{fig:riemann}
\end{figure}

The 2-D Riemann problem of configuration~3~\cite{schulz1993numerical} initiates four shocks at the quadrant boundaries and develops Kelvin-Helmholtz instabilities along the resulting slip lines, making it sensitive to numerical dissipation in the shock-adjacent vortical regions. The slip lines are contact discontinuities along which density jumps while velocity and pressure remain continuous, so both the shock-capturing and rank-1 entropy wave correction paths of WA-CR and WA-WENO-CR are tested simultaneously. To clarify the approach used in this study, the computational setup is described below. The initial conditions on $[0,1.2]^2$ are as follows:
\begin{equation}\label{eq:rp_ic}
(\rho,u,v,p)=\begin{cases}
(1.5,\;0,\;0,\;1.5),             & x>1,\;y>1,\\
(0.5323,\;1.206,\;0,\;0.3),      & x<1,\;y>1,\\
(0.138,\;1.206,\;1.206,\;0.029), & x<1,\;y<1,\\
(0.5323,\;0,\;1.206,\;0.3),      & x>1,\;y<1,
\end{cases}
\end{equation}
on a $512\times512$ grid, run to $t=1.1$. Figure~\ref{fig:riemann} shows density contours computed by various schemes.

WA-3 captures the primary shock structure correctly; however, it resolves less fine-scale vortical detail along the slip lines than the fifth-order schemes. In contrast, WA-5 and WA-CR produce nearly identical results, featuring sharp and well-resolved Kelvin-Helmholtz roll-up along the slip lines with no spurious oscillations at the shocks. Notably, base WENO5 produces the least resolved result of all schemes tested: the slip-line vortices are heavily smeared, and even WA-3 resolves more fine-scale structure. This confirms that the wave-appropriate design with optimized $\eta_a^*$ is more effective than increasing the reconstruction order alone. Furthermore, WA-WENO-CR recovers the resolution of WA-5 and WA-CR, demonstrating that the rank-1 entropy correction compensates for the excess dissipation of the baseline WENO scheme. Overall, this result provides the clearest illustration in the paper of the benefit of the rank-1 correction approach.

\subsection{Shock-bubble interaction}
\label{sec:sbi}

\begin{figure}[H]
\centering
\begin{subfigure}{0.35\textwidth}
  \includegraphics[width=\textwidth]{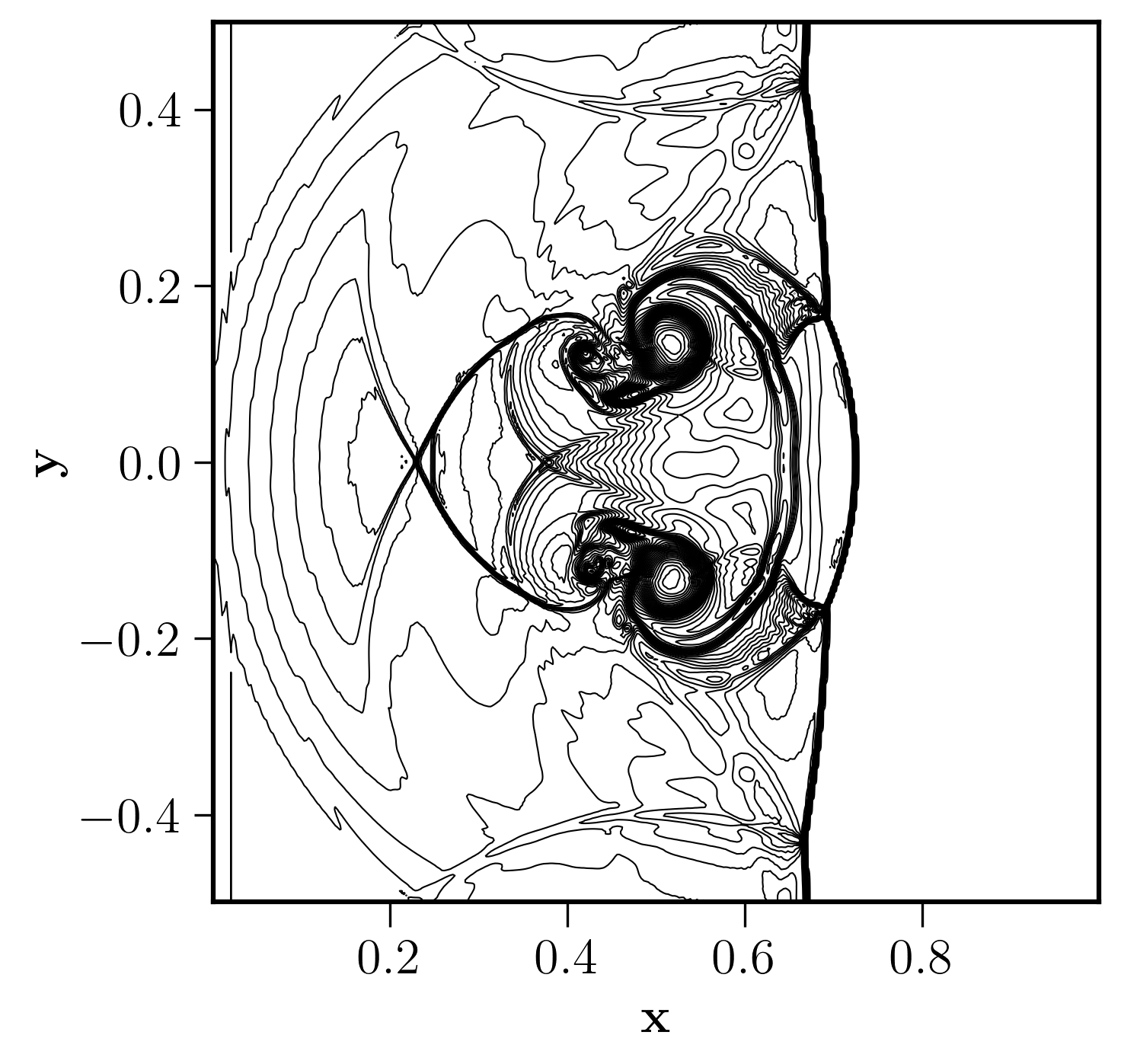}
  \caption{\textcolor{black}{WA-3}}
  \label{fig:sbi_a}
\end{subfigure}%
\begin{subfigure}{0.35\textwidth}
  \includegraphics[width=\textwidth]{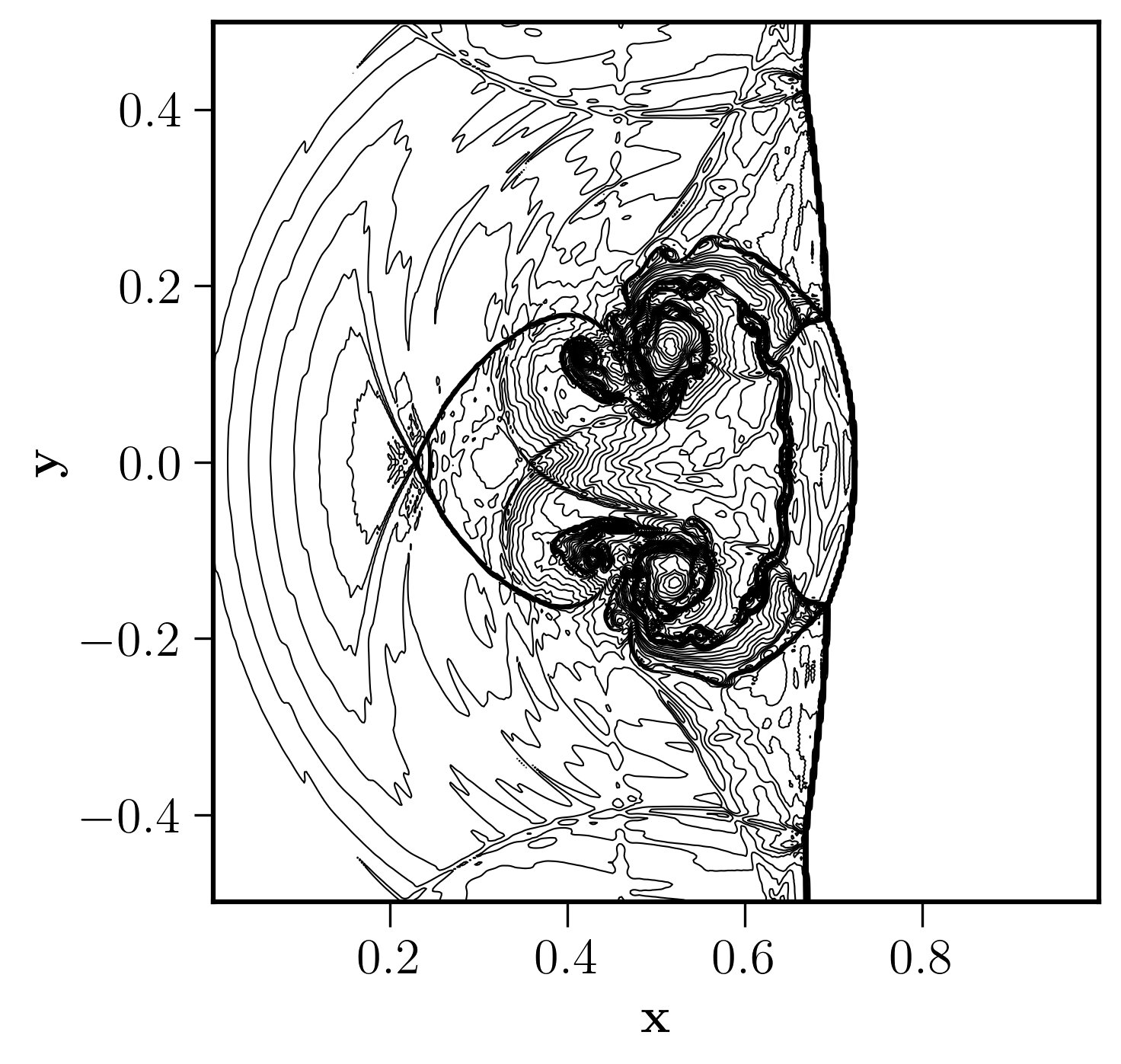}
  \caption{\textcolor{black}{WA-5}}
  \label{fig:sbi_b}
\end{subfigure}
\begin{subfigure}{0.35\textwidth}
  \includegraphics[width=\textwidth]{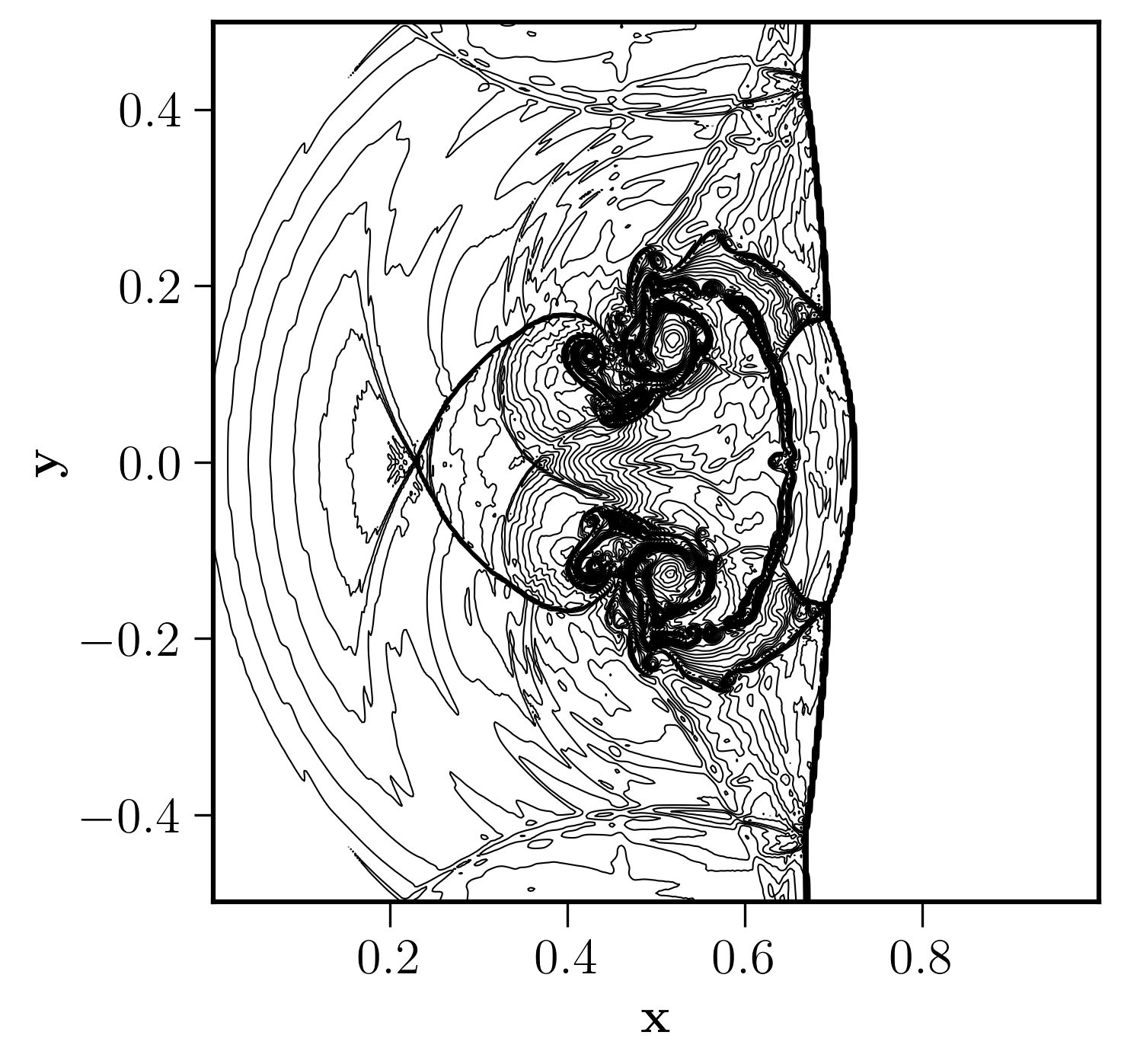}
  \caption{\textcolor{black}{WA-CR}}
  \label{fig:sbi_c}
\end{subfigure}
\begin{subfigure}{0.35\textwidth}
  \includegraphics[width=\textwidth]{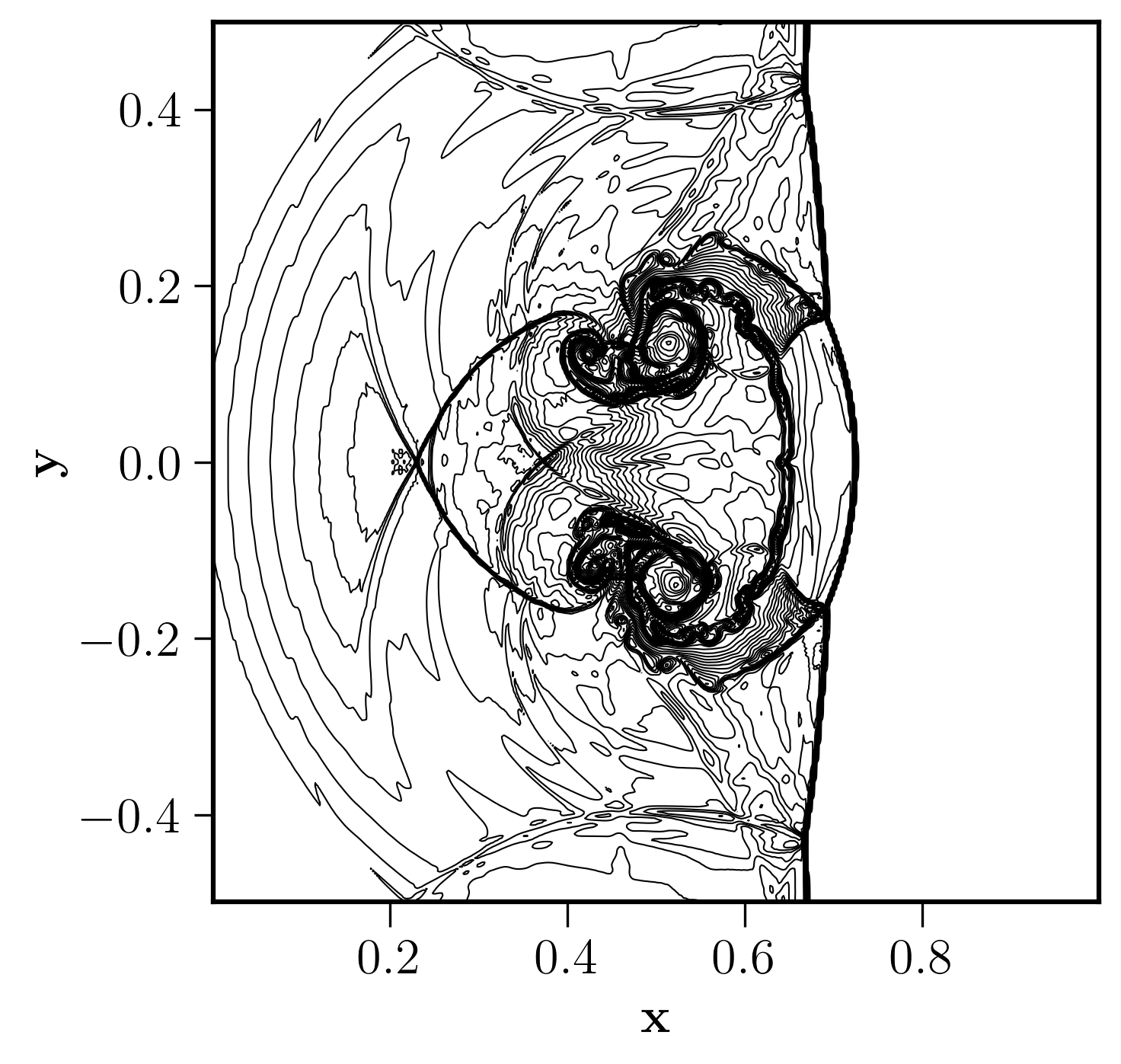}
  \caption{\textcolor{black}{WA-WENO-CR}}
  \label{fig:sbi_d}
\end{subfigure}
\caption{Shock-bubble interaction ($\mathrm{Ma}=6$ shock, helium bubble,
$400\times400$ grid, Sec.~\ref{sec:sbi}): density contours at the final time. WA-WENO-CR confirms the limiter-agnostic property of the rank-1 correction.}
\label{fig:sbi}
\end{figure}

The shock-bubble interaction tests the combined handling of a strong shock, a material interface, and the Richtmyer-Meshkov roll-up vortices that develop at the bubble boundary. A Mach~6 shock in air interacts with a helium bubble of radius $0.15$ centred at $(0.25,0)$ in the domain $[0,1]\times[-0.5,0.5]$:
\begin{equation}\label{eq:sb_ic}
(\rho,u,v,p)=\begin{cases}
(1.0,\;-3,\;0,\;1), & \text{pre-shocked air},\\
\bigl(\tfrac{216}{41},\;\tfrac{1645}{286}-3,\;0,\;\tfrac{251}{6}\bigr),
& \text{post-shocked air},\\
(0.138,\;-3,\;0,\;1), & \text{helium bubble}.
\end{cases}
\end{equation}
Simulations use a $400\times400$ grid and the HLL Riemann solver. Inflow and outflow conditions are imposed at the left and right boundaries; Neumann conditions apply elsewhere. Figure~\ref{fig:sbi} shows density contours at the final time, $t=0.15$.

This configuration tests the algorithm for both reconstruction paths simultaneously: the Ducros sensor activates in the shock region, where the full characteristic path is used, while the rank-1 entropy wave correction handles the helium/air material interface in the smooth post-shock region. WA-3 captures the primary shock and broad bubble deformation correctly. However, it resolves less fine-scale Richtmyer-Meshkov roll-up at the bubble boundary than the fifth-order schemes. By comparison, WA-5 and WA-CR produce results with sharp vortical structures at the interface, confirming that the rank-1 correction accurately handles the material interface without degrading shock-region accuracy. Likewise, WA-WENO-CR closely matches WA-CR, again confirming the limiter-agnostic property of the correction.

\subsection{Viscous shock tube}
\label{sec:vst}

\begin{figure}[H]
\centering
\begin{subfigure}{0.45\textwidth}
  \includegraphics[width=\textwidth]{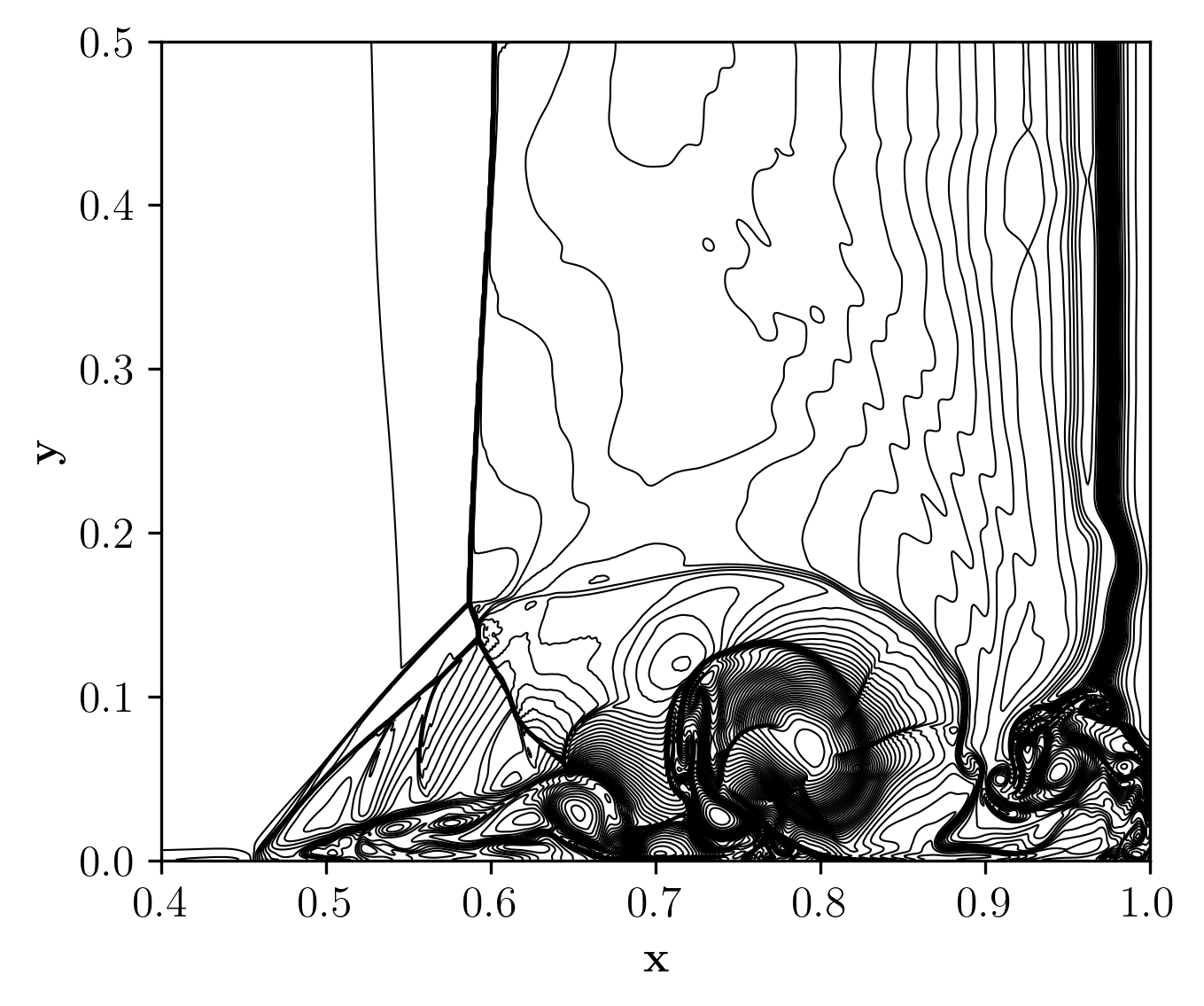}
  \caption{WA-3}
  \label{fig:vst1000_a}
\end{subfigure}%
\begin{subfigure}{0.45\textwidth}
  \includegraphics[width=\textwidth]{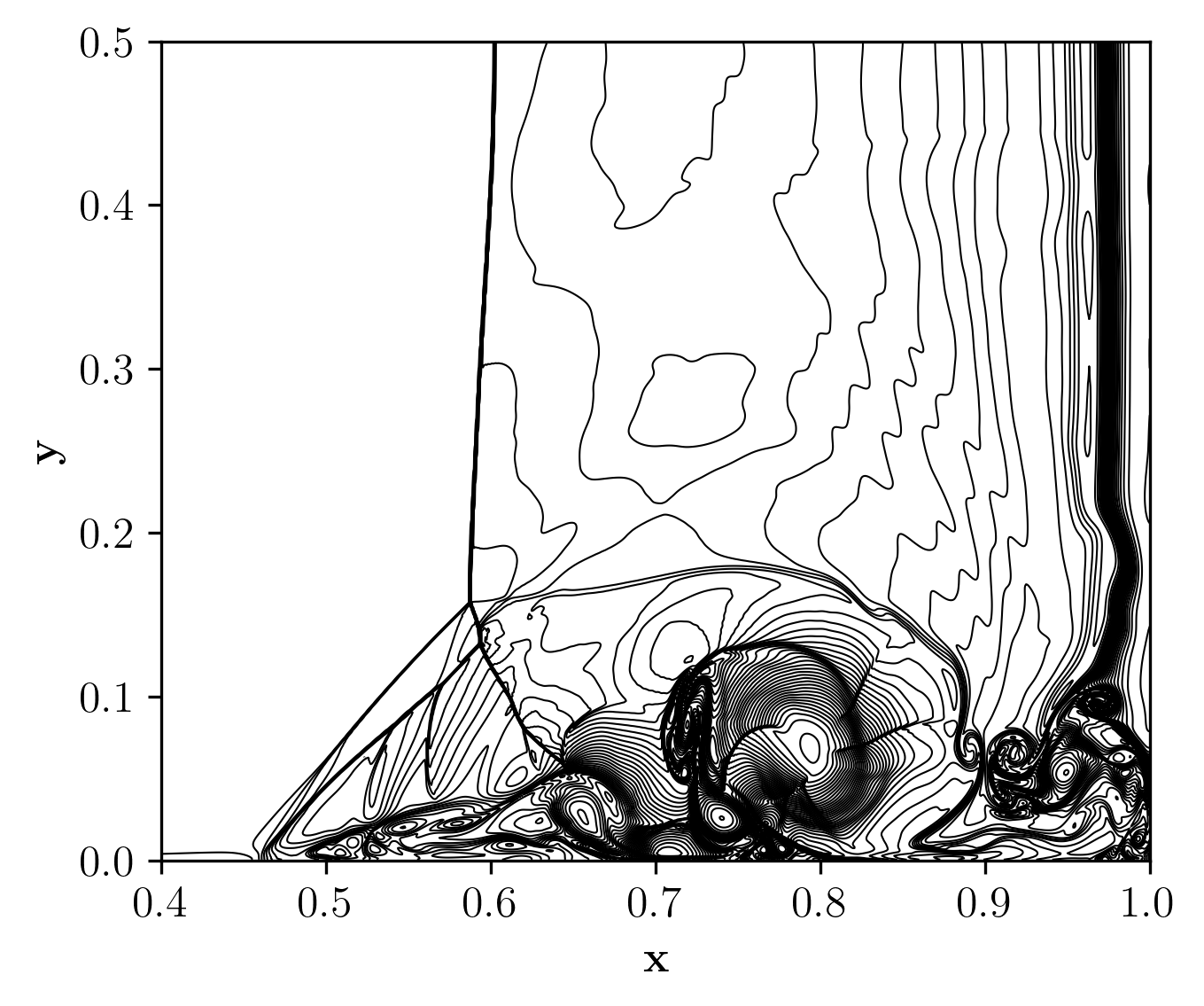}
  \caption{WA-5}
  \label{fig:vst1000_b}
\end{subfigure}
\begin{subfigure}{0.45\textwidth}
  \includegraphics[width=\textwidth]{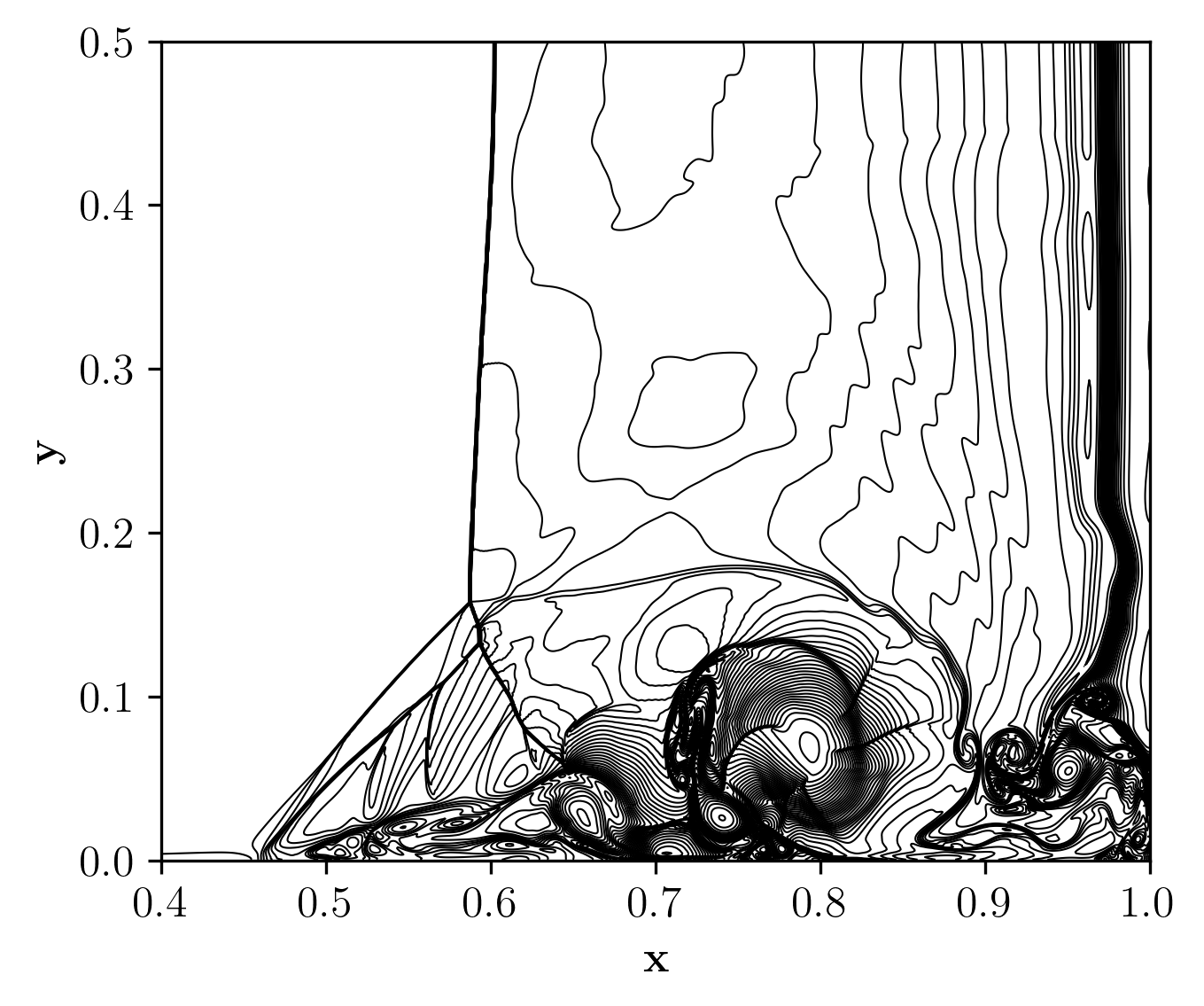}
  \caption{WA-CR}
  \label{fig:vst1000_c}
\end{subfigure}%
\begin{subfigure}{0.45\textwidth}
  \includegraphics[width=\textwidth]{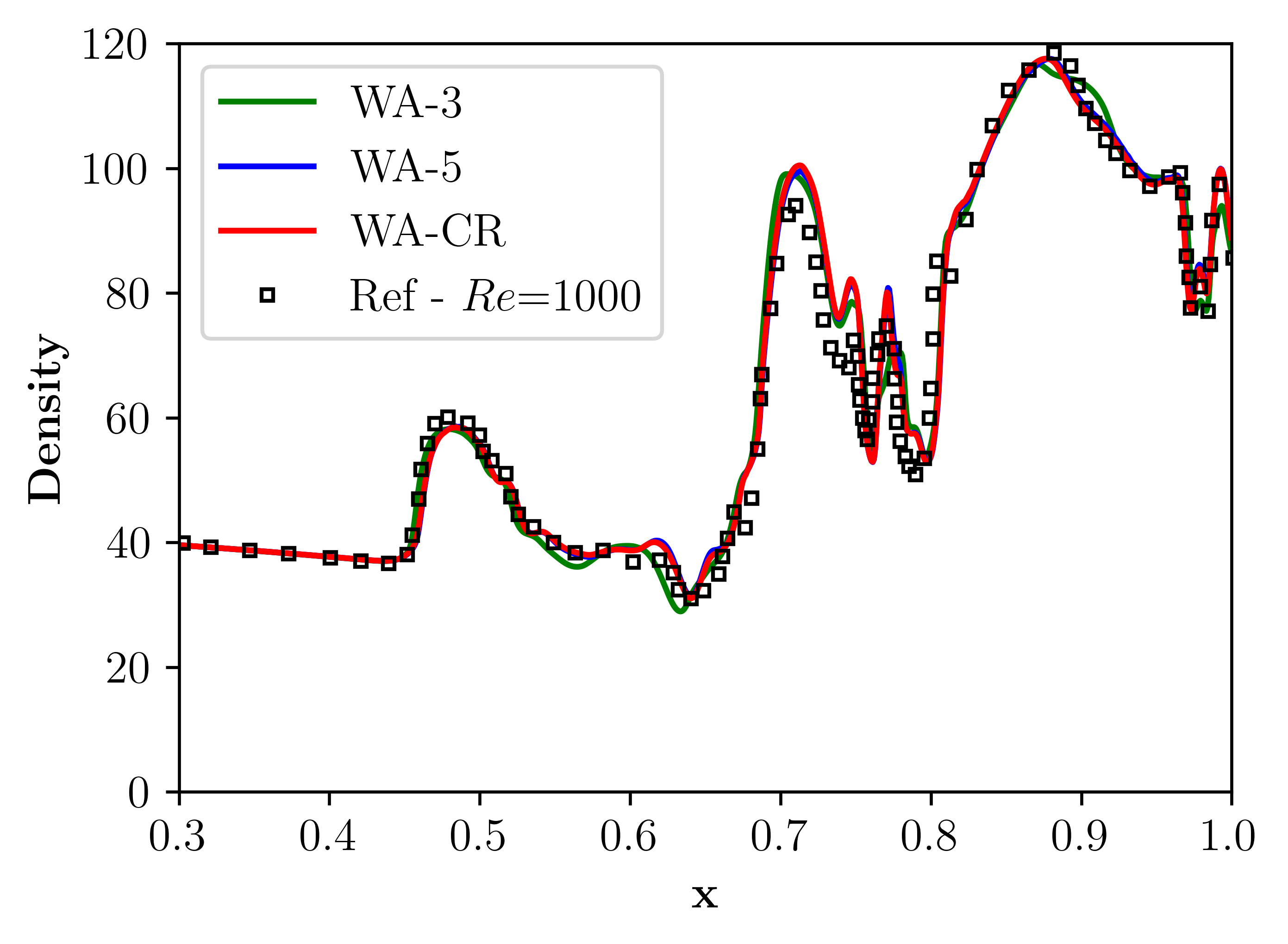}
  \caption{Wall density profile, $Re=1000$}
  \label{fig:vst1000_d}
\end{subfigure}
\caption{Viscous shock tube ($Re=1000$, $1280\times640$ grid, $t=1$, Sec.~\ref{sec:vst}): flow-field density contours and density profile along the wall ($y=0$). WA-CR matches WA-5.}
\label{fig:vst_re1000}
\end{figure}

The viscous shock-tube problem of Daru and Tenaud~\cite{daru2009numerical} involves the propagation of a Mach~2.37 shock wave and contact discontinuity that form a thin boundary layer at the bottom wall. The shock-boundary layer interaction produces a complex vortex system, a separation region, and a lambda-shaped shock pattern, making it a demanding test for schemes that must handle both viscous boundary layers and strong shocks simultaneously \cite{chamarthi2022}. The initial conditions are
\begin{equation}\label{eq:vst_ic}
(\rho,u,v,p)=\begin{cases}
(120,\;0,\;0,\;120/\gamma), & 0<x<0.5,\\
(1.2,\;0,\;0,\;1.2/\gamma), & 0.5\leq x<1,
\end{cases}
\end{equation}
and the simulation runs to $t_f=1$. Two Reynolds numbers are considered: $Re=1000$ on a $1280\times640$ grid and $Re=2500$ on a $2500\times1250$ grid. For reference, Kundu et al.~\cite{kundu2021investigation} used 109 million cells for the $Re=2500$ case; the present results are obtained on a grid nearly 35 times coarser. Figure~\ref{fig:vst_re1000} shows flow-field contours and the wall-normal density profile at $Re=1000$.

WA-3 captures the primary shock structure and the lambda-shock pattern correctly, but provides less fine-scale vortical detail in the separation region compared to the fifth-order schemes. WA-5 and WA-CR yield nearly identical contours, both offering sharper vortical structures and better-resolved secondary shock features. The wall-normal density profile in panel (d) demonstrates that all three schemes match the reference solution well, with WA-5 and WA-CR overlapping and WA-3 displaying slight differences at the density peaks. This case relies heavily on the full characteristic reconstruction path due to the strong shock and lambda structure, whereas the rank-1 correction consistently addresses the contact discontinuity. Notably, the shock wave thickness differs between WA-3 and WA-5, with WA-5 producing a thinner shock, indicating lower dissipation.

\begin{figure}[H]
\centering
\begin{subfigure}{0.45\textwidth}
  \includegraphics[width=\textwidth]{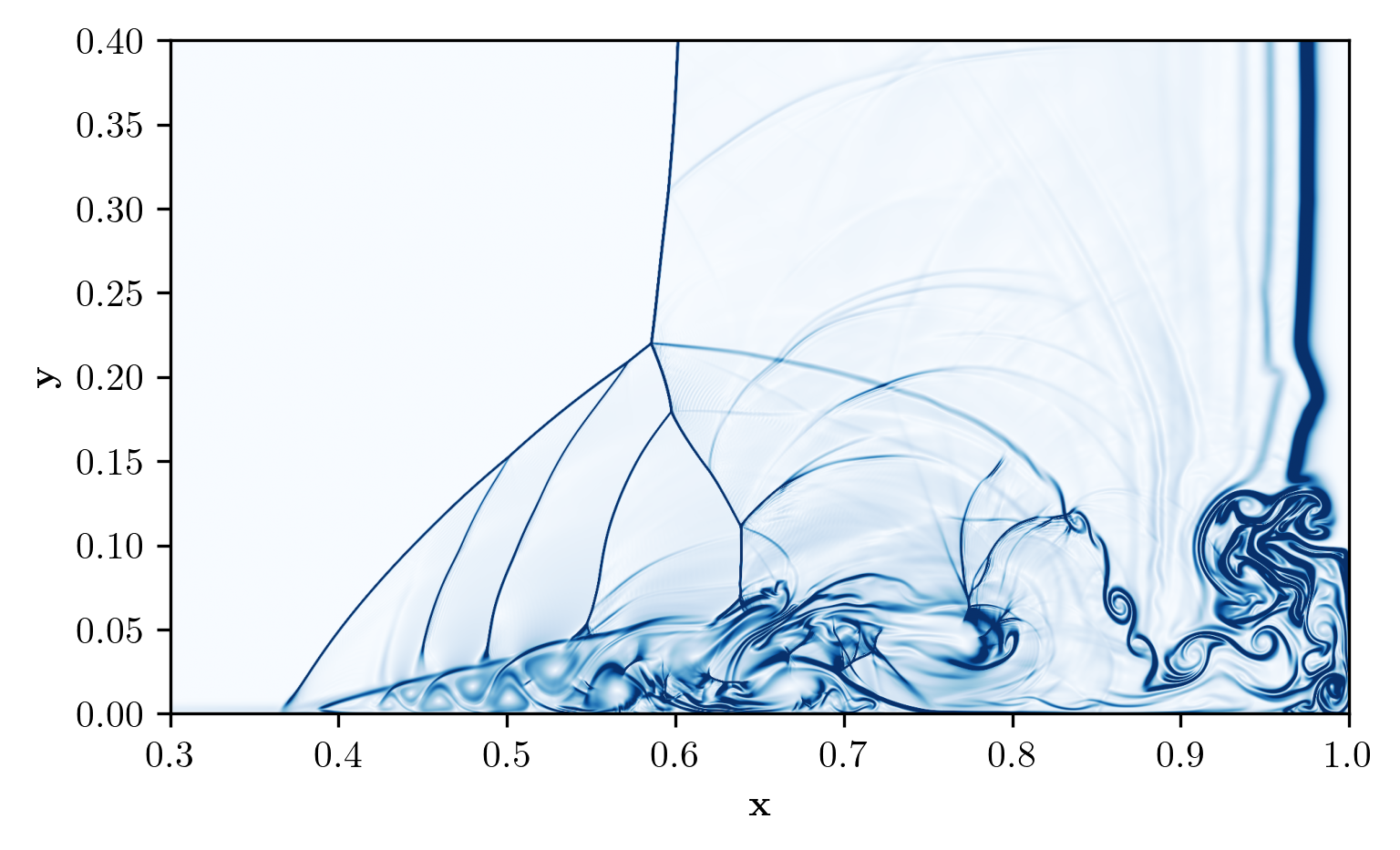}
  \caption{WA-3}
  \label{fig:vst2500_a}
\end{subfigure}%
\begin{subfigure}{0.45\textwidth}
  \includegraphics[width=\textwidth]{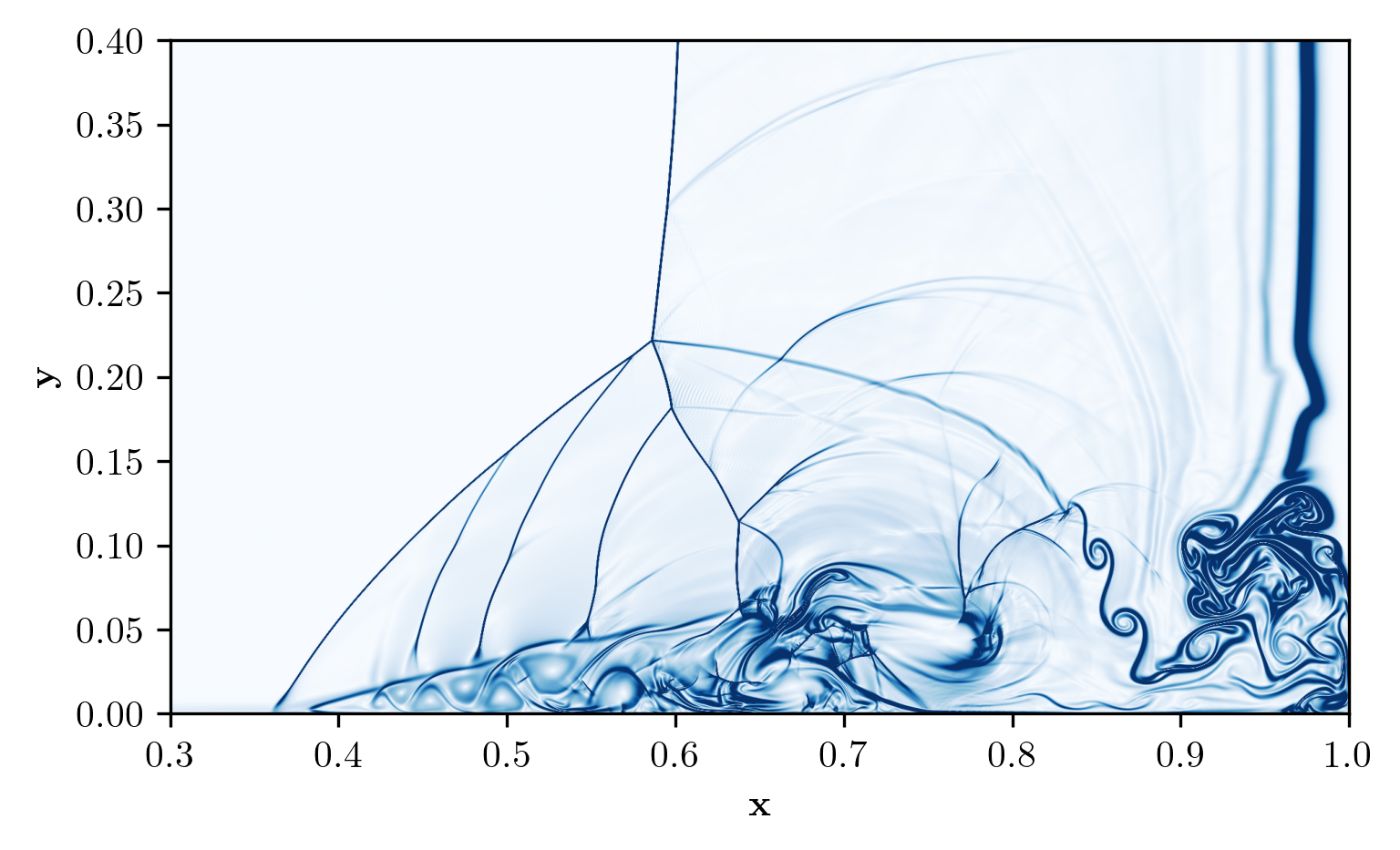}
  \caption{WA-5}
  \label{fig:vst2500_b}
\end{subfigure}
\begin{subfigure}{0.45\textwidth}
  \includegraphics[width=\textwidth]{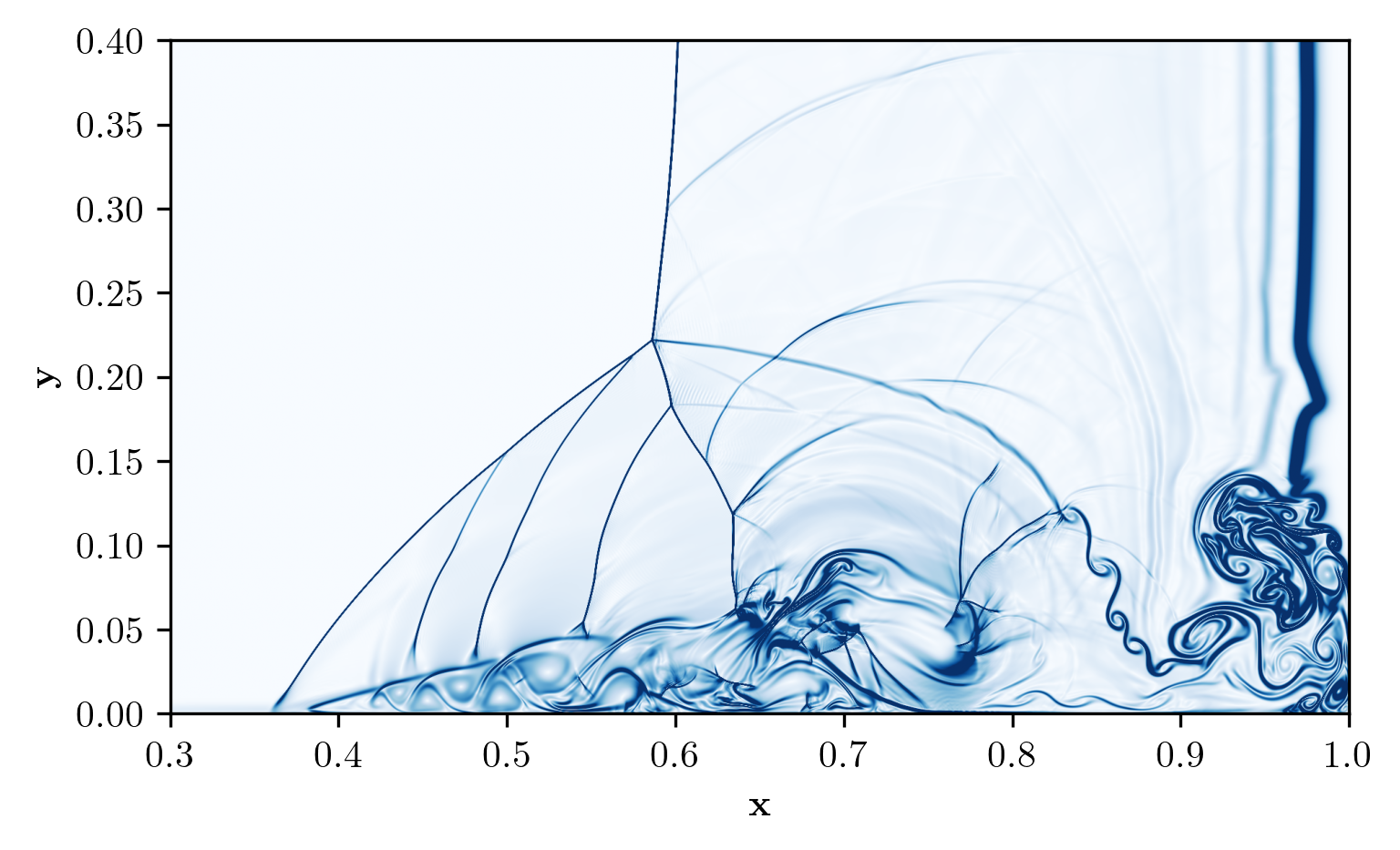}
  \caption{WA-CR}
  \label{fig:vst2500_c}
\end{subfigure}%
\begin{subfigure}{0.45\textwidth}
  \includegraphics[width=\textwidth]{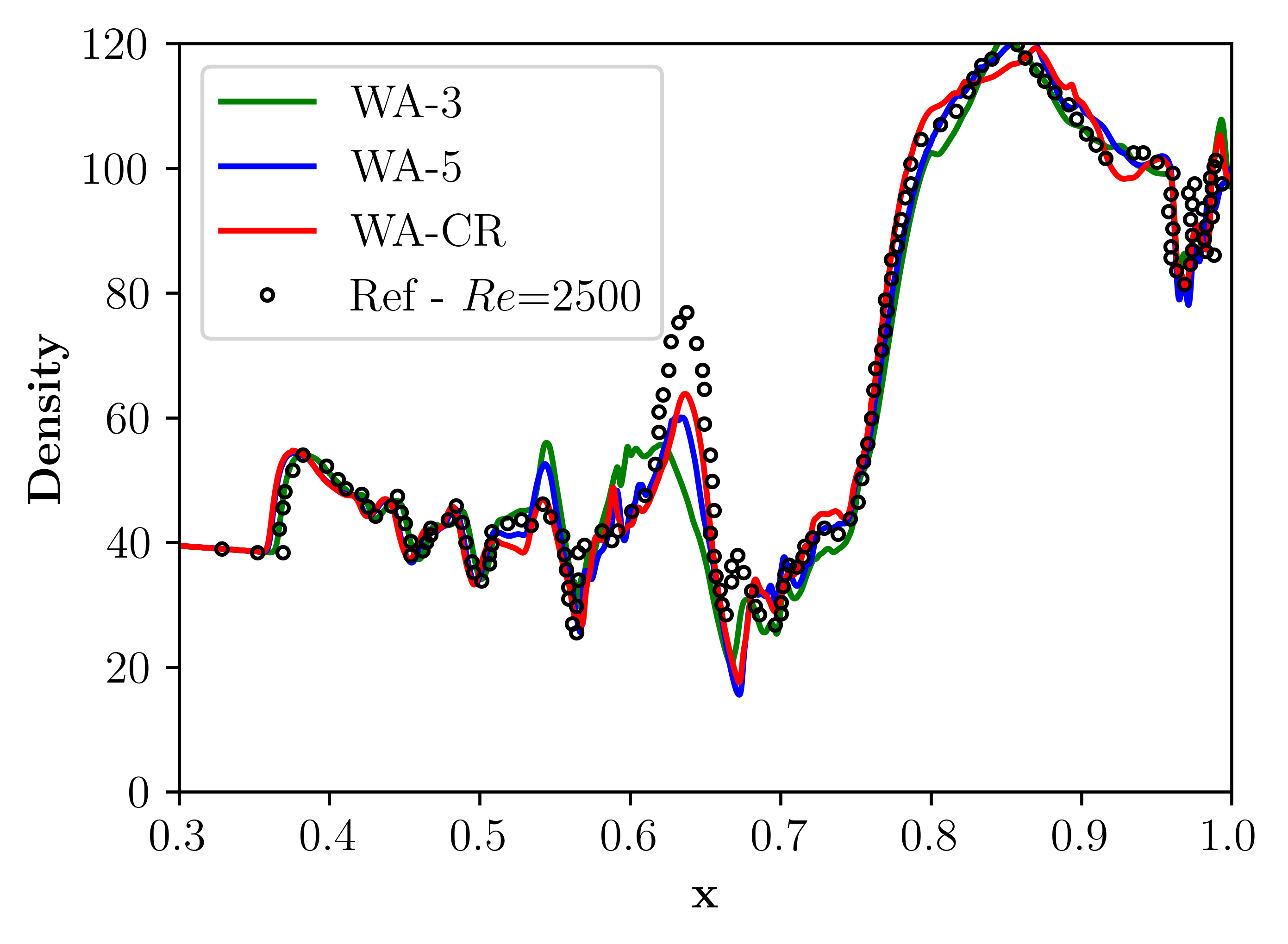}
  \caption{Wall density profile, $Re=2500$}
  \label{fig:vst2500_d}
\end{subfigure}
\caption{Viscous shock tube ($Re=2500$, $2500\times1250$ grid, $t=1$, Sec.~\ref{sec:vst}):
flow-field density contours and density profile along the wall ($y=0$).
WA-5 and WA-CR resolve finer vortical structures in the separation
region than WA-3 at this higher Reynolds number. WA-CR matches WA-5
in accuracy.}
\label{fig:vst_re2500}
\end{figure}

Figure~\ref{fig:vst_re2500} shows results at the higher Reynolds number $Re=2500$. Under these conditions, the separation region contains finer vortical structures that are more sensitive to numerical dissipation. WA-3 captures the primary lambda shock and the broad separation bubble correctly, but it resolves the fine-scale Kelvin-Helmholtz roll-up in the shear layer near $x=0.8$--$1.0$ less clearly than WA-5 or WA-CR. Both WA-5 and WA-CR resolve these structures clearly and are visually indistinguishable. This confirms that the rank-1 correction maintains full accuracy at higher Reynolds numbers, where the flow structures are finer, and the contact discontinuity is sharper. The wall-normal density profile in panel (d) shows that all three schemes closely match the reference, with WA-5 and WA-CR overlapping throughout and WA-3 exhibiting minor deviations at the density peaks.

\begin{figure}[H]
\centering
\begin{subfigure}{0.48\textwidth}
  \includegraphics[width=\textwidth]{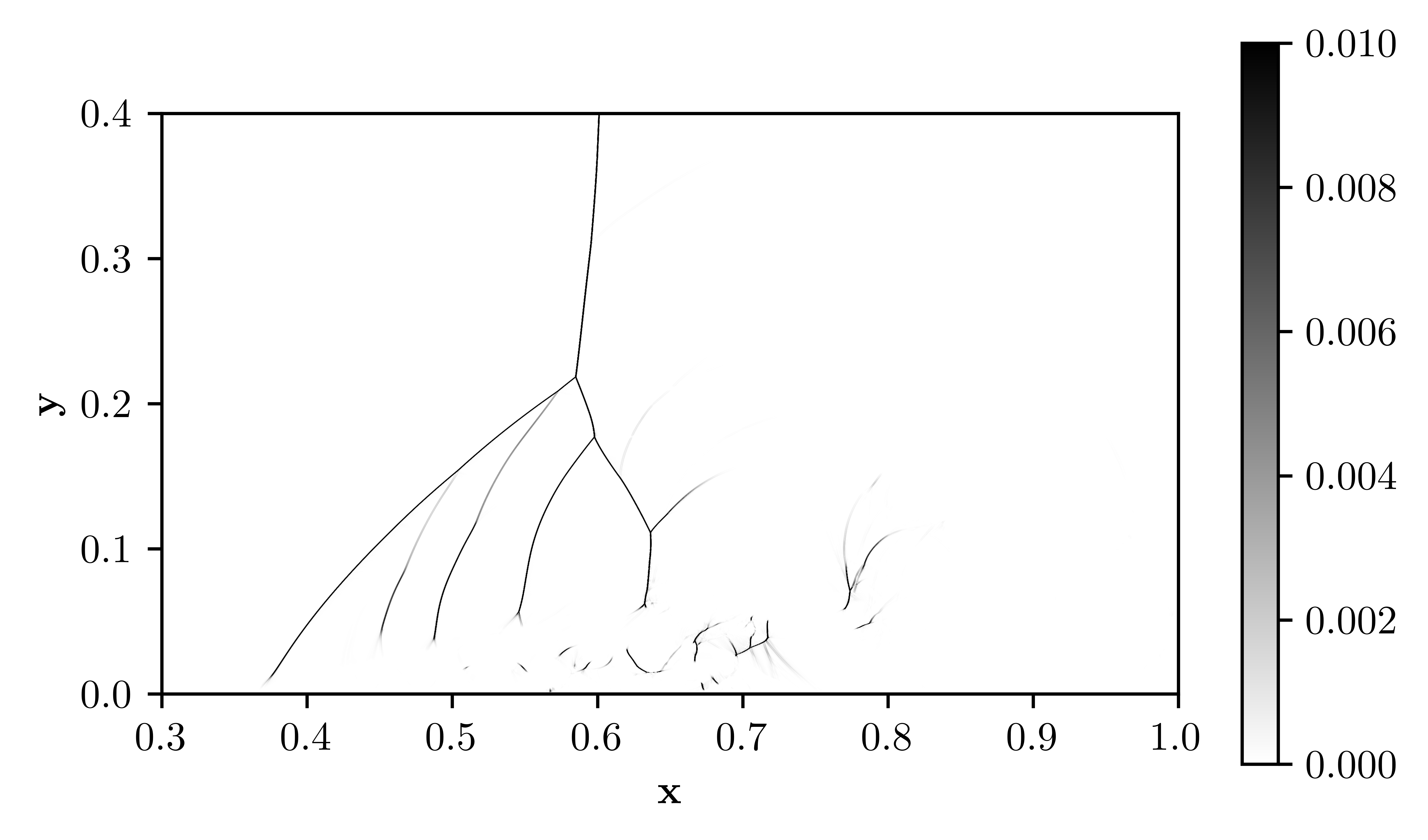}
  \caption{Ducros sensor, $x$-direction}
  \label{fig:vst_duc_x}
\end{subfigure}
\begin{subfigure}{0.48\textwidth}
  \includegraphics[width=\textwidth]{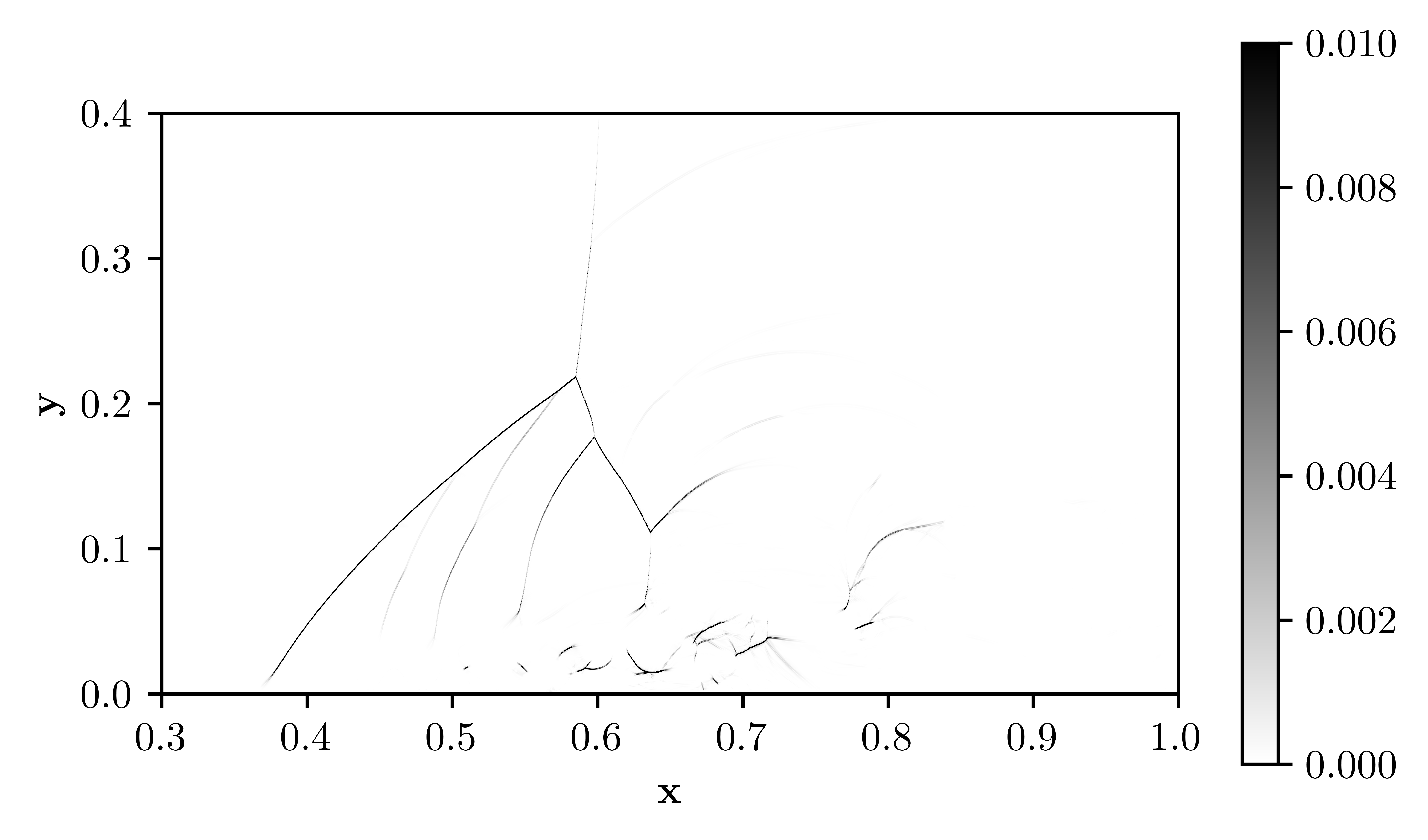}
  \caption{Ducros sensor, $y$-direction}
  \label{fig:vst_duc_y}
\end{subfigure}
\begin{subfigure}{0.5\textwidth}
  \includegraphics[width=\textwidth]{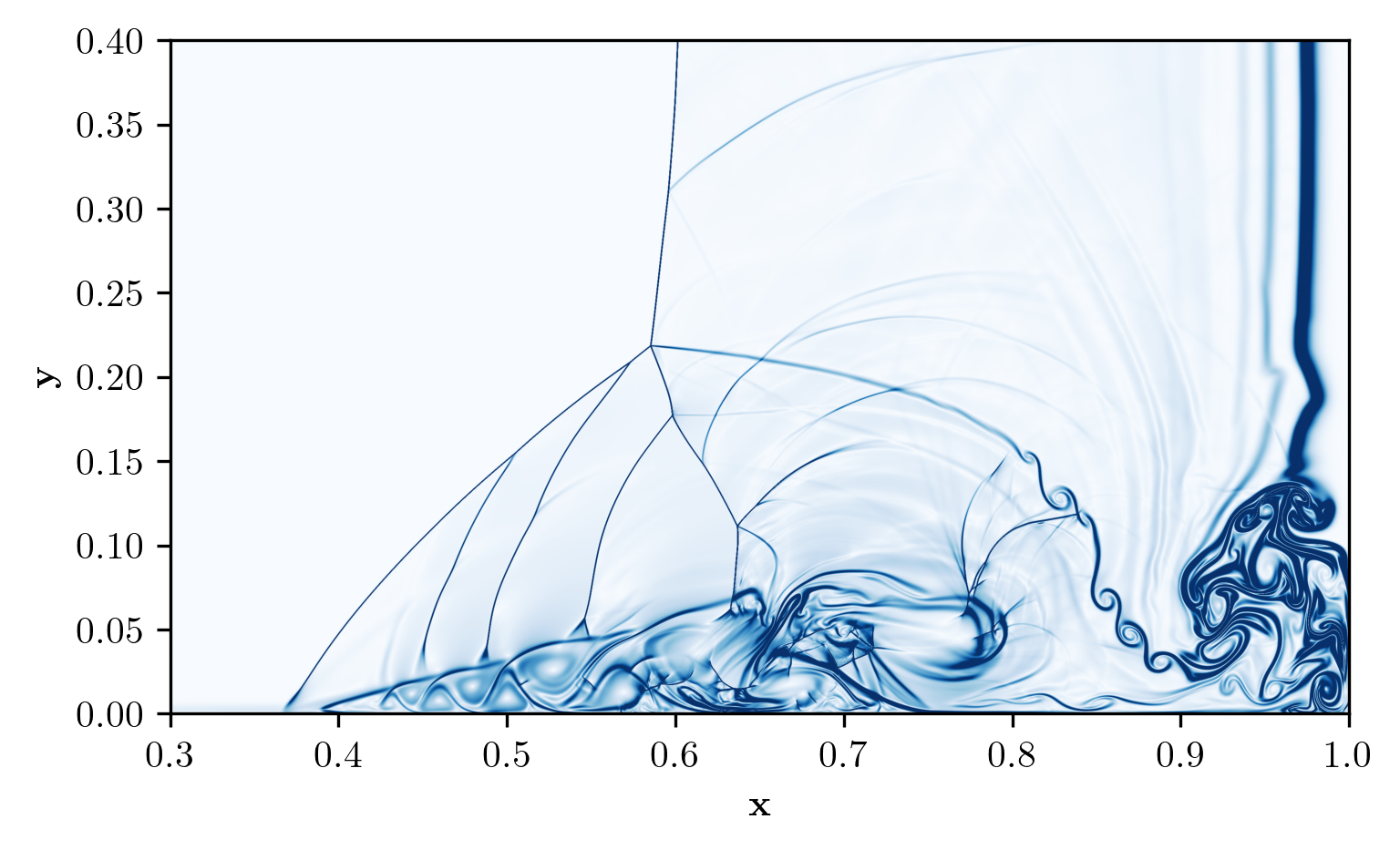}
  \caption{Fine-grid reference, WA-CR}
  \label{fig:vst_fine}
\end{subfigure}
\caption{Viscous shock tube ($Re=2500$), Sec.~\ref{sec:vst}: Ducros sensor fields in the
(a) stream-wise and (b) wall-normal directions, and (c) fine-grid
reference solution computed with WA-CR. The sensor activates only
in the shock and shock/boundary-layer interaction regions; the
remainder of the domain uses the conservative reconstruction path,
accounting for the cost reduction of WA-CR relative to WA-5.}
\label{fig:vst_ducros}
\end{figure}

Figure~\ref{fig:vst_ducros} shows the Ducros sensor fields in the stream-wise and wall-normal directions. These are presented alongside a fine-grid reference solution, $4000\times2000$. The sensor activates only in the narrow shock and shock/boundary-layer interaction regions. The bulk of the boundary layer, the separation zone, and the subsonic region are handled entirely by the conservative reconstruction path with rank-1 entropy wave correction. This localization explains the wall-time reduction reported in Table~\ref{tab:cost}. The characteristic path is invoked at only a small fraction of the domain interfaces. As noted in Remark~\ref{rem:ducros}, the complete two-component formulation of Ducros et al.~\cite{ducros1999large} is used throughout this work:
\begin{equation}
    \Omega_d = \underbrace{\frac{\left| -p_{i-2} + 16 p_{i-1} - 30 p_{i} 
    + 16 p_{i+1} - p_{i+2} \right|}{\left| p_{i-2} + 16 p_{i-1} + 30 p_{i} 
    + 16 p_{i+1} + p_{i+2} \right|}}_{\text{pressure sensor}} 
    \cdot
    \underbrace{\frac{(\nabla\cdot\mathbf{u})^2}
    {(\nabla\cdot\mathbf{u})^2+|\nabla\times\mathbf{u}|^2}}_{\text{dilatation-to-vorticity ratio}}.
\end{equation}
As discussed in Section~\ref{sec:intro}, some implementations in the literature employ only the dilatation-to-vorticity component. Feng et al.~\cite{feng2024general} are among those who omit the pressure-based indicator. Sciacovelli et al.~\cite{sciacovelli2021assessment} demonstrated that the dilatation-to-vorticity ratio alone can misidentify intense vortical regions as shocked. In contrast, the combined formulation provides reliable localization of genuine shock regions, as shown in Figure~\ref{fig:vst_ducros}.

\subsection{Computational cost summary}

Table~\ref{tab:cost} summarises the wall time reduction achieved by WA-CR relative to WA-5 across all test cases. Simulations are carried out on an Apple Mac Mini (M1 chip) using the Intel Fortran compiler (ifort) 2021 under Rosetta 2 emulation. All runs are single-threaded. The reported times reflect serial wall-clock performance. The savings range from 29\% to 41\% and correlate with the fraction of interfaces where the Ducros sensor is inactive. In these cases, the conservative variable reconstruction is used. The largest reductions occur in the shock-bubble interaction (41\%) and the 2-D Riemann problem (36\%), where the shock occupies only a small fraction of the domain. Here, the conservative path handles the bulk of the interfaces. The double shear layer (29\%) and the double Mach reflection (31\%) show smaller but still substantial savings. The supersonic Taylor-Green vortex (35\%) falls between these, as shocks develop and interact throughout the domain, but occupy only a minority of interfaces at any given time. In all cases, the cost reduction comes at no accuracy penalty. This is demonstrated by the results in the preceding sections. \textcolor{black}{The timings are reported for the hardware available to the author, and the exact speedup should therefore be interpreted as hardware dependent. The same reduction in characteristic projections may also benefit GPU implementations, although the magnitude of the gain will depend on the specific implementation and hardware. A dedicated GPU performance study is beyond the scope of the present paper.}

\begin{table}[H]
\centering
\caption{Computational wall time for WA-5 and WA-CR across all test
cases. The cost reduction reflects the fraction of interfaces where
the conservative variable reconstruction replaces the full characteristic
projection.}
\label{tab:cost}
\begin{tabular}{lccc}
\hline
Test case & WA-5 & WA-CR & Cost reduction \\
\hline
Taylor-Green vortex ($\mathrm{Ma}=1.25$) & 14{,}251\,s & 9{,}237\,s  & $\approx35\%$ \\
Double shear layer                        &    743\,s   &   531\,s    & $\approx29\%$ \\
Rayleigh-Taylor instability               &    210\,s   &   145\,s    & $\approx31\%$ \\
Double Mach reflection                    &    718\,s   &   499\,s    & $\approx31\%$ \\
2-D Riemann problem                       & 1{,}201\,s  &   763\,s    & $\approx36\%$ \\
Shock-bubble interaction                  &    424\,s   &   251\,s    & $\approx41\%$ \\
Viscous shock tube ($Re=1000$)            & 36{,}571\,s & 24{,}532\,s & $\approx34\%$ \\
\hline
\end{tabular}
\end{table}

\section{Conclusions \textcolor{black}{and future work}}
\label{sec:conclusions}

This paper identifies \textcolor{black}{an optimized} acoustic upwind bias that yields stable, accurate results across subsonic to hypersonic flow regimes in the wave-appropriate reconstruction framework and introduces a cheaper, conservative-characteristic reconstruction algorithm with a rank-1 entropy wave correction.

\begin{itemize}

\item The acoustic upwind bias $\eta_a$, previously fixed at $1.0$ as a conservative default~\cite{chamarthi2023wave,hoffmann2024centralized}, is identified as the sole free parameter in the wave-appropriate framework once all other dissipation sources are eliminated by design. A physics-constrained scalar optimization using Brent's method converges in approximately 25 CFD evaluations, far fewer than multi-parameter Bayesian frameworks, because the wave-appropriate design reduces the problem to a bounded scalar minimization.

\item The optimization yields $\eta_a^*=0.54$ for WA-3 and $\eta_a^*=0.6010$ for WA-5. \textcolor{black}{These values minimize the stated kinetic-energy objective among the evaluated feasible trials and transfer without retuning across the tested range, from subsonic turbulence to hypersonic flows, including discontinuities. They should not be interpreted as a separate proof of the smallest possible stable value of $\eta_a$.} The optimized nonlinear $N$th-order scheme consistently matches or outperforms the standard linear $(N\!+\!2)$th-order scheme at full acoustic upwinding, at no additional computational cost.

\item The prior conservative-characteristic scheme~\cite{chamarthi2024generalized} required two sensors, a Ducros shock sensor and a density-based contact detector, to switch between reconstruction paths. It is shown that the contact detector is unnecessary. Near a contact discontinuity, the deficiency of a conservative variable reconstruction is entirely a rank-1 perturbation in the entropy characteristic amplitude $C_2$, correctable by a scalar entropy projection, one MP5 pass, and a rank-1 update of five multiply-add operations per interface side.

\item The resulting algorithm, WA-CR, reduces wall time by 29 to 41\% relative to the full characteristic scheme across various benchmark configurations while matching or exceeding its accuracy. The contact is handled structurally rather than by detection, eliminating the need for a sensor that would otherwise require empirical calibration. The rank-1 correction is limiter-agnostic, as the WENO variant WA-WENO-CR also produces results without oscillations for cases with contact discontinuities.

\item The wave-appropriate principle is not restricted to reconstruction-based schemes. Applied to the KEP scheme, where the base discretization provides zero dissipation for all fluxes, a controlled acoustic upwind bias of $\eta_a = 0.56$ introduced exclusively through the normal-momentum component of the dissipative flux suppresses the spurious vortices observed in the unmodified KEP scheme. The mechanism is direction-dependent dissipation of normal momentum without any eigenvector projection. \textcolor{black}{The value $\eta_a = 0.56$ is consistent with the optimized values for reconstruction-based schemes, but it is used here as an illustrative stabilizing choice rather than a separately optimized universal parameter.}
\end{itemize}

It is important to note that all the algorithms presented in this paper require \textbf{conservative} variable reconstruction, which was also the choice of reconstruction in \cite{chamarthi2023wave}. The wave-appropriate framework and the rank-1 entropy correction are derived from the characteristic structure of the conservative-variable system and do not apply to primitive-variable reconstruction (see Ref.~\cite{hoffmann2024centralized}, where it has been shown that there will be significant differences between results obtained by conservative and primitive variable reconstruction). If pressure must be considered in place of total energy, then the recommended variable set is $[\rho,\ \rho u,\ \rho v,\ p]$; this means the reconstruction operates directly on momentum components ($\rho u,\ \rho v$) and not on velocity components ($u,v$), which would be the case in primitive variable reconstruction.

WA-5, WA-CR, and WA-WENO-CR with $\eta_a^*=0.6010$ are the recommended schemes. The framework can be straightforwardly extended to curvilinear coordinates, and the rank-1 correction principle may apply to any scheme employing conservative-characteristic reconstruction.

\subsection{\textcolor{black}{Implications of the results and future work}}

\textcolor{black}{The spectral diagnostics included in this work should be interpreted as a complement to the nonlinear CFD tests, not as a replacement for them. Fourier analysis is useful because it shows the local modified-wavenumber effect of reducing $\eta_a$: the dispersive behavior of the underlying upwind reconstruction is largely retained, while the dissipative branch is reduced. However, the modified-wavenumber result by itself does not determine the behavior of the full nonlinear shock-capturing solver. The Taylor--Green vortex calculations and the $64^3$ and $128^3$ energy spectra are therefore used to show how the reduced acoustic dissipation appears in an under-resolved turbulent calculation.}

\textcolor{black}{The results also clarify why nominal order alone is not a sufficient description of a numerical scheme. In the literature, schemes are often named by the stencil family, reconstruction polynomial, or one-dimensional flux formulation. This convention is useful, but the effective behavior of the complete multidimensional solver also depends on whether the method is implemented through flux reconstruction or finite-volume variable reconstruction, how fluxes are integrated, how nonlinear limiting is activated, and which characteristic wave families receive dissipation. The comparisons in this paper therefore emphasize complete-solver behavior on under-resolved turbulent grids, rather than only the named reconstruction order.}

\textcolor{black}{The optimized values of $\eta_a$ should be viewed with their intended scope. They were obtained from Taylor--Green vortex tests and then verified across the benchmark problems considered in this paper. The resulting $\eta_a^*$ is therefore the optimal feasible value for the stated reference-matching objective, not a universal lower stability bound. In a more general formulation, the optimal acoustic bias may be flow- and space-dependent rather than a single constant. Wall-bounded turbulence, strongly anisotropic turbulence, and grid-stretched configurations therefore require separate analysis and are natural directions for future work. Any optimized numerical scheme inherits assumptions from the problem used for the optimization. This is true for CFD-objective optimization, Fourier-space dispersion/dissipation optimization, and data-driven training objectives. Recent ML-ILES work illustrates the same point: neural-network reconstruction weights trained on compressible decaying homogeneous isotropic turbulence can perform well for unseen trajectories and grid resolutions within that flow family, while extension to shock--turbulence interaction and wall-bounded flows remains a separate validation step~\cite{bezgin2025ml}.}

\textcolor{black}{This interpretation is also consistent with the classical modal viewpoint of Chu and Kovasznay~\cite{chu1958non}, who decomposed compressible disturbances into acoustic, vortical, and entropy modes, with nonlinear effects arising through interactions among these modes. The present method transfers this modal viewpoint to finite-volume reconstruction: the flow information is projected onto characteristic wave families, and the numerical dissipation is assigned according to the physical role of each family. Acoustic waves retain a controlled upwind bias for stability, shear/vortical waves are centralized in smooth regions to avoid unnecessary damping, and entropy/contact errors are handled through the entropy-wave treatment used by the corresponding algorithm, including the rank-1 entropy/contact correction in WA-CR, rather than by applying the same dissipation uniformly to all variables.}

\textcolor{black}{The same modal viewpoint also explains why wall-bounded extensions require separate validation. At a solid boundary, the imposed conditions act on the resultant conservative or primitive variables, not on independent acoustic, vortical, and entropy modes. Thus, even if the interior reconstruction is wave-aware, wall closures, grid stretching, and wall models can locally couple the modal components and alter the effective dissipation balance.}

\textcolor{black}{The distinction in the present framework is that the optimization is not applied blindly to a global stencil dissipation coefficient. The characteristic structure of the Euler equations is used first: the non-acoustic waves are assigned their dissipation level from physical arguments, and only the acoustic upwind bias is left as an optimization parameter. Thus, the relevant question is not only how much dissipation a stencil contains, but also which wave family receives that dissipation. In this sense, the present approach is physics-informed at the discretization-design level: the governing wave structure fixes the dominant dissipation-placement choices before optimization is used. Accordingly, future work may focus on theoretically characterizing the acoustic stability and accuracy tradeoff, potentially deriving a spatially varying $\eta_a$ from the interaction between acoustic waves, vortical structures, and discontinuities.}

\section*{Appendix}\label{appen}
A two-dimensional implementation of the solver is provided as supplementary material. It is implemented using the NVIDIA Warp \cite{macklin2022warp}. The code can be executed after installing the dependency via \texttt{pip install warp-lang}. It reproduces the results obtained by the WA-3 scheme, shown in the figure. \ref{fig:rp_a}.

\bibliographystyle{elsarticle-num}
\bibliography{lasty}

\begin{thebibliography}{10}
\expandafter\ifx\csname url\endcsname\relax
  \def\url#1{\texttt{#1}}\fi
\expandafter\ifx\csname urlprefix\endcsname\relax\def\urlprefix{URL }\fi
\expandafter\ifx\csname href\endcsname\relax
  \def\href#1#2{#2} \def\path#1{#1}\fi

\bibitem{chamarthi2023wave}
A.~S. Chamarthi, N.~Hoffmann, S.~Frankel, A wave appropriate discontinuity
  sensor approach for compressible flows, Physics of Fluids 35~(6) (2023).

\bibitem{hoffmann2024centralized}
N.~Hoffmann, A.~S. Chamarthi, S.~H. Frankel, Centralized gradient-based
  reconstruction for wall modelled large eddy simulations of hypersonic
  boundary layer transition, Journal of Computational Physics (2024) 113128.

\bibitem{chamarthi2025wave}
A.~S. Chamarthi, Wave-appropriate multidimensional upwinding approach for
  compressible multiphase flows, Journal of Computational Physics 538 (2025)
  114157.

\bibitem{chamarthi2024generalized}
A.~S. Chamarthi, A generalized adaptive central-upwind scheme for compressible
  flow simulations and preventing spurious vortices, arXiv preprint
  arXiv:2409.02340 (2024).

\bibitem{chamarthi2025physics}
A.~S. Chamarthi, Physics appropriate interface capturing reconstruction
  approach for viscous compressible multicomponent flows, Computers \& Fluids
  303 (2025) 106858.

\bibitem{ducros1999large}
F.~Ducros, V.~Ferrand, F.~Nicoud, C.~Weber, D.~Darracq, C.~Gacherieu,
  T.~Poinsot, Large-eddy simulation of the shock/turbulence interaction,
  Journal of Computational Physics 152~(2) (1999) 517--549.

\bibitem{sandham2014transitional}
N.~Sandham, E.~Schuelein, A.~Wagner, S.~Willems, J.~Steelant, Transitional
  shock-wave/boundary-layer interactions in hypersonic flow, Journal of Fluid
  Mechanics 752 (2014) 1--33.

\bibitem{xiao2005simple}
F.~Xiao, Y.~Honma, T.~Kono, A simple algebraic interface capturing scheme using
  hyperbolic tangent function, International Journal for Numerical Methods in
  Fluids 48~(9) (2005) 1023--1040.
\newblock \href {https://doi.org/10.1002/fld.975} {\path{doi:10.1002/fld.975}}.

\bibitem{xiao2011revisit}
F.~Xiao, S.~Ii, C.~Chen, Revisit to the thinc scheme: a simple algebraic vof
  algorithm, Journal of Computational Physics 230~(19) (2011) 7086--7092.

\bibitem{batchelor1967introduction}
G.~K. Batchelor, An introduction to fluid dynamics, Cambridge university press,
  1967.

\bibitem{meng2018numerical}
J.~C. Meng, T.~Colonius, Numerical simulation of the aerobreakup of a water
  droplet, Journal of Fluid Mechanics 835 (2018) 1108--1135.

\bibitem{de2020sharp}
F.~De~Vanna, F.~Picano, E.~Benini, A sharp-interface immersed boundary method
  for moving objects in compressible viscous flows, Computers \& Fluids 201
  (2020) 104415.

\bibitem{subbareddy2009fully}
P.~K. Subbareddy, G.~V. Candler, A fully discrete, kinetic energy consistent
  finite-volume scheme for compressible flows, Journal of Computational Physics
  228~(5) (2009) 1347--1364.

\bibitem{van2022immersed}
W.~van Noordt, S.~Ganju, C.~Brehm, An immersed boundary method for wall-modeled
  large-eddy simulation of turbulent high-mach-number flows, Journal of
  Computational Physics 470 (2022) 111583.

\bibitem{de2023uranos}
F.~De~Vanna, F.~Avanzi, M.~Cogo, S.~Sandrin, M.~Bettencourt, F.~Picano,
  E.~Benini, Uranos: A gpu accelerated navier-stokes solver for compressible
  wall-bounded flows, Computer Physics Communications 287 (2023) 108717.

\bibitem{ghate2023finite}
A.~Ghate, S.~K. Lele, Finite difference methods for turbulence simulations, in:
  Numerical Methods in Turbulence Simulation, Elsevier, 2023, pp. 235--284.

\bibitem{sciacovelli2021assessment}
L.~Sciacovelli, D.~Passiatore, P.~Cinnella, G.~Pascazio, Assessment of a
  high-order shock-capturing central-difference scheme for hypersonic turbulent
  flow simulations, Computers \& Fluids 230 (2021) 105134.

\bibitem{schranner2013physically}
F.~S. Schranner, X.~Y. Hu, N.~A. Adams, A physically consistent weakly
  compressible high-resolution approach to underresolved simulations of
  incompressible flows, Computers \& Fluids 86 (2013) 109--124.

\bibitem{schranner2016optimization}
F.~S. Schranner, V.~Rozov, N.~A. Adams, Optimization of an implicit large-eddy
  simulation method for underresolved incompressible flow simulations, AIAA
  Journal 54~(5) (2016) 1567--1577.

\bibitem{winter2016iterative}
J.~M. Winter, F.~S. Schranner, N.~A. Adams, Iterative bayesian optimization of
  an implicit les method for under-resolved simulations of incompressible
  flows, in: en. In: 10th International Symposium on Turbulence and Shear Flow
  Phenomena, TSFP, Vol.~10, 2016.

\bibitem{feng2022multi}
Y.~Feng, F.~S. Schranner, J.~Winter, N.~A. Adams, A multi-objective bayesian
  optimization environment for systematic design of numerical schemes for
  compressible flow, Journal of Computational Physics 468 (2022) 111477.

\bibitem{feng2024general}
Y.~Feng, J.~Winter, N.~A. Adams, F.~S. Schranner, A general multi-objective
  bayesian optimization framework for the design of hybrid schemes towards
  adaptive complex flow simulations, Journal of Computational Physics 510
  (2024) 113088.

\bibitem{feng2023deep}
Y.~Feng, F.~S. Schranner, J.~Winter, N.~A. Adams, A deep reinforcement learning
  framework for dynamic optimization of numerical schemes for compressible flow
  simulations, Journal of Computational Physics 493 (2023) 112436.

\bibitem{tam1993dispersion}
C.~K.~W. Tam, J.~C. Webb, Dispersion-relation-preserving finite difference
  schemes for computational acoustics, Journal of Computational Physics 107~(2)
  (1993) 262--281.
\newblock \href {https://doi.org/10.1006/jcph.1993.1142}
  {\path{doi:10.1006/jcph.1993.1142}}.

\bibitem{lele1992compact}
S.~K. Lele, Compact finite difference schemes with spectral-like resolution,
  Journal of Computational Physics 103~(1) (1992) 16--42.

\bibitem{pirozzoli2006spectral}
S.~Pirozzoli, On the spectral properties of shock-capturing schemes, Journal of
  Computational Physics 219~(2) (2006) 489--497.

\bibitem{li2022class}
Y.~Li, C.~Chen, Y.-X. Ren, A class of high-order finite difference schemes with
  minimized dispersion and adaptive dissipation for solving compressible flows,
  Journal of Computational Physics 448 (2022) 110770.

\bibitem{huang2023fivepoint}
H.~Huang, T.~Liang, L.~Fu, A five-point teno scheme with adaptive dissipation
  based on a new scale sensor (2023).
\newblock \href {http://arxiv.org/abs/2303.10020} {\path{arXiv:2303.10020}}.

\bibitem{bezgin2025ml}
D.~A. Bezgin, A.~B. Buhendwa, S.~J. Schmidt, N.~A. Adams, Ml-iles: End-to-end
  optimization of data-driven high-order godunov-type finite-volume schemes for
  compressible homogeneous isotropic turbulence, Journal of Computational
  Physics 522 (2025) 113560.

\bibitem{van2006upwind}
B.~Van~Leer, Upwind and high-resolution methods for compressible flow: From
  donor cell to residual-distribution schemes, in: 16th AIAA Computational
  Fluid Dynamics Conference, 2003, p. 3559.

\bibitem{roe1986discrete}
P.~L. Roe, Discrete models for the numerical analysis of time-dependent
  multidimensional gas dynamics, Journal of Computational Physics 63~(2) (1986)
  458--476.

\bibitem{chandrashekar2013kinetic}
P.~Chandrashekar, Kinetic energy preserving and entropy stable finite volume
  schemes for compressible euler and navier-stokes equations, Communications in
  Computational Physics 14~(5) (2013) 1252--1286.

\bibitem{deng2019fifth}
X.~Deng, Y.~Shimizu, F.~Xiao, A fifth-order shock capturing scheme with
  two-stage boundary variation diminishing algorithm, Journal of Computational
  Physics 386 (2019) 323--349.

\bibitem{fu2019low}
L.~Fu, A low-dissipation finite-volume method based on a new teno
  shock-capturing scheme, Computer Physics Communications 235 (2019) 25--39.

\bibitem{Nishikawa2010}
H.~Nishikawa, {Beyond Interface Gradient: A General Principle for Constructing
  Diffusion Schemes}, 40th Fluid Dynamics Conference and Exhibit (2010).

\bibitem{toro1994restoration}
E.~F. Toro, M.~Spruce, W.~Speares, Restoration of the contact surface in the
  hll-riemann solver, Shock waves 4~(1) (1994) 25--34.

\bibitem{harten1983upstream}
A.~Harten, P.~D. Lax, B.~v. Leer, On upstream differencing and godunov-type
  schemes for hyperbolic conservation laws, SIAM review 25~(1) (1983) 35--61.

\bibitem{chamarthi2023gradient}
A.~S. Chamarthi, Gradient based reconstruction: Inviscid and viscous flux
  discretizations, shock capturing, and its application to single and
  multicomponent flows, Computers \& Fluids 250 (2023) 105706.

\bibitem{chamarthi2023efficient}
A.~S. Chamarthi, Efficient high-order gradient-based reconstruction for
  compressible flows, Journal of Computational Physics 486 (2023) 112119.

\bibitem{chamarthi2023implicit}
A.~S. Chamarthi, N.~Hoffmann, H.~Nishikawa, S.~H. Frankel, Implicit gradients
  based conservative numerical scheme for compressible flows, Journal of
  Scientific Computing 95~(1) (2023) 17.

\bibitem{shu2006essentially}
C.-W. Shu, Essentially non-oscillatory and weighted essentially non-oscillatory
  schemes for hyperbolic conservation laws, in: Advanced Numerical
  Approximation of Nonlinear Hyperbolic Equations: Lectures given at the 2nd
  Session of the Centro Internazionale Matematico Estivo (CIME) held in
  Cetraro, Italy, June 23--28, 1997, Springer, 2006, pp. 325--432.

\bibitem{godunov1959}
S.~K. Godunov, A difference method for numerical calculation of discontinuous
  solutions of the equations of hydrodynamics, Matematicheskii Sbornik 89~(3)
  (1959) 271--306.

\bibitem{van1977towards}
B.~Van~Leer, Towards the ultimate conservative difference scheme. iv. a new
  approach to numerical convection, Journal of Computational Physics 23~(3)
  (1977) 276--299.

\bibitem{van1979towards}
B.~Van~Leer, Towards the ultimate conservative difference scheme. v. a
  second-order sequel to godunov's method, Journal of computational Physics
  32~(1) (1979) 101--136.

\bibitem{suresh1997accurate}
A.~Suresh, H.~Huynh, Accurate monotonicity-preserving schemes with runge-kutta
  time stepping, Journal of Computational Physics 136~(1) (1997) 83--99.

\bibitem{lusher2021assessment}
D.~J. Lusher, N.~D. Sandham, Assessment of low-dissipative shock-capturing
  schemes for the compressible taylor--green vortex, AIAA Journal 59~(2) (2021)
  533--545.

\bibitem{virtanen2020scipy}
P.~Virtanen, R.~Gommers, T.~E. Oliphant, M.~Haberland, T.~Reddy, D.~Cournapeau,
  E.~Burovski, P.~Peterson, W.~Weckesser, J.~Bright, et~al., Scipy 1.0:
  fundamental algorithms for scientific computing in python, Nature methods
  17~(3) (2020) 261--272.

\bibitem{Borges2008}
R.~Borges, M.~Carmona, B.~Costa, W.~S. Don, {An improved weighted essentially
  non-oscillatory scheme for hyperbolic conservation laws}, Journal of
  Computational Physics 227~(6) (2008) 3191--3211.

\bibitem{Jiang1995}
G.-S. Jiang, C.-W. Shu, {Efficient Implementation of Weighted ENO Schemes},
  Journal of Computational Physics 126~(126) (1995) 202--228.

\bibitem{brachet1983small}
M.~E. Brachet, D.~I. Meiron, S.~A. Orszag, B.~Nickel, R.~H. Morf, U.~Frisch,
  Small-scale structure of the taylor--green vortex, Journal of Fluid Mechanics
  130 (1983) 411--452.

\bibitem{hickel2006adaptive}
S.~Hickel, N.~A. Adams, J.~A. Domaradzki, An adaptive local deconvolution
  method for implicit les, Journal of Computational Physics 213~(1) (2006)
  413--436.

\bibitem{liang2024new}
T.~Liang, L.~Fu, A new high-order shock-capturing teno scheme combined with
  skew-symmetric-splitting method for compressible gas dynamics and turbulence
  simulation, Computer Physics Communications 302 (2024) 109236.

\bibitem{grinstein2007implicit}
F.~F. Grinstein, L.~G. Margolin, W.~J. Rider, Implicit large eddy simulation,
  Vol.~10, Cambridge university press Cambridge, 2007.

\bibitem{minion1997performance}
M.~L. Minion, D.~L. Brown, Performance of under-resolved two-dimensional
  incompressible flow simulations, ii, Journal of Computational Physics 138~(2)
  (1997) 734--765.

\bibitem{fleischmann2019numerical}
N.~Fleischmann, S.~Adami, N.~A. Adams, Numerical symmetry-preserving techniques
  for low-dissipation shock-capturing schemes, Computers \& Fluids 189 (2019)
  94--107.

\bibitem{toro2009riemann}
E.~Toro, Riemann Solvers and Numerical Methods for Fluid Dynamics: A Practical
  Introduction, Springer Berlin Heidelberg, 2009.

\bibitem{acker2016improved}
F.~Acker, R.~d.~R. Borges, B.~Costa, An improved weno-z scheme, Journal of
  Computational Physics 313 (2016) 726--753.

\bibitem{woodward1984numerical}
P.~Woodward, P.~Colella, The numerical simulation of two-dimensional fluid flow
  with strong shocks, Journal of Computational Physics 54~(1) (1984) 115--173.

\bibitem{schulz1993numerical}
C.~W. Schulz-Rinne, J.~P. Collins, H.~M. Glaz, Numerical solution of the
  riemann problem for two-dimensional gas dynamics, SIAM Journal on Scientific
  Computing 14~(6) (1993) 1394--1414.

\bibitem{daru2009numerical}
V.~Daru, C.~Tenaud, Numerical simulation of the viscous shock tube problem by
  using a high resolution monotonicity-preserving scheme, Computers \& Fluids
  38~(3) (2009) 664--676.

\bibitem{chamarthi2022}
A.~S. Chamarthi, S.~Bokor, S.~H. Frankel, On the importance of high-frequency
  damping in high-order conservative finite-difference schemes for viscous
  fluxes, Journal of Computational Physics (2022) 111195.

\bibitem{kundu2021investigation}
A.~Kundu, M.~Thangadurai, G.~Biswas, Investigation on shear layer instabilities
  and generation of vortices during shock wave and boundary layer interaction,
  Computers \& Fluids 224 (2021) 104966.

\bibitem{chu1958non}
B.-T. Chu, L.~S. Kov{\'a}sznay, Non-linear interactions in a viscous
  heat-conducting compressible gas, Journal of fluid mechanics 3~(5) (1958)
  494--514.

\bibitem{macklin2022warp}
M.~Macklin, Warp: A high-performance python framework for gpu simulation and
  graphics, in: NVIDIA GPU Technology Conference (GTC), Vol.~3, 2022.

\end{thebibliography}
\end{document}